\documentstyle[emulateapj,psfig]{article}

\makeatletter

\newenvironment{inlinefigure}{%
\def\@captype{figure}%
\noindent\begin{minipage}{0.999\linewidth}\begin{center}}
{\end{center}\end{minipage}\smallskip}
\makeatother

\newcommand{\hal}{H$\alpha$}%
\newcommand{\lya}{Ly$\alpha$}

\def\gs{\mathrel{\raise0.35ex\hbox{$\scriptstyle >$}\kern-0.6em
\lower0.40ex\hbox{{$\scriptstyle \sim$}}}}
\def\ls{\mathrel{\raise0.35ex\hbox{$\scriptstyle <$}\kern-0.6em
\lower0.40ex\hbox{{$\scriptstyle \sim$}}}}
\newcommand{\mum}{$\,\mu$m}

\newcommand{\ratio}{S$_{\rm 850 \mu m}$/S$_{\rm 1.4 GHz}$}
\newcommand{\msun}{{\rm\,M_\odot}}
\newcommand{\sfr}{{\rm\,M_\odot\,yr^{-1}}}
\newcommand{\lsun}{\,{\rm L}_\odot}
\newcommand{\td}{$T_{\rm d}$}
\newcommand{\fir}{$L_{\rm FIR}$}
\newcommand{\tir}{$L_{\rm TIR}$}

\lefthead{Chapman et al.}

\begin{document}

\title{A Redshift Survey of the Submillimeter Galaxy Population}
\author{S.\,C.\ Chapman,$\!$\altaffilmark{1}
A.\,W.\ Blain,$\!$\altaffilmark{1}
Ian Smail,$\!$\altaffilmark{2}
R.\,J.\ Ivison\altaffilmark{3,4}
}
\altaffiltext{1}{California Institute of Technology, Pasadena, CA,
91125, U.S.A.}
\altaffiltext{2}{Institute for Computational Cosmology, University of Durham, South Rd, Durham DH1 3LE, UK}
\altaffiltext{3}{Astronomy Technology Centre, Royal Observatory, Blackford Hill, Edinburgh EH9 3HJ, UK}
\altaffiltext{4}{Institute for Astronomy, University of Edinburgh, Blackford Hill, Edinburgh EH9 3HJ, UK}
\slugcomment{Submitted to Astrophysical Journal}

\begin{abstract} 
We have obtained spectroscopic redshifts using the Keck-I telescope for
a sample of 73 submillimeter (submm) galaxies, with a median 850\mum\
flux density of 5.7\,mJy, for which precise positions are available
through their faint radio emission.  The galaxies lie at redshifts out
to $z=3.6$, with a median redshift of 2.2 and an interquartile range
$z=1.7$--2.8.  Modeling a purely submm flux-limited sample, based on
the expected selection function for our radio-identified sample,
suggests a median redshift of 2.3 with a redshift distribution
remarkably similar to the optically- and radio-selected Quasars.  The
observed redshift distributions are similar for the AGN and starburst
sub-samples.  The median $R_{\rm AB}$=24.6 for the sample.  However,
the {\it dust-corrected} ultraviolet (UV) luminosities of the galaxies
rarely hint at their huge bolometric luminosities indicated by their
radio/submm emission, underestimating the true luminosity by a median
factor of $\sim100$ for SMGs with pure starburst spectra.  Radio and
submm observations are thus essential to select the most luminous,
high-redshift galaxies.  The 850\mum, radio, and redshift data is used
to estimate the dust temperatures, and characterize photometric
redshifts.  Using 450\mum\ measurements for a subset of our sample we
confirm that the median dust temperature of \td=36$\pm$7\,K, derived
assuming the local FIR--radio correlation applies at high redshift, is
reasonable.  Individual 450\mum\ detections are consistent with the
local radio-FarIR relation holding at $z\sim2$.  This median \td\ is
lower than that estimated for similarly luminous {\it IRAS} 60$\mu$m
galaxies locally.  We demonstrate that dust temperature variations make
it impossible to estimate redshifts for individual submm galaxies using
simple long-wavelength photometric methods to better than $\Delta z
\simeq 1$.  We calculate total infrared and bolometric luminosities
(the median infrared luminosity estimated from the radio is
$8.5^{+7.4}_{-4.6}\times10^{12}$\,L$_\odot$), construct a luminosity
function, and quantify the strong evolution of the submm population
across $z=0.5$--3.5, relative to local {\it IRAS} galaxies.  We use the
bolometric luminosities and UV-spectral classifications to determine a
lower limit to the active galactic nucleus (AGN) content of the
population, and measure directly the varying contribution of
highly-obscured, luminous galaxies to the luminosity density history of
the Universe for the first time.  We conclude that bright submm
galaxies contribute a comparable star formation density to Lyman-break
galaxies at $z=2-3$ and including galaxies below our submm flux limit
this population may be the dominant site of massive star formation at
this epoch.  The rapid evolution of submm galaxies and QSO populations
contrasts with that seen in bolometrically lower luminosity galaxy
samples selected in the restframe UV, and suggests a close link between
submm galaxies and the formation and evolution of the galactic halos
which host QSOs.
\end{abstract}


\section{Introduction} \label{secintro}

The submillimeter (submm) galaxy population was first detected seven
years ago with the Submillimetre Common User Bolometer Array (SCUBA --
Holland et al.\ 1999) on the JCMT (Smail, Ivison \& Blain 1997; Hughes
et al.\ 1998; Barger et al.\ 1998; Eales et al.\ 1999). Their discovery
motivated a variety of surveys using both SCUBA and a similar
instrument, MAMBO (Bertoldi et al.\ 2000; Kreysa et al.\ 2002), on the
IRAM 30-m telescope.  Several surveys were undertaken of blank fields,
using different strategies to determine their depths and area coverage
(Barger et al.\ 1999a, 2002; Eales et al.\ 2000; Scott et al.\ 2002;
Borys et al.\ 2003; Serjeant et al.\ 2003; Webb et al.\ 2003a;
Dannerbauer et al.\ 2004; Greve et al.\ 2004).  These surveys detect
sources at a rate of approximately one source per night. A second class
of survey utilized massive clusters to provide a mild gravitational
lensing boost to aid in the detection and study of SMGs. These surveys
uncovered sources at a rate of about two per night, and examples of
such surveys include Smail et al.\ (2002), Chapman et al.\ (2002a),
Cowie, Barger \& Kneib (2002), and Knudsen\ (2004).

These surveys have gradually built up a large enough sample of
Sub-Millimeter Galaxies (SMGs) to produce a statistically useful count
(e.g., Blain et al.\ 2002). However, until very recently most of our
detailed knowledge of the properties of SMGs came from a handful of
SMGs identified in lensing cluster fields (e.g., Ivison et al.\ 1998,
2000, 2001; Frayer et al.\ 1998, 1999, 2004), as follow-up of blank
field sources without the lensing boost required significantly more
resources (e.g., Gear et al.\ 2000; Lutz et al.\ 2001; Dannerbauer et
al.\ 2002; Dunlop et al.\ 2004).

A breakthrough in our understanding of the properties of SMGs came from
exploiting ultradeep 20-cm radio maps.  A strong correlation exists
between the far-infrared (FIR) and radio flux densities of galaxies,
both locally and at high redshifts (e.g., Helou et al.\ 1985; Condon
1992; Garrett 2002), and so deep radio imagery with the Very Large
Array (VLA)\footnote{The National Radio Astronomy Observatory is a
facility of the National Science Foundation operated under cooperative
agreement by Associated Universities, Inc.}  can be used to help
pinpoint and study SMGs (Ivison et al.\ 1998; Smail et al.\ 2000;
Barger, Cowie \& Richards 2000; Chapman et al.\ 2001, 2002b).  These
radio maps provide a $\sim1.5$\arcsec\ beam and $\sim0.5$\arcsec\
astrometric precision relative to the optical frame, sufficient to
accurately locate the counterpart of a submm source.  The use of radio
imaging has culminated in the successful identification of at least
$65$\% of SMGs brighter than S$_{\rm 850 \mu m}>5$\,mJy and their
photometric characterization in the optical/near-infrared waveband
(Ivison et al.\ 2002; Chapman et al.\ 2003a; Wang et al.\ 2004; Borys
et al.\ 2004; Greve et al.\ 2004).

The ability to precisely locate the position of a submm emitting source
is essential if we wish to study their properties in any detail.  In
particular, this is a necessary first step when trying to derive
redshifts and luminosities for these systems.  It had been hoped that
long-wavelength observations of the dust emission spectrum of these
galaxies might be prove a reliable route to derive their redshifts and
luminosities.  The submm/radio flux ratio was first used by Carilli \&
Yun (1999) in this manner to estimate the typical redshift of SMGs,
however the technique was recognized immediately to have limited
accuracy ($\sim50$\% redshift errors) if there was a range of dust
temperatures (\td) present.  This uncertainty is particularly important
when deriving luminosities and related properties from submm
observations as the submm flux density is $S_{\rm 850\mu m} \propto
T_{\rm d}^{\simeq 3.5}$ for a fixed FIR luminosity at $z \simeq 2$.  In
addition, there is a strong degeneracy between \td\ and $(1+z)$ (Blain
1999), which limits the usefulness of simple photometric redshifts for
estimating luminosities for the SMG population.  Refinement of the
modeling and fitting techniques appears not to have overcome this basic
source of uncertainty (e.g., Aretxaga et al.\ 2003; Wiklind 2003).
Indeed, even surveys at several submm wavelengths (e.g., Hughes et al.\
2002) cannot completely overcome the degeneracy between dust
temperature and redshift (Blain 1999; Blain, Barnard \& Chapman 2003 --
see also Aretxaga et al.\ 2004 for a contrary view).

As a consequence, precise redshifts are crucial for interpreting almost
every aspect of SMGs. Prior to 2002, only a handful of spectroscopic
redshifts were available for unambiguously-identified SMGs (Ivison et
al.\ 1998, 2000; Barger et al.\ 1999b; Lilly et al.\ 1999).  Recent
attempts to measure redshifts for SMGs have met with more success
(Chapman et al.\ 2002c, 2003a; Barger et al.\ 2002; Ledlow et al.\
2002; Smail et al.\ 2003a, 2003b; Kneib et al.\ 2004).  However, the
resulting sample is still restricted in size and unrepresentative of
the general properties of the SMG population (with a bias towards
optically-bright counterparts and a preponderance of strong-lined AGN).
A redshift survey of a large and representative sample of submm
galaxies is therefore urgently required.

Chapman et al.\ (2003b -- hereafter C03) demonstrated that
spectroscopic redshifts can be obtained for even the optically faintest
SMGs, spanning a factor of $100\times$ in $I$-band flux, allowing a
much more representative sample of the population to be studied.  Their
approach involved constructing densely-packed distributions of submm
galaxies across $\sim 10'$ fields (matched to the area coverage of
multi-object spectrographs on 10-m telescopes) with precise positions
from radio counterparts.  These samples could then be efficiently and
effectively targeted using deep spectroscopy in the UV/blue spectral
region.  With a large, unbiased sample of SMGs constructed in this
manner it is possible to address questions about the SMG population
with more certainty: including their dust temperatures (T$_{\rm d}$)
and SED properties, their luminosities at various wavelengths and
luminosity evolution, their contribution to the FIR background, and
their relation to other populations of galaxies and AGN at high
redshift.

In this paper, we present an expanded sample from the ten SMGs with
robust spectroscopic redshifts described by C03, to provide a total
sample of 73 redshifts for unambiguously-identified SMG's.  We discuss
the properties and observations of this sample, along with selection
effects in \S2. We present the basic observational results obtained for
this sample, including the redshift distributions, variation in SEDs
with redshift as characterized by the submm/radio flux ratio, and
optical properties in \S3. In \S4 we then use basic assumptions to
derive dust temperatures and bolometric luminosities for our sample,
compare the UV properties of the galaxies with their radio/submm
emission, assess their contribution to the luminosity and star
formation histories of the Universe and the FIR background (FIRB) and
discuss their evolutionary connections with other high-redshift
populations.  Finally, in \S5 we give our main conclusions.  All
calculations assume a flat, $\Lambda$CDM cosmology with
$\Omega_\Lambda=0.7$ and $H_0=71$\,km\,s$^{-1}$\,Mpc$^{-1}$.

%
%
\section{Sample Definition, Observations and Analysis}

The parent sample of SMG's used for our survey consists of 150 sources
detected at 850\mum\ ($>3\sigma$) with SCUBA/JCMT, lying in seven
separate fields (field centers listed in Table~1): CFRS03, Lockman
Hole, HDF, SSA13, Westphal-14, ELAIS-N2, and SSA22.  104 of these
SMGs have radio identifications from deep VLA radio maps at 1.4\,GHz.
This radio-identified subset are the focus of this paper.

In all fields the SCUBA submm data was retrieved from the JCMT
archive\footnote{The JCMT is operated by the Joint Astronomy Centre on
behalf of the United Kingdom Particle Physics and Astronomy Research
Council, the Netherlands Organisation for Scientific Research, and the
National Research Council of Canada. The JCMT archive is provided
through the Canadian Data Archive Center.} and reduced in a consistent
manner using the {\it SURF} reduction tools (Jenness et al.\ 1998) and
our own software to extract beam-weighted submm fluxes.\footnote{Map
fluxes are obtained by extracting the effective beam imprint, $(-0.5,
1, -0.5)\times\delta$ SCUBA chopping/nodding profile (e.g., Scott et
al.\ 2002; Borys et al.\ 2003).}  In some cases, additional radio
sources were targeted in SCUBA's {\it photometry} mode (Holland et al.\
1999) to efficiently construct large samples of SMGs to target in
contiguous regions around fields mapped by SCUBA (e.g., Chapman et al.\
2001a, 2002b, 2003a). In addition, follow-up SCUBA photometry was used
to verify the reality and submm flux densities of 11 of the sources
detected in SCUBA maps.  These new SCUBA observations were obtained
during JCMT observing runs in 2002 and 2003, with sky opacity at
225\,GHz, $\tau_{225}=0.04$--0.09.  The observing strategy was to
integrate for a fixed length of time (1.0\,hrs) on all targeted
galaxies, with additional time allocated to targets which did not
achieve our nominal RMS sensitivity goal of dS$_{850}$=1.5\,mJy due to
weather conditions.  We note the observational mode used to identify
sources in Table~3, based on whether their submm detection was obtained
entirely in {\it photometry} mode ($'P'$), entirely in mapping mode
($'M'$) or through a combination of the two modes ($'MP'$).

Radio data for these fields either existed from previous work by
members of our group (Lockman Hole, ELAIS-N2), was rereduced for the
purpose of this study (CFRS-03, SSA22, Westphal-14), or obtained from
the public release (HDF, Richards 2000).  The SSA13 radio data were
obtained from E.A.\ Richards (private communication), and is so far
unpublished (a subsequent reduction of the SSA13 data is described in
Fomalont et al.\ 2004).  The radio data for the Lockman Hole and
ELAIS-N2 fields come from Ivison et al.\ (2002) who identified
counterparts to the submm galaxies in these regions from Scott et al.\
(2002).  For those fields which we rereduced the radio data was
retrieved from the VLA archive when available and combined with new
data obtained by our group in SSA22 (36~hrs, A-array), Westphal-14
(24~hrs, B-array), and CFRS-03 (16~hrs, B-array). The radio maps were
reduced in an identical manner to those described in Ivison et al.\
(2002).  The resulting radio maps have depths range from 4$\mu$Jy to
15$\mu$Jy rms; see Table~1.

Deep optical imaging in the $B$, $R$, and $I$ passbands is available
for all of our fields. This consists of several hour integrations with
mosaic CCD cameras on 4-m and 8-m class telescopes, taken either from
public archives, or obtained by our group during observing runs
throughout 2000--2002.  HDF ($BRI$), SSA13 ($I$), and Lockman ($RI$)
imaging was obtained with the SUPRIME camera on the Subaru
telescope. The HDF data was retrieved from the public release presented
in Capak et al.\ (2004).  The SSA13 and Lockman data was retrieved from
the Subaru archive and reduce by our group.  The $BR$ imaging in SSA13
was obtained with the Kitt~Peak 4m telescope and the MOSAIC camera, and
reduced with the {\sc MSCRED} tasks in {\sc IRAF}.  The $B$ imaging in
Lockman and the ELAIS-N2 imaging ($BRI$) was obtained with the
wide-field camera on the William Herschel Telescope (WHT) and reduced
in a standard manner using {\sc IRAF}.  $g,Rs,I$ imaging in SSA22 and
Westphal-14 was obtained from the public release in Steidel et al.\
(2003), and the details are described therein.  Additional SSA22
imaging was obtained with the LFC instrument ($g$) on the Palomar 5m
telescope and the CFHT-12k mosaic camera ($R,I$), and was reduced with
{\sc MSCRED} in {\sc IRAF}.  References to all these instruments are
listed in Table~1.  Near-infrared imaging is also available for the
majority of submm sources in our fields from a number of different
instruments and telescopes, typically reaching at least $K=20$ and
$J=22$.  Details of the optical, radio, and submm data in each field is
given in Table~1. The near-IR properties of our SMGs are discussed
fully in Smail et al.\ (2004a).

SMGs with radio identifications allow the position of the rest-frame
FIR emission to be unambiguously identified with a position in the
optical imaging to within the relative astrometric alignment of the
radio/optical frames.  Optical images were distortion corrected, and
tied to the same astrometric grid as the radio data using large numbers
of optically-bright radio sources across the field, resulting in an rms
positional uncertainty of typically $\sim$0.5\arcsec\ (see the detailed
discussions in Richards 2000; Ivison et al.\ 2002).  The $R_{\rm
AB}$-magnitudes (subsequently called $R$) in 2\arcsec\ apertures of the
targets range from $R=18.3$ to $R>27.5$ (Fig.~1).

Targets were selected for spectroscopic followup from the seven fields,
chosen at random and prioritized equally, {\it without preference for
optical brightness}.  Observations of two sources were obtained with
ESI echelle spectrograph on the Keck-II telescope on the night of 2001
July 16 and have been previously discussed by Chapman et al.\ (2002c).
Over the course of seven observing runs between 2002 March and 2004
February we observed 98 of the 104 sources in our radio-SMG sample with
the Low Resolution Imaging Spectrograph (LRIS, Oke et al.\ 1995)
spectrograph on the Keck-I telescope\footnote{The data presented herein
were obtained at the W.M.\ Keck Observatory, which is operated as a
scientific partnership among the California Institute of Technology,
the University of California and the National Aeronautics and Space
Administration. The Observatory was made possible by the generous
financial support of the W.M.\ Keck Foundation.}, obtaining reliable
redshifts for a total of 73 galaxies.  The first ten spectroscopic
identifications from our program were presented in C03.

%
%
\begin{figure*}[htb]
\centerline{\psfig{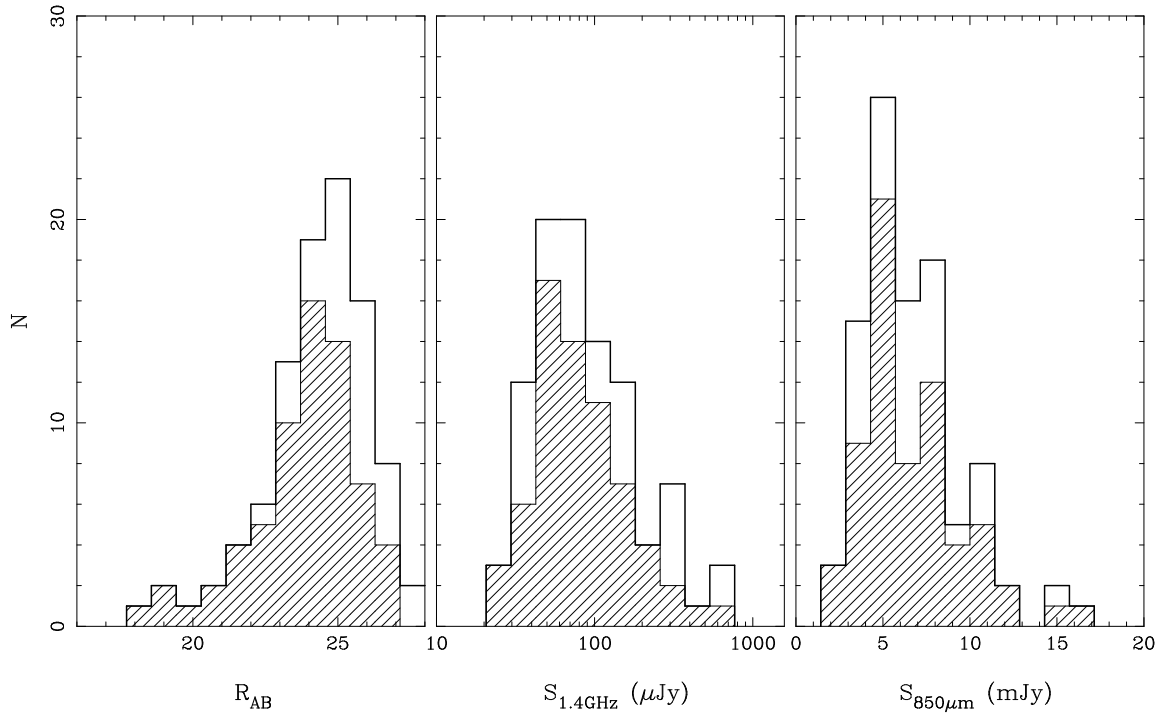}}
\vspace{6pt}
\caption{\small
This figure compares the properties of our
spectroscopically-identified sample to the parent catalog of all submm
galaxies identified in the radio waveband and observed
spectroscopically.  We show the relative distributions of
$R$-magnitude, radio flux, and submm flux (the shaded histograms are
the spectroscopic sample).  As expected the
spectroscopically-unidentified galaxies are typically fainter in the
optical, but have similar 1.4GHz/850\mum\ ratios, consistent with the
suggestion that they are likely to lie at similar redshifts to our
spectroscopically-identified sample, but are on average more obscured
in their restframe UV.
}
\label{figsel}
\addtolength{\baselineskip}{10pt}
\end{figure*}

The details of the spectroscopic configurations for our observing runs
and their success rates are presented in Table~2.  Observations taken
with LRIS using several different settings of gratings and
cameras. Data taken before 2002 March was obtained before the
commissioning of the large mosaic CCD blue camera, and used a smaller
format blue device.  All subsequent data was taken with the larger
format (4k$\times$4k) blue camera (Steidel et al.\ 2004).  Our
observations use either the 5600\AA\ [D560] or 6800\AA\ [D680] dichroic
to divide the light between the red and blue cameras.  The 400\,l/mm
[B400] grism was always used in the blue arm to provide wavelength
coverage from the atmospheric limit out to the dichroic wavelength for
most of the slitlets on the masks. This grism provides reasonable
resolution ($\sim 5$--6\AA) with our adopted 1.2--1.4\arcsec\
slitwidths. Either the 400\,l/mm [R400] or 600\,l/mm [R600] gratings
were used in the red arm, dependent on the dichroic selected.  Spectral
resolutions of $\sim 6$--8\AA\ are achieved in the red.

Integration times were between 1.5--6.0\,hrs in dark or gray
conditions, split into 30-min exposures. Conditions varied from
photometric to light cirrus, and seeing ranged between 0.7\arcsec\ and
1.1\arcsec.  Data reduction followed standard multi-slit techniques
using custom {\sc iraf} scripts.  The spectra typically probe an
observed wavelength range of 3100--8000\AA.

\subsection{Spectroscopic Identifications}
\label{specinc}

To obtain redshifts from our spectroscopic observations,
one-dimensional spectra were compared with template spectra and
emission line catalogs.  Of the 98 radio-SMGs observed, redshifts were
obtained with confidence for 73 galaxies, for a total spectroscopic
completeness of $74$\%.  The distribution of the optical, radio,
and submm fluxes of the parent and spectroscopically identified sample
are shown in Fig.~1.  Representative spectra are shown in
Fig.~\ref{spectra}.  Table~1 lists the number of radio-SMGs observed
with LRIS in each field, and the number of successful redshift
measurements.  Field to field variations reflect weather quality, as
well as intrinsic source properties (e.g., Ly$\alpha$ line strength).

Twelve SMGs from our sample have previously published redshifts from
other groups: SMM\,J141741.90+522823.6, SMM\,J141742.20+523026.0 by Eales
et al.\ (2000); SMM\,J030244.56+000632.3 by Webb et al.\ (2003);
SMM\,J123629.13+621045.8, SMM\,J123632.61+620800.1, SMM\,J123634.51+621241.0,
SMM\,J123635.59+621424.1, SMM\,J123607.53+621550.4, SMM\,J123721.87+621035.3,
SMM\,J131201.17+424208.1, SMM\,J131215.27+423900.9, SMM\,J131225.20+424344.5
by Barger et al.\ (2001a, 2001b, 2003).\footnote{Note that not all
these sources were measured as SMGs, but were listed with spectroscopic
redshifts in catalogs of radio or X-ray sources.}  The $R_{\rm AB}$
mags of these sources are amongst the brightest in our sample, with an
average $R=22.3\pm2.2$.  All these redshifts from the literature are in
agreement with our measurements within errors.

Three further sources were tentatively identified by Simpson et al.\
(2004) using the Subaru-OHS spectrograph.  SMM\,J163658.19+410523.8
agrees with our redshift as noted in Simpson et al.\ (2004).  However,
SMM\,J105158.02+571800.3 and SMM\,J141809.00+522803.8 disagree with our
measured redshifts by d$z$=0.20 and d$z$=0.16 respectively.  Our
redshift for SMM\,J105158.02+571800.3 ($z=2.239$) is derived from two
UV-absorption features and the detection of \hal\ (Swinbank et al.\
2004), and we regard our redshift as a more robust identification.  Our
redshift for SMM\,J141809.00+522803.8 is derived primarily from strong
Ly$\alpha$ in emission, but lies at a redshift ($z=2.71$) that makes it
difficult to followup in nebular lines using near-IR spectrographs,
casting some doubt on the reality of the Simpson et al.\ redshift
(Simpson et al.\ 2004 in fact suggest that their redshift is likely to
be spurious based on the weakness of the features and the residuals
present from sky-line subtraction).

The primary criteria for considering a redshift as robust is the
identification of multiple emission/absorption lines.  Our redshift
identifications are confirmed by the detection of other lines and
continuum features in 75\% (55) of the identified spectra: AGN lines
(C{\sc iv}$\lambda$1549, S{\sc iv}$\lambda$1397, N{\sc
v}$\lambda$1240), as well as weaker stellar and interstellar features,
and continuum breaks.  We also consider redshifts to be robust if we
detected a large equivalent width line ($>20$\AA), and there is
supporting evidence that this line is \lya.  Only one quarter (18) of
the 73 spectroscopic redshifts are single-line identifications.  There
are three items of supporting evidence which are used to support the
single-line \lya\ identifications: the identification of a continuum
break (if the continuum is detected) across the proposed Ly$\alpha$
line from the red to the blue (3/18), the absence of emission lines
which do not match the proposed Ly$\alpha$-derived redshift (all 18),
and an asymmetrical line profile, with the blue-wing truncated, typical
of Ly$\alpha$ emission from high-redshift galaxies (8/18).  In very few
cases were identifications ambiguous using these criteria.

%
%
\begin{inlinefigure}\vspace{6pt}
\psfig{figure=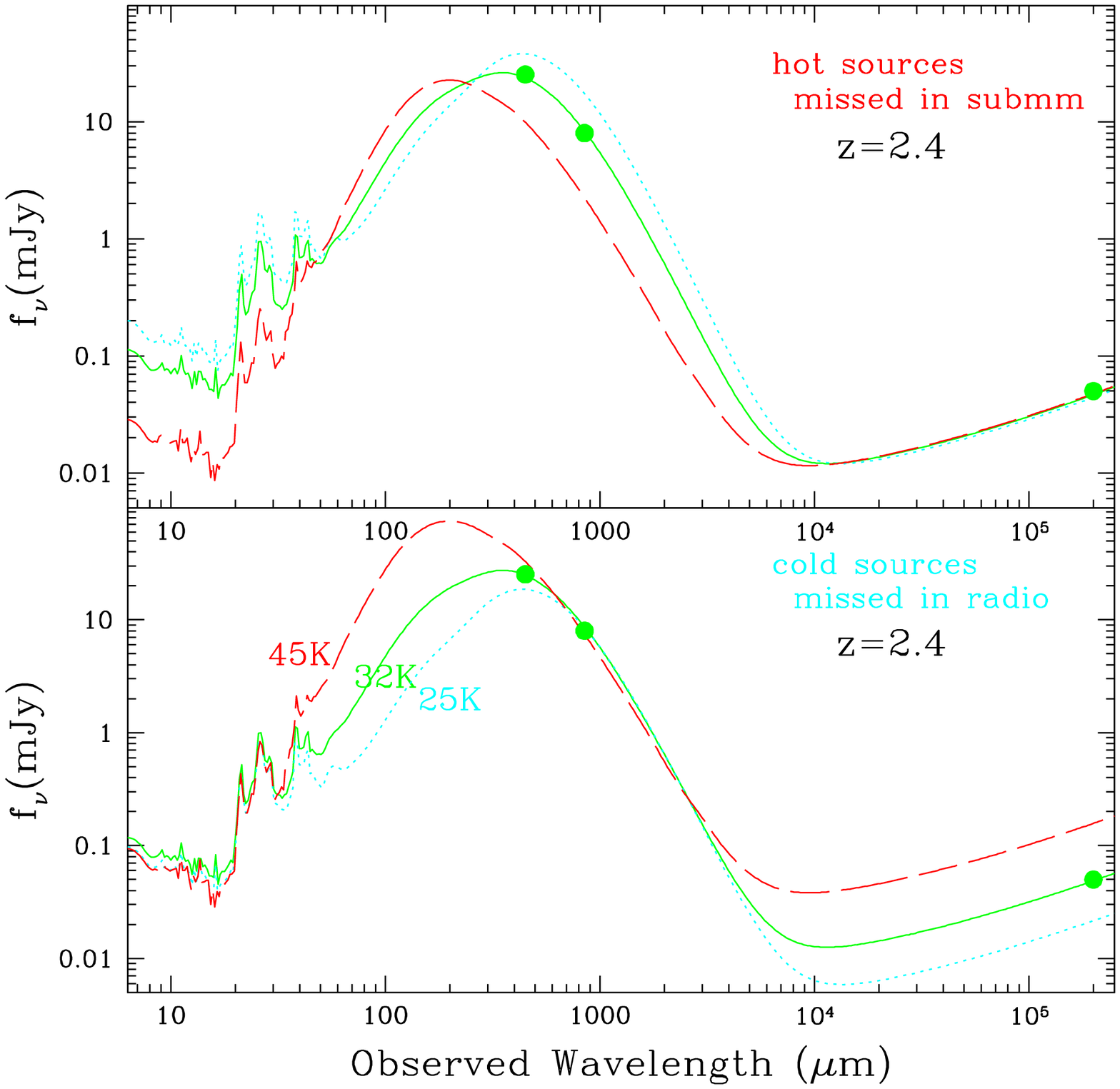,angle=0,width=3.5in}\vspace{6pt}
\caption{\small
We show possible SEDs describing the emission from a typical SMG, lying
near the median redshift for the model-corrected sample ($z=2.4$), with
flux densities of 6\,mJy at 850\mum\ and 50\,$\mu$Jy at 1.4-GHz (near
our radio detection limit).  Overlaid are SED templates at three dust
temperatures (25, 32 \& 45\,K) spanning the typical range observed in
the SMGs.  In the upper panel, the SEDs are normalized to the radio
point to emphasize how sources with hotter characteristic dust
temperatures, and lower implied dust masses, are missed in the submm at
$z\gs 2$.  In the lower panel, the SEDs are normalized to the
850-$\mu$m point, highlighting how sources with cooler characteristic
temperatures are undetectable in the radio at the redshifts higher than
the sample median.
}
\label{seleff}
\addtolength{\baselineskip}{10pt}
\end{inlinefigure}

While single line \lya\ identifications may not be convincing to some
readers, several arguments support our interpretation.  For many of the
single emission line detections, the observed wavelength lies below
4000\AA (sometimes below 3700\AA), precluding a reasonable
identification as [O{\sc ii}]\,$\lambda$3727 at $z<0.07$ given the
optical faintness and submm/radio detection.

We also have two independent tests of the reliability of our redshifts.
Firstly, we have obtained Keck/NIRSPEC and VLT/ISAAC near-IR
spectroscopic observations for a significant fraction of our sample to
probe the nebular line emission to measure the star formation rates,
estimate metallicities and study kinematics.  These observations have
successfully detected restframe H$\alpha$ (and frequently [N{\sc ii}])
emission in 26 cases (Swinbank et al.\ 2004), confirming the UV-based
redshifts in the present paper.  Ten of the 18 single-line
identifications have been confirmed in \hal.  The near-IR \hal\ spectra
were also able to break degeneracies in 5 cases where spectral
identifications based only on UV-absorption lines were consistent with
two similar redshifts.  These \hal/[N{\sc ii}] results were also used
to aid in the spectroscopic classification of our sample, as indicated
in Table~3.

Secondly, 15 of our SMG redshifts have been confirmed with CO line
emission using the IRAM Plateau de Bure Interferometer (Neri et al.\
2003; Greve et al.\ 2005), including 2 single-line
identifications. These detections not only confirm the precision of the
UV-based redshifts for the counterparts we targeted, but equally
importantly they also confirm that these galaxies are gas-rich systems
suitable to be the source of the luminous far-infrared emission
detected in the submm waveband.

The strength of the Ly$\alpha$ lines for the SMGs vary tremendously in
both line flux (L$_\nu$ from 1--60$\mu$Jy) and restframe equivalent
width, which ranges from $-$3\AA\ (absorption) to $>150$\AA\ (we note
that we see no obvious variation in the radio/submm properties of SMGs
as a function of Ly$\alpha$ line strength).  With the generally faint
rest-frame UV continua exhibited by our SMG (65\% are fainter than
$R_{AB}>24.4$), there is a bias in our sample against obtaining
redshifts for the weaker emission line sources at the faintest
continuum fluxes. This is reflected in Fig.~\ref{figsel}, where the
increased failure rate for obtaining spectroscopic redshifts is
apparent for $R_{AB}\gs 24$.  This bias is highlighted by the fact that
sources with identifiable Ly$\alpha$ absorption only appear in our
sample for SMGs with $R_{\rm AB}\ls 25$.  We discuss the spectral
properties and incompleteness of our sample in more detail in \S3.4.

%
%
\section{Results}

%
%
\begin{figure*}
\centerline{
\psfig{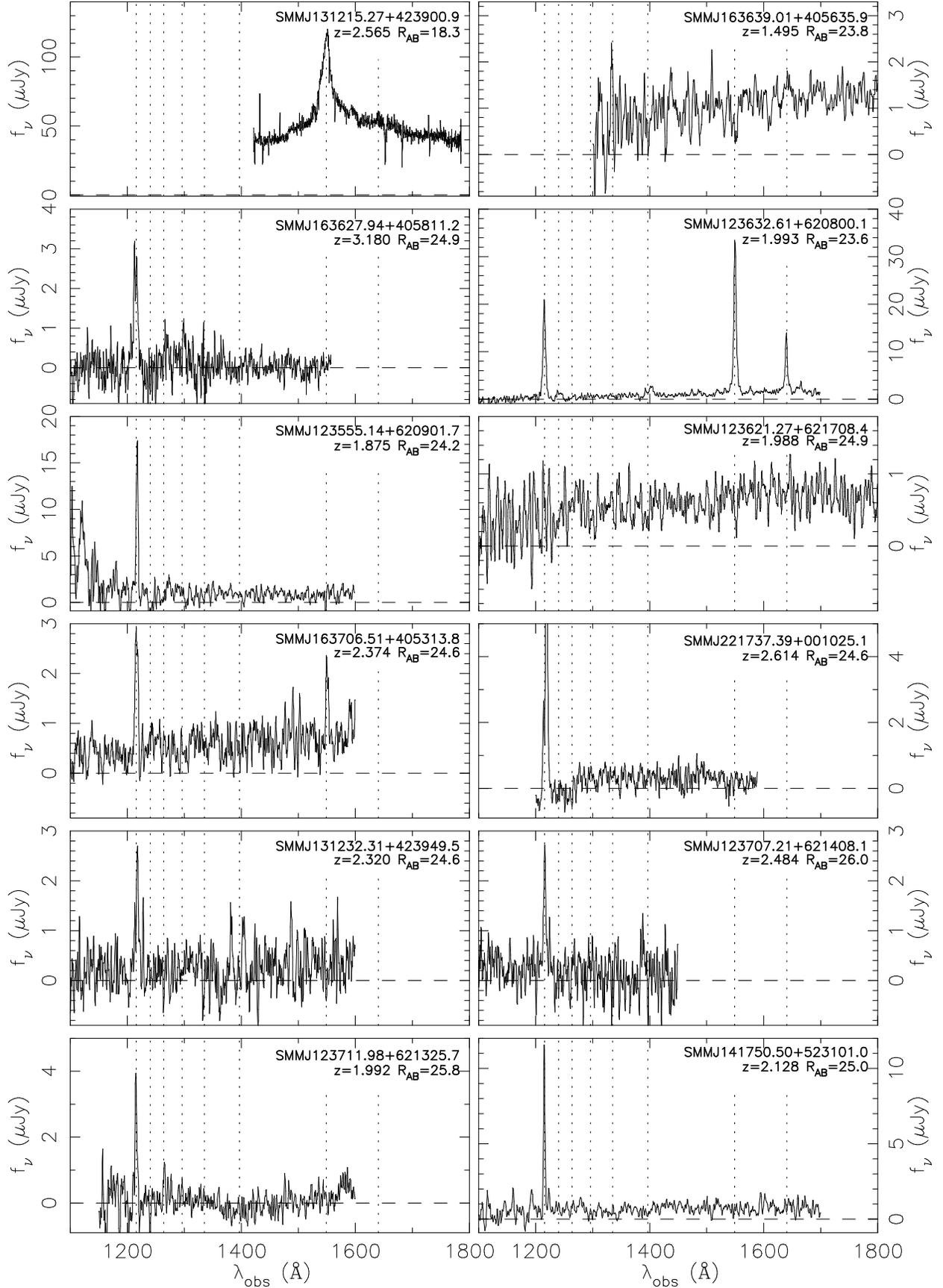}
}
\caption{\small
Representative spectra for twelve SMGs from our complete sample.  The
strongest UV lines used in the redshift identifications are marked by
dashed lines. All spectra have been shifted to a common rest-frame
wavelength scale.
}
\label{spectra}
\addtolength{\baselineskip}{10pt}
\end{figure*}

\subsection{Sample properties and submm-radio selection effects}

To understand the characteristics of the submm population we first have
to quantify how the selection criteria for our sample (e.g., radio,
submm and optical flux limits, spectroscopic incompleteness) may have
influenced the observed properties.

Fig.~\ref{figsel} shows our spectroscopic completeness as a function of
$R$-band magnitude, 1.4-GHz radio flux and 850\,$\mu$m submm flux.  The
median properties of the parent sample are $R_{\rm AB}=25.4\pm 1.8$,
$S_{\rm 1.4 GHz}=75\pm 127\mu$Jy and $S_{\rm 850 \mu m}=6.0\pm 2.9
$\,mJy, while the spectroscopically-identified population has $R_{\rm
AB}=24.6\pm 1.7$, $S_{\rm 1.4 GHz}=78\pm 106\mu$Jy and $S_{\rm 850 \mu
m}=5.7\pm 3.0$\,mJy, As expected, our spectroscopic sample is biased
towards the optically brighter galaxies (the median $R$-band magnitude
for the unidentified spectroscopic targets is $R_{\rm AB}=26.1\pm
1.2$), but there is no discernable difference in the submm or radio
distributions (which are effectively decoupled from the restframe UV
emission).  This suggests that the long-wavelength properties of our
spectroscopic sample are likely to be representative of the more
general submm population.

A crucial feature of the present study is that by analyzing only the
radio-identified SMG we are considering only part of the total SMG
population.  We must therefore determine the influence of the resulting
selection function before drawing wider conclusions about flux-limited
submm samples.

About 65\% of the bright ($>5$\,mJy) SMG population are detectable in
the deepest radio maps obtainable with the VLA (Ivison et al.\ 2002;
Chapman et al.\ 2003a; Wang et al.\ 2004; Borys et al.\ 2004).  Greve
et al.\ (2004) and Ivison et al.\ (2005), have recently suggested that
the fraction of bright SMGs, robustly confirmed at both 850$\mu$m and
1200$\mu$m, which are detected in the radio may be even higher --
$\sim$80\%.  The remaining $\sim35$\% of SMGs {\it not} detected in our
radio maps (and therefore not included in the distributions of
Fig.~\ref{figsel}) could in principle have a wide range of properties
and redshifts.

We can elucidate the effects of our radio pre-selection further by
considering the spectral energy distributions (SEDs) of SMGs in more
detail.  A range of possible SEDs for a canonical 6\,mJy radio-SMG at
$z=2.4$ with a 50-$\mu$Jy radio counterpart is shown in
Fig.~\ref{seleff}.  The upper panel shows SEDs (Dale \& Helou 2002)
spanning a range of dust temperatures, all with the same radio flux,
and therefore comparable FIR luminosities.  The range in dust
temperatures depicted is less than a factor two, but results in close
to a factor ten range in submm flux (see Blain et al.\ 2002, 2004a for
additional details of this selection effect).  As a consequence the
hottest SED (with the lowest implied dust mass) shown in
Fig.~\ref{seleff} (with a characteristic temperature of 45\,K) falls
below the SCUBA detection threshold in our sample of 3\,mJy.

This trend becomes more dramatic at lower redshifts: by shifting the
SED templates to shorter wavelengths, it becomes apparent that only the
coolest sources can be detected above our $\sim$3\,mJy submm flux
limit.  By contrast, the selection bias vanishes at higher redshifts;
increasingly hotter sources become detectable with SCUBA above our
radio limit ($\sim30$\,$\mu$Jy in the typical field).

However, our requirement of a radio detection to pinpoint the
spectroscopic counterpart of the submm source means that the opposite
selection bias comes into play in our sample (Fig.~\ref{seleff} lower
panel); the coldest SMG's at $z=2.4$ lie below our radio flux
limit. This bias become an increasing concern at higher redshifts where
warmer SMG's fall beneath our radio limit.  We will quantify the
influence of our radio selection using knowledge of the range of
observed SMG SED properties in \S4.1.

The observed redshift distribution in Fig.~\ref{nz} illustrates
succinctly how the radio-undetected SMGs may relate to our radio-SMG
sample in this context: we expect to miss sources lying between the
submm and radio model curves due to our requirement of a radio
detection to pinpoint the host galaxies.  The radio-undetected minority
of the submm population is likely to overlap significantly in redshift
with our present radio-SMG sample.  These galaxies would have
characteristic dust temperatures that are typically cooler than the
radio-detected galaxies at their redshifts (via the (1+$z$)--\td\
degeneracy).  Fig.~\ref{nz} shows that the two populations begin to
overlap significantly at $z\sim 2.5$.  A well-studied example of a
radio-undetected SMG in this redshift range is the extremely red
SMM\,J04431+0210 identified by Smail et al.\ (1999), the redshift of
this galaxy was measured as $z=2.51$ by Frayer et al.\ (2003) and
confirmed beyond doubt as the submm source through the detection of
molecular gas in the CO(3--2) line by Neri et al.\ (2003).

\subsection{Ambiguous Radio Counterparts to Submm Sources}

There are a number of other issues associated with the identification
process which need to be considered.  In particular, we note that the
radio positions of a handful of sources lie very near ($<2$\arcsec) to
an obviously low-redshift galaxy.  These galaxies are unlikely to be
related to the submm-emitting source, once the complete range of the
galaxy and submm source properties (radio luminosity and colors) are
considered. Instead we believe that these low-redshift galaxies are in
fact lensing the background submm sources. Detailed discussions of two
of these systems can be found in Chapman et al.\ (2002c) and Dunlop et
al.\ (2004).  There are five such galaxies in our total sample and we
have not include these in the catalog of 73 submm sources with robust
redshifts as we believe that the foreground galaxies are not the
correct identifications for the submm sources.  In three of these cases
there is evidence from near-infrared imaging of faint $K$-band galaxies
lying closer to the radio position than the galaxy which was
spectroscopically observed.  These submm sources are particularly
difficult to study (or even target) optically, and may have to wait for
blind CO lines searches at millimeter wavelengths to determine their
redshifts.  For completeness, we include these galaxies in our
identification table, but flag them as probable lenses and do not
include them in any of our subsequent analysis.

We note that there are examples of low redshift ($z<1$) galaxies where
the radio emission is coincident with the spectroscopically-targeted
counterpart (within the relative astrometric errors), and we consider
these to be the correct identifications and include them in our sample
and analysis.  These galaxies have inferred dust temperatures which
appear cold relative to similarly-luminous, local {\it IRAS} galaxies
(Chapman et al.\ 2002d; Blain et al.\ 2004b).

In addition to the ambiguity in the small number of cases where radio
sources lie close to bright, foreground galaxies, Ivison et al.\ (2002)
have demonstrated that roughly 10\% of radio-SMGs have more than one
radio source within their submm error circle (8\arcsec\ diameter,
derived based on Monte Carlo simulations, and through comparison to the
robust identifications of SMGs in Smail et al.\ 2002).  Taking a
well-studied example from the literature: SMM\,J\,09431+4700 (Ledlow et
al.\ 2002) represents a striking case of an SMG with two probable radio
counterparts (denoted H6 and H7).  Ledlow et al.\ (2002) obtained a
spectroscopic identification of $z=3.35$ for the optically-brighter
radio source, H6.  Nevertheless, both radio sources were detected in
1\,mm continuum from the IRAM PdB (Tacconi et al.\ 2005) -- confirming
that both contribute to the measured 850-$\mu$m flux. A search for CO
emission in the system at the redshift of H6 failed to detect any
molecular emission from this galaxy, but did detect a massive gas
reservoir in the optically fainter sources, H7, confirming that both
galaxies lie at the same redshift (Neri et al.\ 2003).

We find eight examples of multiple radio counterparts to submm sources
in our sample (noted in Table~3).  In these cases we have taken spectra
of all of the radio sources.  In three cases (SMM\,J123621.27+621708.4,
SMM\,J123616.15+621513.7, and SMM\,J163650.43+405734.5) we obtained a
spectroscopic identification for one radio source but failed on the
second. In three cases (SMM\,J105200.22+572420.2,
SMM\,J123707.21+621408.1, and SMM\,J123711.98+621325.7) we confirmed
that both radio sources lie at the same redshift to within
1000\,km\,s$^{-1}$.  While in the final two cases
(SMM\,J105238.26+571651.3 and SMM\,J105225.90+571906.8), we found one
radio source at high-redshift $z>2$, and the other at $z<0.5$.  In
these latter two cases, we assume that the high-redshift source is the
correct identification, since its significantly greater radio
luminosity suggests a dominant contribution to the submm emission.
However, detailed multi-wavelength follow-up may reveal that the low
redshift radio source also contributes significantly to the submm
luminosity.

Even when only a single radio counterpart exists, the optical
identifications are not always unambiguous -- with two optical
counterparts within the radio error-circle, or $\sim$1\arcsec\ offsets
from the radio to optical identification.  This affects only a small
fraction of our sample (four SMGs or $\ls 5$\%).  Again, we have
attempted to obtain redshifts for all components, but the sample is not
complete in this respect (two possible counterparts remain to be
observed spectroscopically).  For the $\sim$10\% of cases where the
optical identification is slightly offset from the radio (noted in
Table~3), the optical source is sometimes extended and overlaps the
radio source in lower surface brightness extensions. In these cases we
are confident that we have identified the correct counterpart, although
we have not necessarily characterized the source of the bolometric
emission in the Keck spectrum.  Chapman et al.\ (2004b) address some of
these issues through higher spatial resolution radio and optical
imagery, while Ivison et al.\ (2005) have examined the detailed
identifications of SMGs in a robust sample within one of our survey
fields.  We have commented on individual objects which are subject to
any of these identification issues in Table~3.

\subsection{Redshift distributions}

The redshift distribution of our submm galaxy sample (Fig.~\ref{nz})
displays a marked peak at $z \sim 2.0$--2.5, with an apparent dip in
the distribution at $z\sim 1.5$ that is almost certainly a result of
our failure to efficiently identify redshifts for SMG's at $z=1.2$--1.8
where no strong spectral features fall into our observational windows.
The decline in the numbers of SMGs at low redshifts is due to a
combination of the submm selection function, and the intrinsic
evolution in the population (e.g., Blain et al.\ 1999a).  At high
redshifts, our requirement of a radio detection (to locate the submm
counterpart) limits the maximum redshift detectable at our radio survey
depths: a radio flux limit of 30\,$\mu$Jy will yield detections of
typical temperature and luminosity SMGs at $z<3.5$.

We describe the selection effects using an evolutionary model which
accounts for the dust properties of local galaxies with a range in
template SEDs and that has been tuned to fit the submm/radio flux
distribution (Chapman et al.\ 2003c; Lewis, Chapman \& Helou\ 2004).
The model takes the local FIR luminosity function and evolves it in
luminosity with increasing redshift following the functional form given
in Blain et al.\ (2002). The functional form of the pure luminosity
evolution function in this model is given as $g(z) = (1+z)^{3/2}
sech^2[{\rm b} \ln(1+z) - {\rm c}] cosh^2{\rm c}$.  The values of $b$
and $c$ in this function which remain consistent with our observed
redshift distribution (after filling in the redshift desert as
described in this section, and assuming $1\sigma$ Poisson error bars on
the N$(z)$ histogram) are $2.10^{+0.30}_{-0.40}$ and
$1.81^{+0.21}_{-0.37}$ respectively.  The range of parameters
effectively shift the $z_{\rm peak}$ of the peak evolution function
such that $z_{\rm peak} = 1.8^{+0.7}_{-0.3}$.

%
%
\begin{inlinefigure}\vspace{6pt}
\vspace{6pt}
\psfig{figure=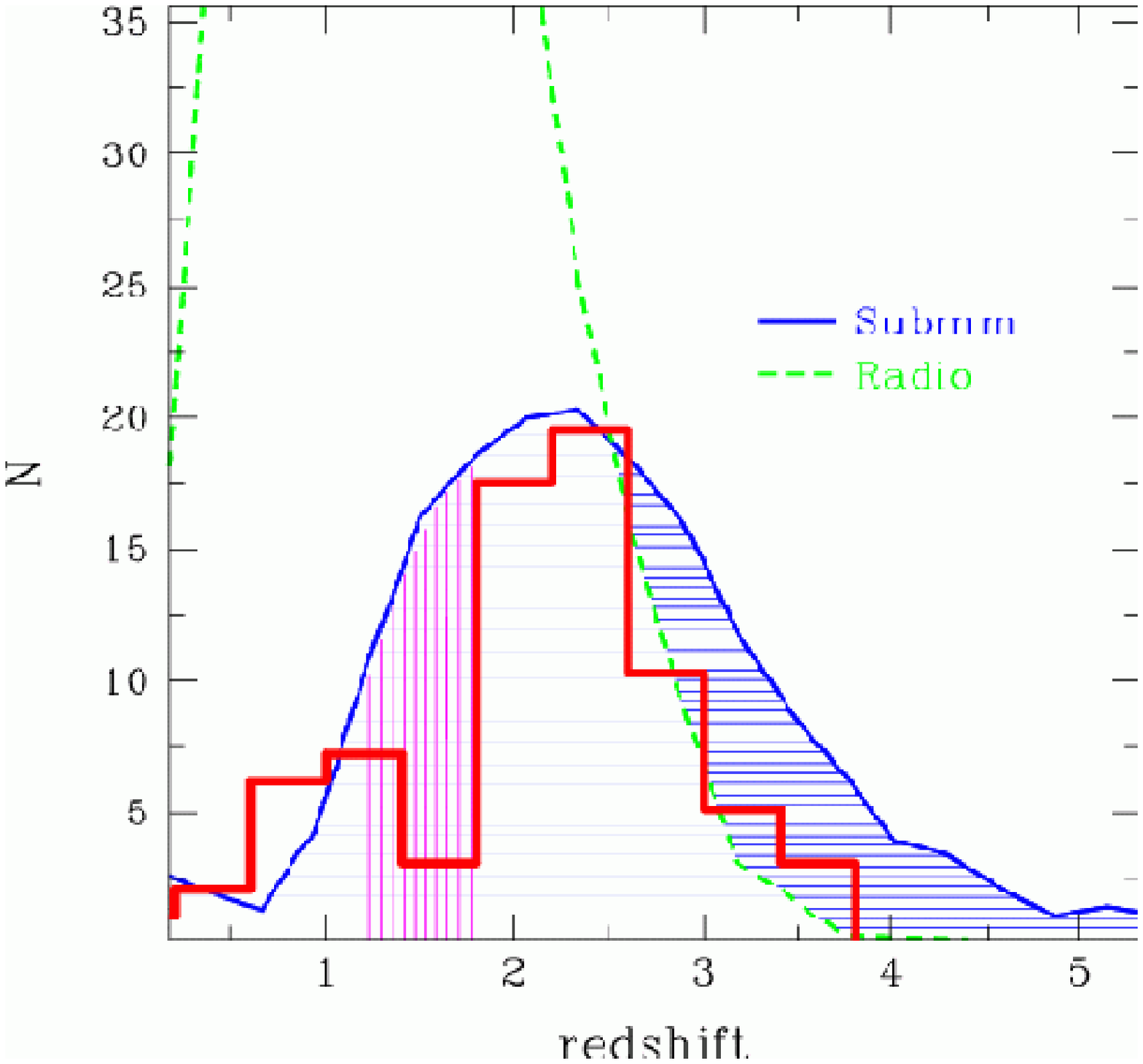,angle=0,width=3.4in}
\vspace{6pt}
\caption{\small
The redshift distribution of our submm galaxy sample (red histogram).
To interpret the likely effects of the sample selection on this
distribution, we plot predicted model redshift distributions for submm
galaxies with $S_{850\,\mu m} >$ 5$\,$mJy (blue solid line) and radio
sources with $S_{1.4\,\rm GHz} >$ 30$\,\mu$Jy (green dashed line) based
on the evolutionary models of Blain et al.\ (2002) and a family of
long-wavelength SEDs tuned to reproduce the distribution of submm/radio
flux ratios (Chapman et al.\ 2003c; Lewis, Chapman \& Helou 2004).  The
magenta shaded region highlights the redshift range where no strong
line features enter the observable wavelength range of LRIS.  The blue
shaded region identifies the distribution of that proportion of the
population we expect to miss due to our radio flux limit.
}
\label{nz}
\addtolength{\baselineskip}{10pt}
\end{inlinefigure}

We plot the predicted redshift distributions for all submm galaxies
with $S_{850\,\mu m} >$ 5$\,$mJy and all radio sources with
$S_{1.4\,\rm GHz} >$ 30$\,\mu$Jy.  We expect to miss sources lying
between the submm and radio model curves due to our requirement of a
radio detection to pinpoint the host galaxies.  The galaxies in this
region represent 35\% of the integral under the submm curve (Fig.~4) in
agreement with the proportion of radio-unidentified SMGs (e.g., Ivison
et al.\ 2002; Chapman et al.\ 2003c).  The combination of selection
functions described by the model curves are clearly in good agreement
with the observed redshift distribution.

%
%
\begin{figure*}
\centerline{
\psfig{figure=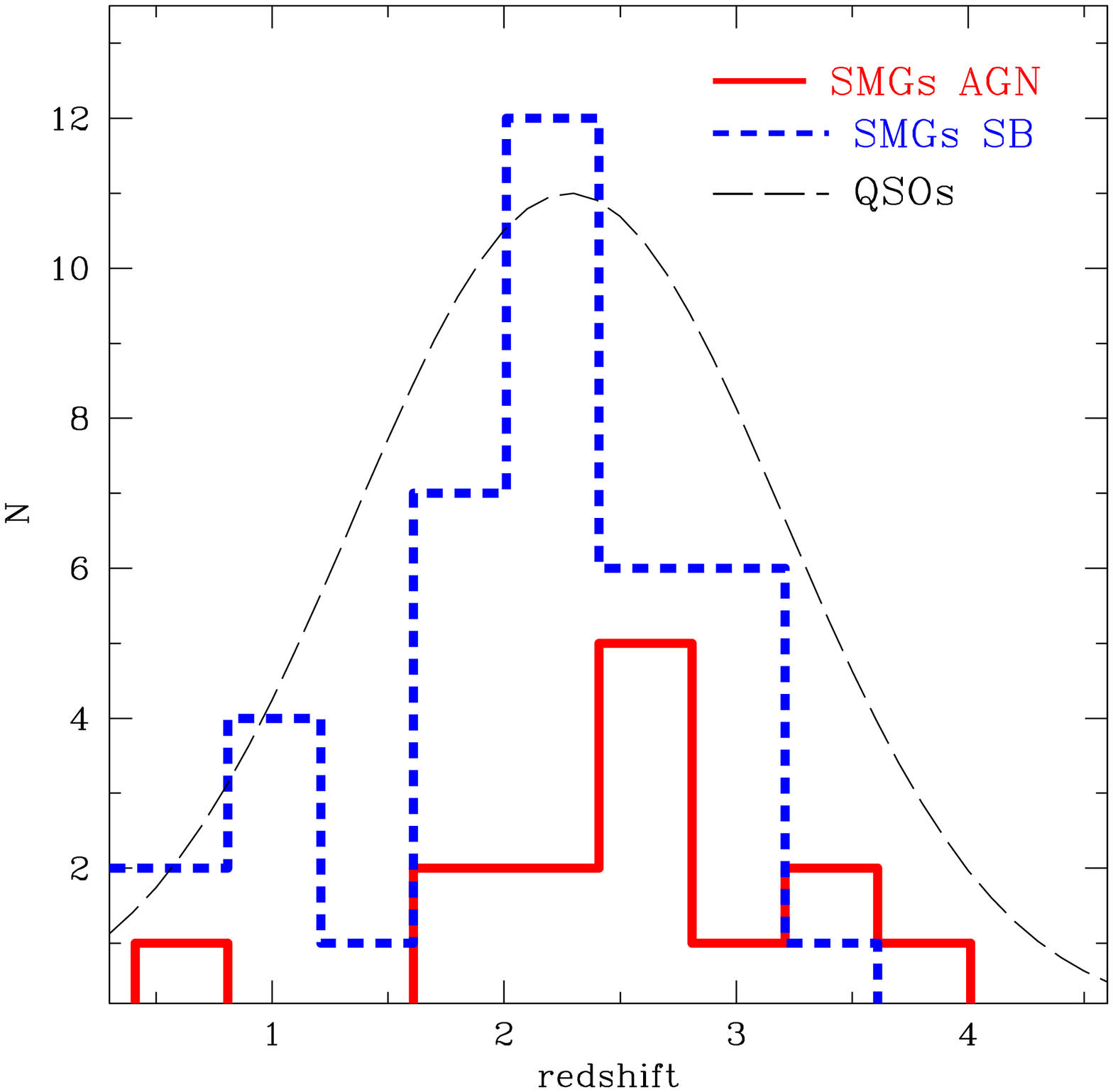,angle=0,width=3.1in}
\psfig{figure=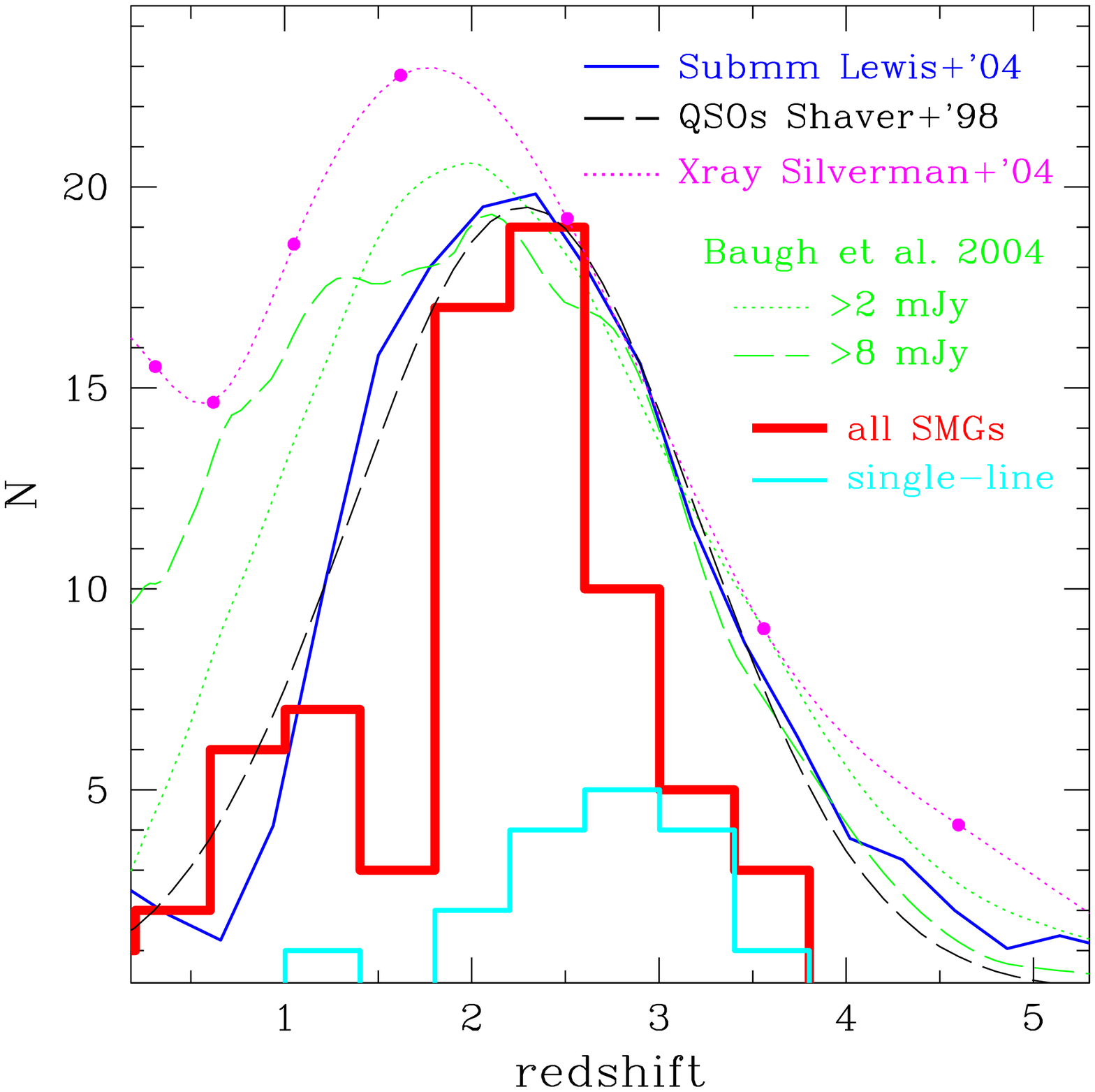,angle=0,width=3.1in}
}
\vspace{6pt}
\caption{\small
{\bf Left panel:} Comparing the N($z$) for radio-SMG with obvious AGN
signatures with those without (which we denote starbursts).  The N($z$)
for a radio-selected sample of Quasars unbiased by extinction (Shaver
et al.\ 1998) is shown as a dashed curve.  {\bf Right panel:} The
N($z$) for radio-SMGs with those SMGs which were first identified in
the UV spectroscopy from a single-line (\lya\ in all cases except one)
highlighted as a separate histogram.  The SMGs are compared with the
Lewis, Chapman \& Helou\ (2004) model for 850$\mu$m sources, the Shaver
et al.\ (1998) radio-selected sample of Quasars unbiased by extinction
(comparable to the 2dF Quasar evolution from Boyle et al.\ 2000 and
Croom et al.\ 2004), and the Silverman et al.\ (2004) sample of Chandra
X-ray selected AGN.  Models of Baugh et al.\ (2004) for $>8$\,mJy SMGs
and $>2$\,mJy SMGs normalized to equal numbers of sources in each
N($z$) are also overlaid.
}
\label{nzSBAGN}
\addtolength{\baselineskip}{10pt}
\end{figure*}

Contrasting the predicted and observed redshift distributions we would
estimate that the dip due to spectroscopic incompleteness at $z\sim
1.5$ would effect around 20\% of our parent sample -- this is close to
the 26\% spectroscopic incompleteness estimated for the total sample in
\S3.3 -- suggesting that most of this arises from the so-called
redshift desert. The influence of the radio selection produces an
increasing incompleteness compared to the original parent sample at
$z\gs 2.5$.  The model predicts that these missing galaxies will lie at
somewhat higher redshifts than our radio-SMG sample, but still overlap
significantly.  Under this model scenario, less than 10\% of SMG's lie
at $z\geq 4$ where we have no identified members of the population in
our survey.  The observed radio-SMG N($z$) (accounting for
incompleteness in the spectroscopic desert) can be reasonably
approximated by a Gaussian with a central redshift of $z=2.1$ and a
sigma of 1.2, while the model submm-only N($z$) is fit by a Gaussian
with a central redshift of $z=2.3$ and a sigma of 1.3.

In \S4.4 and 5, we will attempt to combine all the observational
information on the SMG population to confirm the fraction of SMGs that
could plausibly lie at very high redshifts.  Nevertheless, we reiterate
that our survey gives access to the majority ($\sim 70$\%) of the
bright ($\sim$6\,mJy) SMG population and so allows us to derive
representative properties for the bulk of this important population.

It is interesting to divide the SMG sample based upon their
spectroscopic classifications.  Quasars (Shaver et al.\ 1998; Boyle et
al.\ 2000; Silverman et al.\ 2004) and LBGs (Steidel et al.\ 1999;
Lehnert et al.\ 2003; Giavalisco et al.\ 2004) show very different
evolution histories from $z=1$--6. Similarly, Haarsma et al.\ (2000)
and Cowie et al.\ (2004) have studied the evolution of lower redshift
($z=0$--1) faint $\mu$Jy radio sources, finding strong evolution and
differing evolution for subsets of this population depending on whether
their restframe optical/UV spectra show AGN or star-formation
signatures.  As Fig.~\ref{nzSBAGN}a demonstrates, the redshift
distribution of the SMGs does not depend strongly on the level of AGN
activity apparent in their restframe-UV spectra.  Moreover, the form of
the redshift distribution for the radio-SMG population
(Fig.~\ref{nzSBAGN}b) is very similar to that seen for Quasars (Shaver
et al.\ 1998; Silverman et al.\ 2004), selected either in the optical,
X-ray or radio wavebands, and different from than that seen for
UV-selected galaxies.  We have placed the data from Silverman et al.\
(2004) on Fig.~\ref{nzSBAGN}b by dividing each of their points by the
appropriate comoving volume element in our adopted cosmology, and
connecting the points with a spline fit.  Only by adopting very
contrived redshift distributions for the additional 35\% of the SMG
population which are not currently identified in the radio
(Fig.~\ref{nz}), is it possible to make the SMGs redshift distribution
differ radically from that seen for Quasars.  We shall explore this
point in more detail in \S4.4.  We also highlight in
Fig.~\ref{nzSBAGN}b those SMGs which were first identified in the UV
spectroscopy from a single-line (\lya\ in all cases except one), and
note again that 50\% of these sources have been confirmed at the
correct redshift through near-IR spectroscopy of the \hal\ or
[O{\sc iii}]$\lambda$5007 lines.  These single-line identifications reflect
the higher redshift tail of our measured N($z$), consistent with the
generally brighter optical continuum magnitudes of the $z\sim2$ sources
over the $z\sim3$ sources.

It is also interesting to study the N($z$) as a function of 850$\mu$m
luminosity.  Model predictions of the N($z$) for 850$\mu$m-selected
galaxies have been presented by Baugh et al.\ (2004). In
Fig.~\ref{nzSBAGN}b we have overlaid their model N($z$) for SMGs with
S$_{850}>8$\,mJy and S$_{850}>2$\,mJy, the latter being dominated by
the $\sim$2\,mJy sources because of the steep 850$\mu$m counts.  Their
models are normalized to have the same number of sources in the N($z$)
integral.  Baugh et al.\ (2004) find that the median redshift of the
SMGs does not change significantly over the 2--8\,mJy flux range.

We can divide our observed SMG sample into equal number bins with
S$_{850}>5.5$\,mJy and S$_{850}<5.5$\,mJy.  Our submm {\it brighter}
galaxies (median S$_{850}=7.9$\,mJy) lie preferentially at higher
redshifts with a median redshift $z=2.45$ (interquartile $\pm$0.35),
compared to the submm fainter galaxies (median S$_{850}=4.7$\,mJy) with
median $z=2.01$ (interquartile $\pm$0.61).  At face value, this result
disagrees with the Baugh et al.\ predictions.  As we might expect from
a radio-flux limited survey, the radio properties of the submm bright
and faint samples are indistinguishable (submm bright:
S$_{1.4}=74\pm27\mu$Jy, submm faint: S$_{1.4}=76\pm29\mu$Jy),
suggesting that our radio selection criterion is at the root of the
discrepancy with the Baugh et al.\ model.  For example, if the overall
properties of submm-selected galaxies were similar for
S$_{850}\sim2$\,mJy and S$_{850}\sim8$\,mJy samples (in particular the
FIR-radio correlation) as in the models of Baugh et al.\ (2004), we
would expect to miss more of the S$_{850}\sim2$\,mJy sources at higher
redshifts due to the radio detection criterion.  As we will see in
\S~4.1, these properties imply that the submm bright and faint
sub-samples will coincidentally have indistinguishable distributions in
dust temperature.  Thus while at face value these results appear to
imply that more bolometrically luminous SMGs tend to lie at higher
redshifts, consistent with a strong luminosity evolution (see \S~4.2),
strong selection effects are at present in our submm/radio sample, and
we must exercise caution in our interpretations of these trends.

\subsection{Optical Spectroscopic Characteristics of SMGs}

With 73 spectral identifications, we have the statistics to begin
exploring the range of galaxy types in the SMG population (Table~6).
Eighteen of our galaxies (25\% of the sample) show clear AGN signatures
in their spectra: three of these are broad-line AGN, with the remaining
fifteen being narrow-line AGN.  A larger fraction of SMGs (30/73) have
restframe UV spectroscopic characteristics similar to those of
star-forming galaxies, without any identifiable AGN signatures.  The
remaining 25 spectroscopically-identified SMGs in our sample have
redshifts which are identified primarily through a strong Ly$\alpha$
line, and do not have the continuum signal-to-noise to rule out weak
AGN features.  However, we note that the limits on their Ly$\alpha$ to
[C{\sc iv}]$\lambda$1549 line ratios are generally consistent with
those expected for starbursts.  We stress that the relatively high
completeness of our survey relies in part upon the surprising strength
of the Ly$\alpha$ emission from these supposedly highly-obscured
galaxies.  The flux of the Ly$\alpha$ emission line indicates that
Ly$\alpha$ photons can readily escape from submm-selected galaxies and
suggests that they may have very patchy and inhomogeneous dust
distributions (see Neufeld 1991; Chapman et al.\ 2004b).

The high fraction of UV-emission line galaxies in our sample (38/56 of
those at high enough redshift to detect Ly$\alpha$, or 68\%) is
striking as compared with restframe UV-selected galaxies at $z=2$--3,
given the similarity of their broadband colors (discussed below).  The
Lyman-Break Galaxy population (LBGs, Steidel et al.\ 2003) at $z \sim
3$, show strong Ly$\alpha$ emission in only 25\% of cases (Shapley et
al.\ 2003), and their lower-redshift counterparts, the BX/BM galaxies,
show an even lower fraction with Ly$\alpha$ emission -- (Steidel et
al.\ 2004). In part, however, this might be a a selection effect
resulting from the difficulty of measuring absorption-line redshifts
for the fainter SMGs (from the total spectroscopically observed sample
of 98, there are 38/81 or 47\% Ly$\alpha$ emission-line systems in the
total sample of spectroscopically-observed galaxies, again excluding
the spectroscopically identified SMGs at redshifts too low to detect
Ly$\alpha$, $z<1.7$).  If we only consider the subsample of SMGs with
$R_{\rm AB}<23.5$ (21 galaxies) where our spectroscopic identifications
are effectively complete, Fig.~1, 10/21 are at $z>1.7$ to measure
Ly$\alpha$. 8/10 of these have Ly$\alpha$ in emission.

In addition to the 73 SMGs with robust redshifts there are 25 of the
optically-faintest radio-SMGs for which we have spectroscopic
observations, but failed to measure a redshift.  None of these sources
appear to exhibit strong, narrow emission lines in the observed
UV/optical wavebands.  There are two possible redshift ranges in which
this population could reside. Firstly, they may lie at $z\sim
1.2$--1.8, where no strong emission or absorption features fall in the
sensitive range of the LRIS spectrograph.  Alternatively, they may lie
within the redshift distribution of our spectroscopically-identified
sample, but have spectra which are characterized by absorption lines,
in particular absorbed Ly$\alpha$.  As these tend to be amongst the
faintest sources in the sample (Fig.~1), the signal-to-noise ratio in
their spectra is often lower than for our successful spectroscopic
identifications.

To summarize, from the 98 radio-SMG for which we obtained spectroscopic
observations: 18\% show obvious AGN characteristics; 31\% are
apparently star-forming galaxies; 25\% are difficult to classify, but
remain reasonable candidates to be star-forming galaxies; and 26\% are
spectroscopically unidentified, and so are unlikely to be strong AGN at
$z<1.2$ or $z>1.8$ (but could be star-forming galaxies with weak/absent
emission lines at almost any redshift).  Assuming that the redshift
distributions for the AGN and starburst populations are comparable,
including those spectrally-unclassified SMGs at $z=1.2$--1.8 with AGN
spectral signatures is unlikely to increase the fraction of such
galaxies in the total sample above 25\% (see \S3.5).  This is a lower
AGN fraction than suggested by either the identification of X-ray
counterparts with QSO- or Seyfert-like luminosities ($\gs 36$\%,
Alexander et al.\ 2003) or H$\alpha$ line widths of $\geq
500$\,km\,s$^{-1}$ ($46\pm 14$\%, Swinbank et al.\ 2004).  We conclude
that perhaps a third of SMG's identified as apparently star-forming or
unclassified based on their restframe UV spectral properties are likely
to host unidentified (most likely obscured) AGN.

\subsection{Optical Photometric properties of SMGs}

In the deepest ground-based optical images the majority of SMGs are
detected in most or all wavelengths observed: this is particularly true
in the HDF-N and Westphal-14hr/SSA22 fields for which extremely deep
$UBR$ imaging from Subaru/SuprimeCAM and Kitt Peak/MOSAIC imaging
(Capak et al.\ 2004), and deep $U_n,g,R_s$ (hereafter $UgR$) images
(Steidel et al.\ 2003) exist respectively.

In Fig.~\ref{ugr} we show the $(U-g)$--$(g-R)$ color-color diagram for
those radio-SMGs with robust spectral identifications and redshifts
$z>1.5$.  We compare these to the $z\sim3$ and $z\sim2$ selection
criterion presented in Steidel et al.\ (2004).  Color transformations
were determined between filter bands by matching the catalogs of
$z\sim2$ and $z\sim3$ galaxies lying in the extended-HDF region, and
rederiving the colors from the Capak et al.\ images.  The $UBR$
AB-magnitudes of the Steidel et al.\ (2004) BX/BM galaxies in the HDF
region were first measured using the SUPRIME and MOSAIC images of Capak
et al.\ (2004).  A new BX color-selection box was then defined
empirically for these $UBR$ filters.  The median offset in magnitudes
$U-U_n$, $B-g$, $R-R_s$ (0.22, 0.13, 0.04) was then applied to our
measured $UBR$-mag for the SMGs.

%
%
\begin{inlinefigure}\vspace{6pt}
\psfig{figure=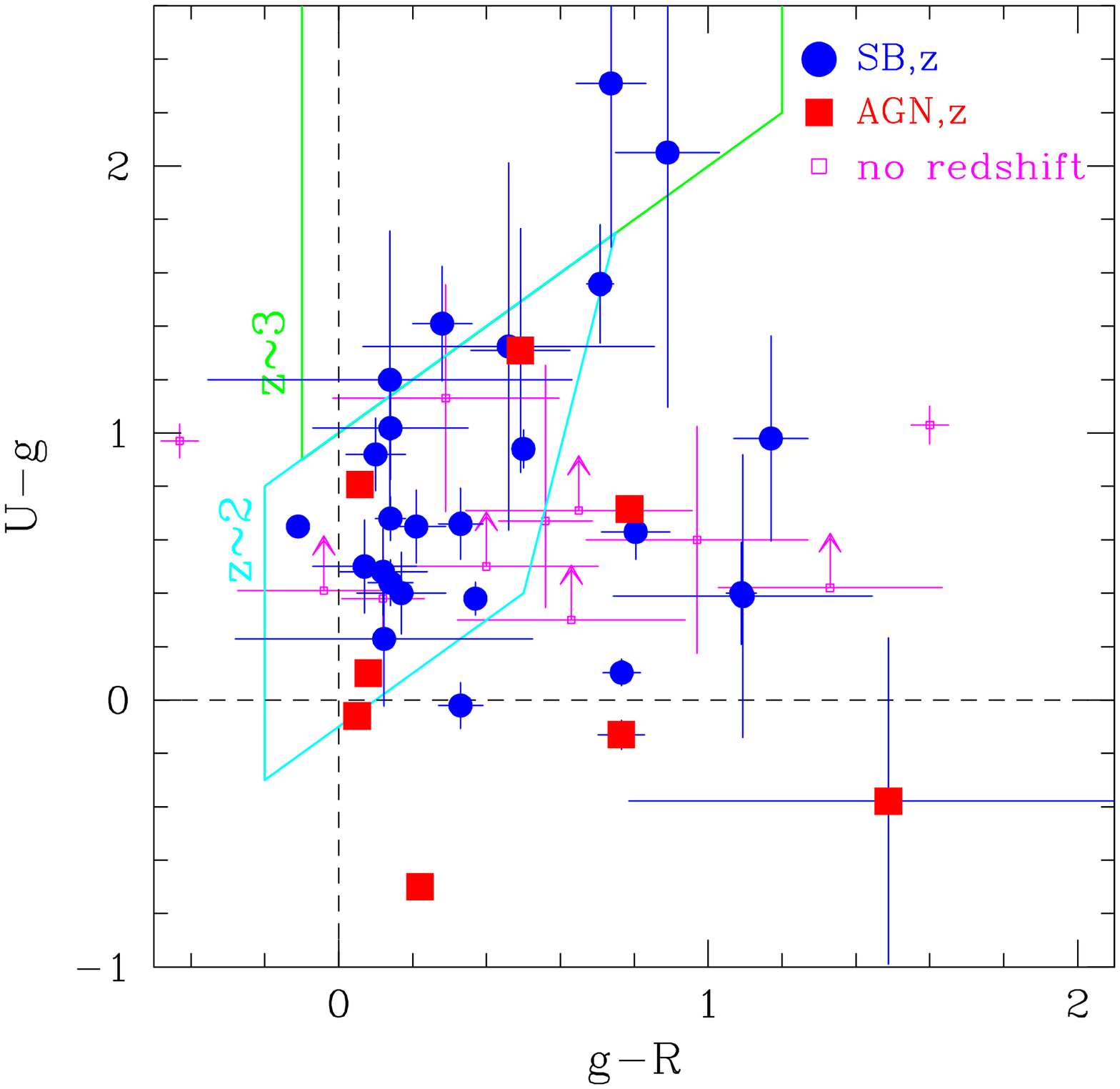,angle=0,width=3.5in}\vspace{6pt}
\caption{\small
The observed frame $(U-g)$--$(g-R)$ color-color diagram for radio-SMGs
in the HDF-N and Westphal-14hr fields, where deep photometry exists.
Circles depict SMGs with redshifts, squares show those galaxies with
AGN spectra, and small squares denote radio-SMGs without spectroscopic
identifications. Most of these latter category have only a lower limit
on their $(U-g)$ colors.  The color regions corresponding to $z\sim3$
(upper region) and $z\sim2$ (lower region) selection are shown (Steidel
et al.\ 2004).  All SMGs lying within the denoted $z\sim2, z\sim3$
regions have spectroscopic redshifts consistent with their colors.  All
colors are on the AB-scale.
}
\label{ugr}
\addtolength{\baselineskip}{10pt}
\end{inlinefigure}

Many of the SMGs ($\sim$65\%) lie well within the $z\sim2$ color
selection region from Steidel et al.\ (2004), in accord with their
spectroscopic redshifts.  Approximately 30\% of the galaxies are too
red in $(g-R)$ to be selected by the restframe-UV criterion, possibly
because of dust extinction. Smail et al.\ (2004a) present a complete
study of the extinction properties of SMGs using near-IR photometry.
The relative classification of AGN and SB populations of SMGs suggest
that the strong emission-line AGN tend towards redder $g-R$ colors with
a median $g-R$=0.5$\pm$0.1, compared with the non-AGN sample
$g-R=0.4\pm0.1$.

We also show the colors of sources for which we failed to obtain
spectroscopic redshifts. Many of these are detected near the limit of
the photometry in the $BR$ passbands, but are undetected at $U$, and
are thus shown as lower limits in $(U-g)$ in Fig.~\ref{ugr}.  Many of
these galaxies also lie within the $z\sim2$ color selection region, but
with much larger uncertainties.  This provides some additional evidence
that the radio-SMGs for which we were unable to obtain spectroscopic
identifications likely span a similar range in redshift to those with
robust identifications.

While the colors of many SMGs appear consistent with the BX-selection,
a significant fraction ($\sim$50\% of the total radio-SMG sample, and
$\sim$30\% of the spectroscopically identified radio-SMGs) are too
faint to be selected in typical BX/LBG samples (typically $R_s<25.5$ --
Steidel et al.\ 2004).  All SMGs lying within the denoted $z\sim2,
z\sim3$ regions have spectroscopic redshifts consistent with their
colors.  SMGs lying outside these boxes typically have $z\sim2$ but
redder $g-R$ than BX galaxies, or have colors which are very blue for
their redshifts, likely a result of their AGN nature.  Of the SMGs
depicted in Fig.~\ref{ugr} (having $z>1.5$ and lying in the HDF,
Westphal-14, SSA22 fields), the median and average quartile
distribution of the total sample is $R=25.3\pm0.8$, while the
distributions for the spectroscopic successes and failures are
respectively, $R=24.9\pm0.9$ and $R=25.7\pm0.4$.

This exercise suggests that deep optical imaging may provide reliable
photometric redshifts, immune from the temperature uncertainty that
plagues simple radio/submm photometric redshift estimates (see \S~3.6).
Using the {\sc hyper-z} software (Bolzonella et al.\ 2000) for this
sample and their $UgRiK$ magnitudes, we derive photometric redshifts
with a median error of $\sim30$\% (see also Smail et al.\ 2004a and
Pope et al.\ in preparation).

\subsection{Submm/Radio indices and redshifts}

In Fig.~\ref{cy}, we show the ratio of submm to radio flux as a
function of redshift for our SMG sample.  We also assess the locations
of the eight radio-identified submm galaxies from the literature with
robust redshift identifications.  The variation of the submm/radio flux
ratio with redshift is the basis for the Carilli-Yun redshift indicator
(Carilli \& Yun 1999, 2000).  However, this diagram also depicts the
variation in SED properties spanned by the SMGs, although these are
subject to considerable selection effects.  This figure is in essence a
depiction of the joint evolution in the FIR/radio correlation and the
range in dust properties in luminous galaxies (see \S~4.1).

We overlay the tracks of two galaxy classes on Fig.~\ref{cy},
representative of quiescent spirals, such as the Milky Way, and
ultra-luminous galaxies (L$_{bol}\sim10^{12}$\,L$_\odot$), such as
Arp220.  We also show the track derived by Yun \& Carilli (2002) using
a compilation of low and high redshift galaxy samples.  This latter
track, and other similar ones from the literature (Dunne et al.\ 2000;
Barger, Cowie \& Richards 2000), have been widely used to estimate SMG
redshifts.  If the SMGs were to represent a population with a narrow
range of spectral energy distributions, then we would expect them, on
average, to trace a well-defined track in the CY diagram. As they
appear to be widely scattered at a fixed redshift (an order of
magnitude range in the flux ratio at $z\gs 1$), a natural
interpretation is that they span a range in SED shapes (characterized
by T$_{\rm d}$, \S~4.1).  We note that our radio selection criteria may
bias us towards identifying objects with enhanced radio-to-FIR emission
(similar to the local {\it IRAS} galaxy, Mrk\,231, whose AGN
contributed a comparable radio luminosity to the starburst in this
galaxy).  The preferential inclusion of these systems in our
radio-identified sample would lower the typical flux ratio of the high
redshift SMGs in our \ratio\ diagram by about 0.3 dex. Yun, Reddy \&
Condon (2001) suggest that such galaxies could make up a few to 10\% of
submm galaxies, implying at most a small bias.

%
%
\begin{inlinefigure}
\vspace{6pt}
\vspace{6pt}
\psfig{figure=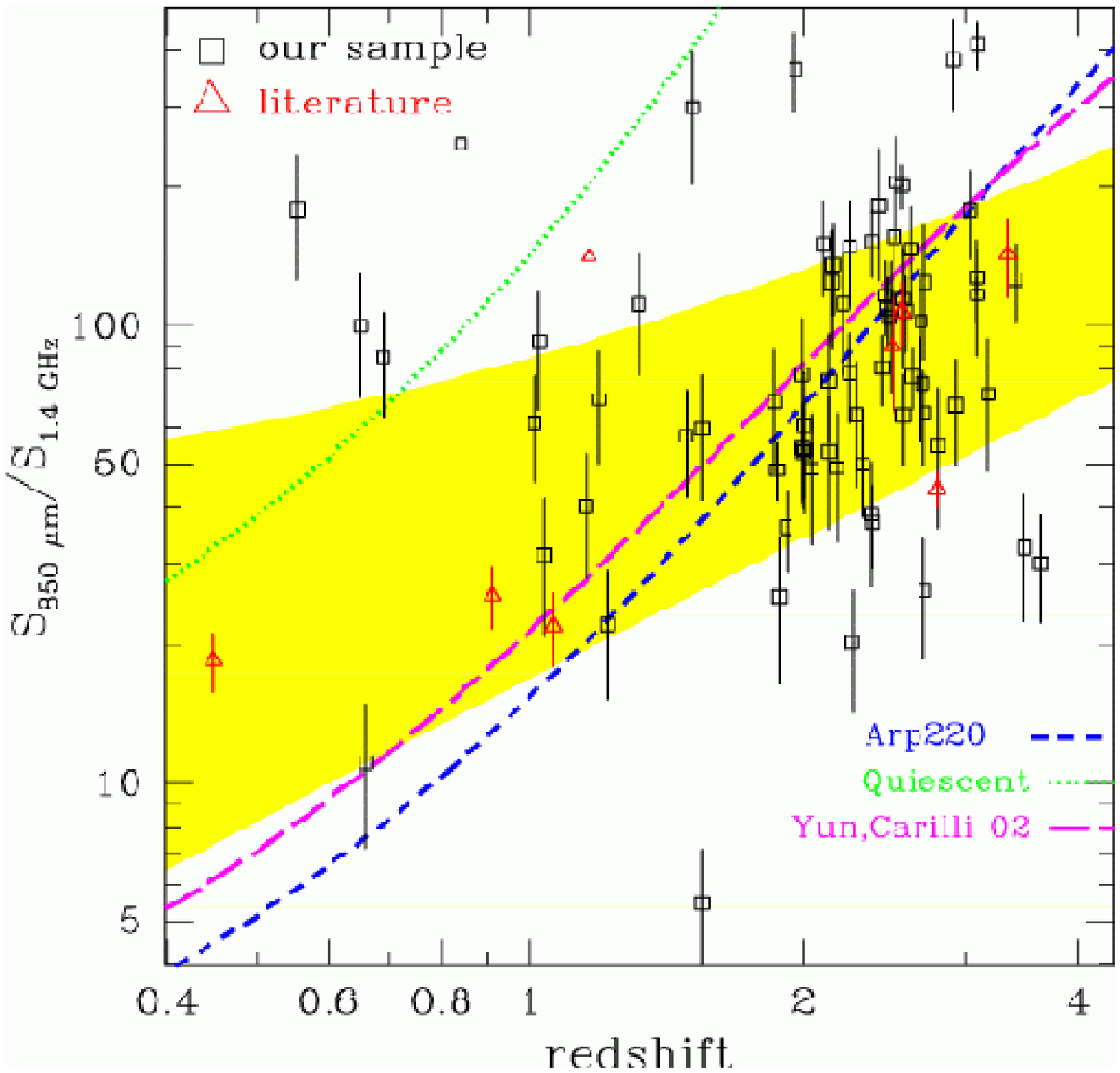,angle=0,width=3.4in}\vspace{6pt}
\vspace{6pt}
\figurenum{7}
\caption{\small
A plot of \ratio\ versus redshift for the radio-SMGs with spectroscopic
redshifts from our sample and from literature sources not lying within
our fields (Smail et al.\ 2002, 2003b; Chapman et al.\ 2002d).  The
shaded region shows the $\pm1\sigma$ envelope of the rms dispersion
calculated in three redshift bins containing equal numbers of SMGs.
The predicted variation in flux ratio for three SEDs are overlaid:
Arp~220, a quiescent spiral galaxy (such as the Milky Way) and an
empirically-derived track from the literature (Yun \& Carilli 2002).}
\label{cy}
\addtolength{\baselineskip}{10pt}
\end{inlinefigure}

To quantify the dispersion of \ratio\ we divide the sample in redshift
into three bins containing equal numbers of SMGs and plot the
1-$\sigma$ dispersion envelope on Fig.~\ref{cy}.  Clearly a large
dispersion in SED properties is spanned by our SMG sample, with a
preference for the warmer templates (the Arp220, rather than quiescent
models). The trend of submm/radio flux ratio with redshift is much
flatter than the rapidly rising template tracks, and can be
parameterized by:
$$ S_{850 \mu m}/S_{1.4 GHz} = 11.1 + 35.2 z, $$ with an average RMS
dispersion in the range $z=1$--4 of $\sigma($\ratio)\,$\sim 40$.  The
shallow slope of this trend greatly increases the error in the redshift
estimate for a given submm/radio ratio.  This behaviour can be
partially explained by the fact that lower redshift sources tend to be
lower luminosity and cooler, following the weak local correlation
between temperature and luminosity (Chapman et al.\ 2003c). However,
this should be only a slight effect.  The selection criteria in the
submm/radio are mainly responsible for the narrow range of submm/radio
values observed, and almost no discrimination is achieved for
individual source redshifts, where $\Delta z\sim1$ (see also Blain et
al.\ 2004).  It should be noted that a purely submm-selected sample
should show an even wider range in submm/radio flux ratios than our
sample -- which already demonstrates that the range of SED properties
in the SMG population render simple photometric redshift estimates too
imprecise to be useful for predicting redshifts for individual galaxies
(e.g., Aretxaga et al.\ 2003; Efstathiou \& Rowan-Robinson 2003;
Wiklind 2003).

The Carilli \& Yun redshift estimator has been extensively employed
(Smail et al.\ 2000; Barger, Cowie \& Richards 2000; Chapman et al.\
2001, 2002b; Ivison et al.\ 2002), but until now it has not proved
possible to critically compare it to a large sample of SMG's with
precise redshifts.  Perhaps surprisingly, the median redshift predicted
for the SMG population using \ratio\ ($z\sim 2.5$) has turned out to be
remarkably close to the value we derive.  In part this is because, on
average, the SMGs have SEDs similar to local ULIRGs and so the average
properties derived for the population are not too far wrong.  However,
this effect is further aided by the relatively narrow redshift
distribution seen for SMGs -- which limits the intrinsic dispersion in
the median redshift for the population.

%
%
\section{Discussion}\label{secdiscuss}

\subsection{Dust temperatures and Luminosities}\label{sectd}

Having constrained the redshift distribution and basic observable
properties of the SMGs, our next goal is to study the distribution of
their SED properties and bolometric luminosities.

Studies of local and moderate redshift galaxies in the radio and submm
wavebands suggest that variations in both the dust properties and the
empirical relation between the FIR luminosity ($L_{\rm FIR}$) and
1.4\,GHz radio (Helou et al.\ 1985; Condon 1992) will affect the
observed radio flux from a SMG. The dust properties in luminous IR
galaxies such as SMGs are potentially very complicated, requiring many
parameters for a complete characterization.  However, Blain et al.\
(2002) have demonstrated that the characteristic dust temperature (\td
-- the best single temperature grey-body fit to the SED) is the main
parameter influencing the ratio of submm to radio emission and hence
the observed radio flux.  Since we only have a redshift and two
photometric points in the relevant region of the SED (850\mum\ and
1.4\,GHz) from which to disentangle the dust properties, we will thus
concentrate only on the dust temperature, fixing the remainder of the
dust properties at canonical values for local Ultra-Luminous Infra-Red
Galaxies (ULIRGs).  In the absence of additional information for most
of our sources, we adopt a dust emissivity of $\beta=1.5$.

We therefore proceed by translating our observables (submm and radio
fluxes, and redshift) into intrinsic physical properties, \td\ and
total infrared luminosity $L_{\rm TIR}$ (defined as the integral
between 8\mum\ and 1100\mum, and derivable from FIR with a small
color-correction term -- e.g., Dale et al.\ 2001).  We calculate \td\
explicitly by adopting a suite of dust SED templates from Dale \& Helou
(2002), spanning an equivalent range of \td\ from 15--90\,K (e.g.,
Fig.~\ref{seds}).

%
%
\begin{inlinefigure}\vspace{6pt}
\psfig{figure=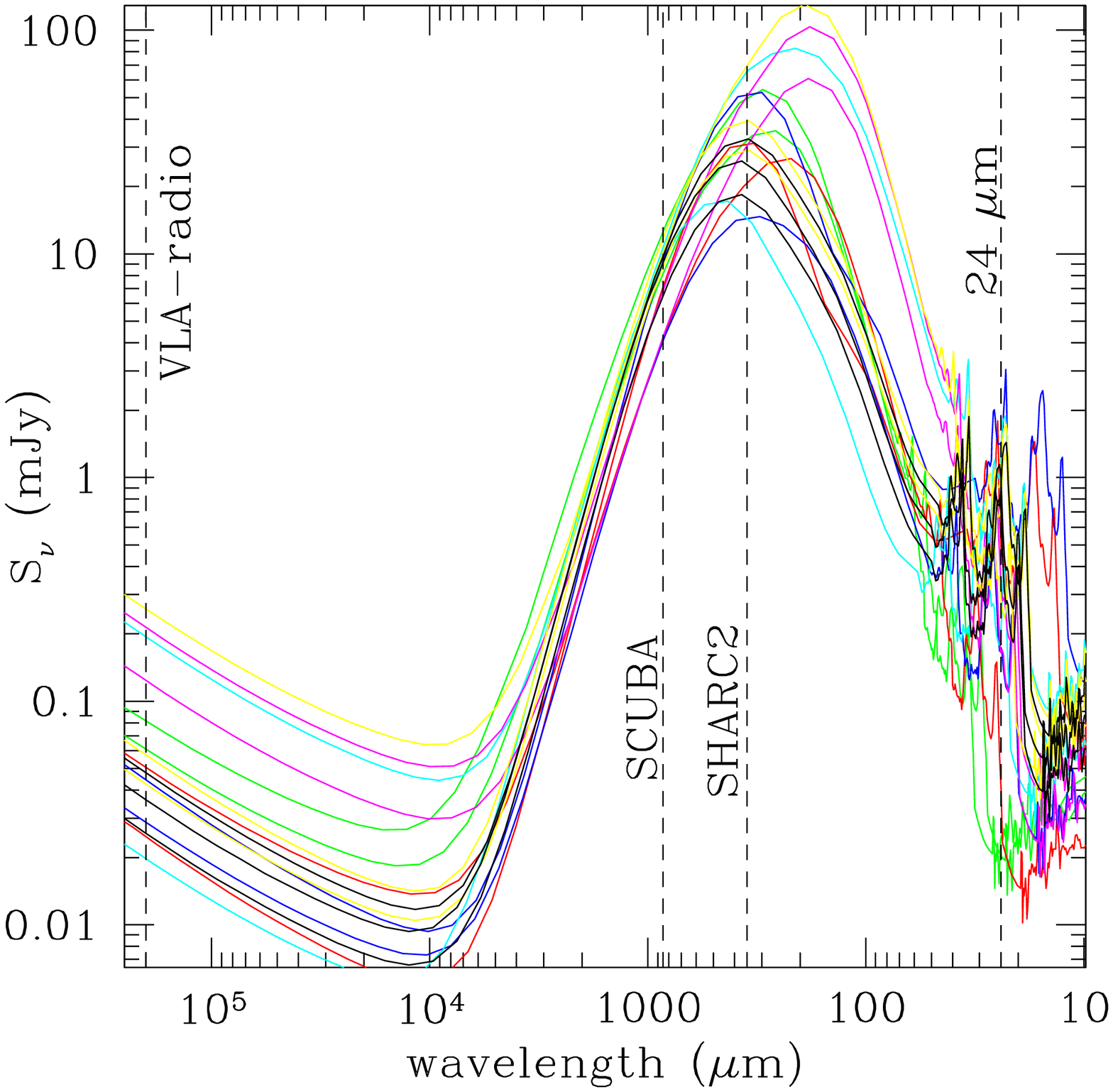,angle=0,width=3.5in}
\figurenum{8}
\vspace{6pt}
\caption{\small
Redshifted SED templates for galaxies from Dale \& Helou (2002) that
are consistent with the radio and 850\mum\ fluxes of 16 representative
SMGs from our sample.  The galaxies span a range in dust properties,
translating into characteristic temperatures from $\sim$20\,K to
$\sim$60\,K if the FIR to radio correlation holds at high redshift.  We
mark on the wavelengths corresponding to observations with SCUBA at
$850\mu$m, VLA at 20cm, SHARC-2 at 350$\mu$m and {\it Spitzer}/MIPS at
24$\mu$m.
}
\label{seds}
\addtolength{\baselineskip}{10pt}
\end{inlinefigure}

The templates assume the median value of the local FIR/radio relation
(Helou et al.\ 1985) allowing a one-to-one mapping of \ratio\ flux
ratio to \td.  These template fits to a representative selection of
galaxies from our sample are shown in Fig.~\ref{seds}.  We also assume
the low-redshift FIR/radio relation to calculate \tir\ for our sources.
Radio luminosities are calculated by first K-correcting the observed
fluxes to the restframe using a synchrotron spectrum with an index of
$\alpha=-0.75$.  Our adopted spectral index is supported by the small
number of galaxies in our sample which have measured radio spectral
indices (Ivison et al.\ 2002; Richards 2000).  In this parametrization,
the location of a galaxy with a measured redshift in Fig.~\ref{cy} is
uniquely described by its value of \td.  In Fig.~\ref{tdl} we plot
T$_{\rm d}$ versus \tir\ for our SMGs; the median \td\ is 36\,K (12\,K
interquartile range) and the median
\tir$=8.5^{+7.4}_{-4.6}\times10^{12}$\,L$_\odot$, but note the strong
correlation between observed luminosity and temperature.  With our
assumption of the FIR--radio relation, and fitting to the observed
$S_{850 \mu m}/S_{1.4 GHz}$ distribution for our SMGs, the dust
temperature is approximately proportional to
$$T_{\rm d} \propto \frac{1 + z_{\rm spec}}{(S_{850 \mu m}/S_{1.4
GHz})^{0.26}}$$ Interestingly, the SMGs in our sample show no
temperature dependence with submm flux density: the submm bright and
faint sub-samples (as described in \S~3.5) have indistinguishable
distributions in $T_{\rm d}$: $36\pm4$\,K versus $37\pm6$\,K
(interquartile error ranges).

Two strong selection effects are at work in Fig.~\ref{tdl}.  The first
is the selection effect imposed by the sensitivity limit of the submm
observations, this is shown as a shaded region in Fig.~\ref{tdl}:
galaxies with hotter dust temperatures are excluded from our sample
unless they exceed a minimum luminosity.  Similarly the requirement for
a radio detection removes the coldest galaxies from our sample,
selectively removing objects from the region in the lower right of
Fig.~\ref{tdl}.  However, the dominant effect clearing out the lower
right region of Fig.~\ref{tdl} is the steeply declining high$-L$ tail
of the luminosity function (see Blain et al.\ 2004b for a detailed
analysis).  The requirement for a radio detection is the origin of the
difference between the high-redshift tail of the observed redshift
distribution and that predict for a purely submm flux-limited sample in
Fig.~\ref{nz}.  The model suggests that a reasonable fraction of SMGs
with radio fluxes below our detection limits should have redshifts in a
similar range ($z\sim2.5$--3.5) to the radio-SMG presented here; but
that these SMGs have a slightly cooler dust temperature (at a given
luminosity) than found in our radio-SMG sample.

Both our inferred dust temperatures and luminosities depend on the
FIR--radio relation holding at high redshifts.  Based on empirical
relations of local IR galaxies, the scatter in the FIR--radio (0.2~dex)
and T$_{\rm d}$-\tir\ relations ($\sim 0.4$~dex) are expected to vary
the observed submm/radio flux ratio in SMGs, producing a random
uncertainty in our SMG calculations.  However, it is systematic errors
due to evolution of these two relations (neither of which has been
measured explicitly at high redshift) that are of more concern when
comparing the median properties of the SMG population with luminous,
obscured galaxies in the local Universe.  We now discuss possible
sources of error and the constraints which are available on the form of
the relations at high redshifts.

One possible source of error in our estimates would come if excess
radio-loud emission from an AGN was boosting the observed radio flux,
this would result in \tir\ being over-estimated.  This emphasises that
correct interpretation of Fig.~\ref{tdl} is more to do with the range
of SED types spanned by our sample as a function of luminosity.  The
SED variation between our SMGs could in principle arise entirely from a
larger scatter in the FIR--radio correlation than observed locally.  In
the extreme case that our entire sample were at the same dust
temperature of \td=36\,K, the FIR--radio correlation would have to have
a dispersion of 0.8\,dex to reproduce the variation in SEDs observed in
Fig.~\ref{tdl}.  There are physical reasons why there could be such
deviations from the correlation at high redshift.  For instance, ages,
differences in dust properties, magnetic field strength or the initial
mass function, all are conceivable in very luminous, active young
galaxies at high redshift, and all would act to increase the dispersion
in the relation (see also Eales et al.\ 2003).

Garrett (2002), has used the radio measurements of 15\mum\ {\it ISO}
galaxies to infer that $z<1$ IR galaxies broadly follow the local
FIR--radio relation.  A more reliable test of our \td--$L$ estimates is
the measurement of SMG SEDs at shorter submm wavelengths.
Fig.~\ref{seds} suggests that the peak of the SEDs for our SMGs will
lie at observed wavelengths around 350--450\mum\ and hence flux
measurements at these wavelengths will be particularly useful for
determining the characteristic dust temperatures of the galaxies.
Although the SCUBA instrument simultaneously measures 450\mum\ fluxes
(in parallel with our 850$\mu$m observations), our 450\mum\ data is
typically of insufficient quality (due to the weather conditions) to
usefully constrain the SEDs, failing to individually detect the vast
majority of SMGs.  However, stacking our 450\mum\ observations does
verify that the FIR/radio and 850\mum\ points appear to predict \td\
reasonably well, although calibration uncertainties at 450\,$\mu$m make
this a difficult measurement (see also Dunne et al.\ 2002).  The
average (inverse variance-weighted) measured flux is $S_{450 \mu
m}$=32$\pm$6\,mJy for our SMG sample, while the median FIR--radio ratio
predicted $S_{450 \mu m}$=41$\pm$12\,mJy.  As we have noted, we do not
have sufficient signal-to-noise on an object-to-object basis at
450\mum\ to determine how tight the correlation is.

%
%
\begin{inlinefigure}\vspace{6pt}
\psfig{figure=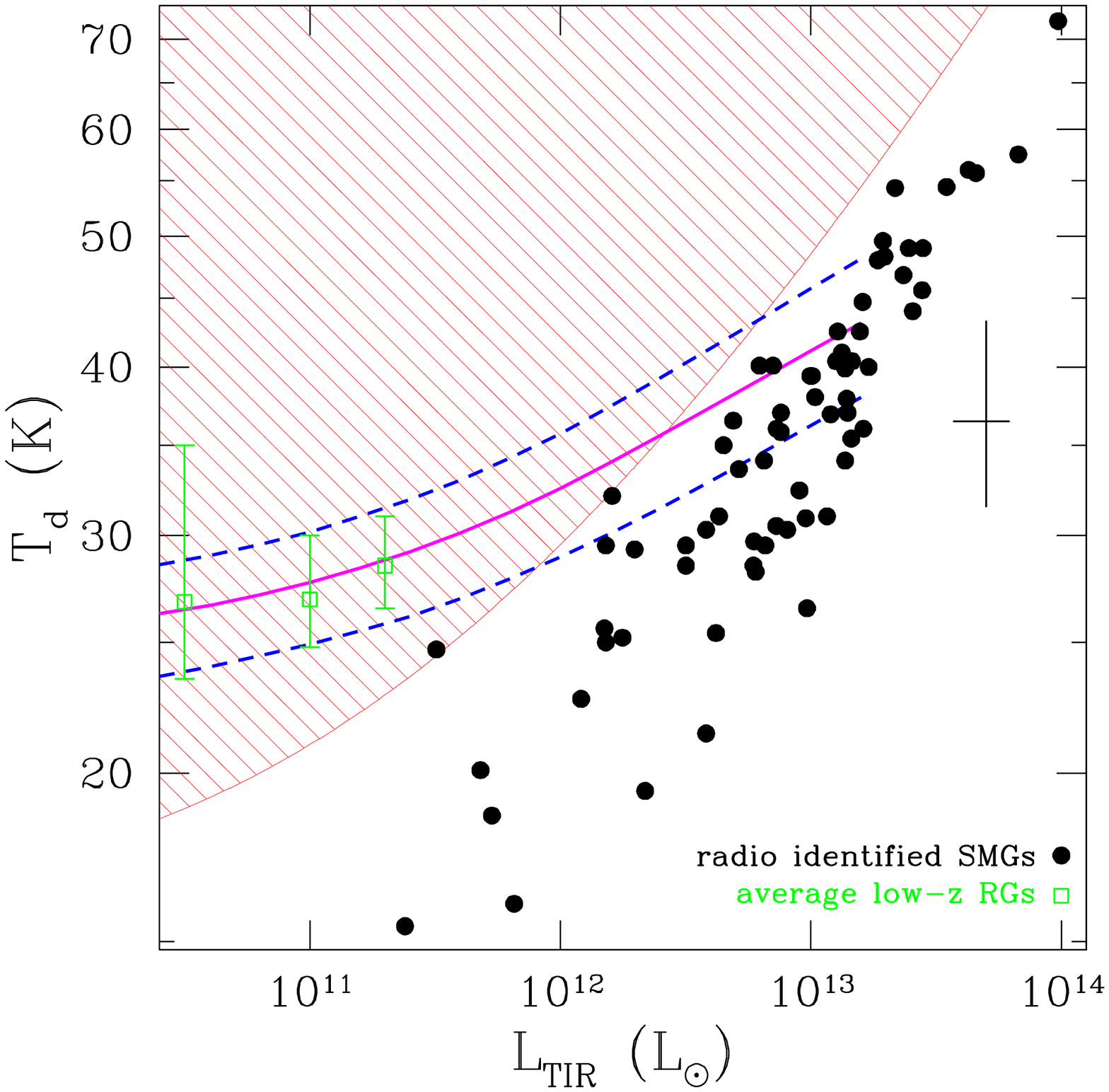,angle=0,width=3.5in}\vspace{6pt}
\figurenum{9}
\caption{\small
Characteristic dust temperature (T$_{\rm d}$) versus the log of the
total IR luminosity for radio-SMGs with spectroscopic redshifts from
our sample.  We see a tight trend of inferred temperature with IR
luminosity -- but propose this mostly arises from the selection
criteria for our sample.  The Chapman et al.\ (2003c) derivation of the
median and interquartile range of local {\it IRAS} galaxies from the
1.2-Jy 60\mum\ catalog are shown, linearly extrapolated to
10$^{13}$\,L$_\odot$.  The average \td\ for $z=0.3$--1 $\mu$Jy radio
sources from Chapman et al.\ (2003a) are shown and agree well with the
trend from {\it IRAS} distribution.  The submm flux limit precludes
detection of sources in the shaded region (shown for 3\,mJy, the
weakest source flux in our sample). The average error bar for galaxies
in the SMG sample is shown at the median temperature of the sample,
offset arbitrarily in luminosity.
}
\label{tdl}
\addtolength{\baselineskip}{10pt}
\end{inlinefigure}

At somewhat shorter wavelengths, Kovacs et al.\ (2004) have detected
more than 10 of our sample using the 350-$\mu$m SHARC-2 camera on the
Caltech Submillimeter Observatory.  These results suggest that our \td\
predictions are accurate to $\sim$20\% (without considering systematic
calibration effects of SHARC-2), corresponding to an uncertainty in the
total IR luminosity of $\sim$85\% (\tir\ $\propto$ \td$^{3.5}$ for
$z\sim2.2$ sources).  This implies that the FIR/radio correlation can
be used to predict the dust temperature of a typical SMG with
reasonable precision, and therefore that radio observations alone could
be used to estimate FIR luminosity of a star-forming galaxy for many
purposes, once the spectroscopic redshift has been measured.

Fig.~\ref{tdl} provides a complementary view of the \ratio--redshift
diagram in Fig.~\ref{cy}, with the redshift information now hidden in
the luminosity and differences in \ratio\ manifested as changes in
T$_{\rm d}$. Fig.~\ref{tdl} does however allow for a direct consistency
check with the properties of local and moderate redshift {\it IRAS}
galaxies, of which our radio-SMG population could represent
high-redshift analogs.  To provide a stepping-stone to the
high-redshift SMG population, we have also calculated the \td\ values
for $z=0.3$--1 $\mu$Jy radio galaxies (Fig.~\ref{tdl}) from Chapman et
al.\ (2003a) in the same manner as for our radio-SMG sample, using the
\ratio\ ratio.  These allow us to confirm that the form of the T$_{\rm
d}$--$L$ relation from the local {\it IRAS} samples appears to hold out
to $z\sim 1$.  Note however that these lower redshift radio sources are
not individually detected in the submm, and so the error bars are
dominated by measurement errors, rather than a true distribution of the
\td\ values.  The average submm flux densities of these galaxies are
$\sim$1\,mJy, and so they fall within the shaded submm flux selection
region of the diagram.

Fig.~\ref{tdl} demonstrates that the distribution of the radio-SMG
population appears inconsistent with the expected local T$_{\rm
d}$--$L$ relation characterized by Chapman et al.\ (2003c) (see also
Blain et al.\ 2004b).  The median \td\ of our SMGs (36\,K) is lower
than the locally predicted \td\ for galaxies of these luminosities (at
\tir=$10^{13}\lsun$, the local median is 42\,K).

As we noted in \S3.6, our radio detection criterion might be expected
to bias us towards galaxies with SEDs similar to Mrk~231, with radio
excess over the local FIR/radio relation, which make up 10\% of local
{\it IRAS} galaxies (Yun, Reddy \& Condon 2001).  Such galaxies would
be offset to {\it hotter} temperatures in Fig.~\ref{tdl}, in the
opposite sense to the offset we see and are therefore not likely to
represent a large fraction of our SMGs.  Nevertheless, the decreased
median \td\ (relative to the local prediction) could still be a
selection effect, since we have potentially missed luminous galaxies
with both hotter and colder dust temperatures through our combined
radio/submm selection function.  Chapman et al.\ (2004b) have uncovered
a sample of apparently hot and luminous galaxies (lying above our submm
selection boundary) whose inclusion would increase the median \td\
significantly.  In contrast, Ivison et al.\ (2005) have suggested
likely identifications for submm galaxies in our sample that lack radio
detections; these galaxies would appear on Fig.~\ref{tdl} with colder
\td\ than our radio-SMGs.  Both of these {\it missing} populations of
luminous galaxies suggest that the true \td-scatter in high-redshift,
luminous galaxies is larger than observed locally. The scatter in the
observed temperatures of our SMGs (12\,K interquartile range) is
already larger than observed locally (8\,K interquartile range).  While
we note that some broadening of the local distribution is suggested for
the highest luminosity galaxies, any additional hotter or colder SMGs
in the high-redshift sample would only increase the this difference in
the dispersion of the two populations.

On their own our spectroscopic redshifts and radio/submm photometry for
radio-SMGs can not provide a complete picture of the range in SEDs
spanned by the most luminous galaxies in the Universe.  However, the
available evidence suggests that some caution should be exercised when
assuming that the \td--$L$ properties of luminous, high-redshift dusty
galaxies are identical to those at low redshifts.

\subsection{The bolometric luminosity function}

Estimates of radio-SMG dust temperatures using the FIR--radio
correlation allow the calculation of bolometric luminosities, and we
can use the effective volume of our submm redshift survey to construct
a bolometric luminosity function in two redshift ranges at
$z=0.9\pm0.3$ and $z=2.5\pm0.5$ (Tables~4, 5).

The 850-\mum\ fluxes of the galaxies in our sample span a factor of 5
range, translating into a similar range in submm luminosities.  By
including the temperature information we can estimate bolometric
luminosities, which span a slightly larger range, $\sim10\times$.  We
note again that our bolometric luminosity calculations assume that the
\td--$L$ properties of luminous, high-redshift dusty galaxies are
similar to those at low redshifts, and some caution should be exercised
when interpreting the results.

Computing the raw volume densities of the radio-SMGs is accomplished
using the radio luminosity and an accessible volume technique as
described in Avni \& Bahcall (1980).  We adopt a general form for the
luminosity function
$$ \Phi(L)\Delta L = \sum_i \frac{1}{V_i}$$ with $\Phi(L)\Delta L$ as
the number density of sources (Mpc$^{-3}$) in the luminosity range
$L$--$L+\Delta L$.  The accessible volume, $V_i$, represents the {\it
i}-th source in the sample, the maximum volume in which the object
could be located and still be detected in our 1.4-GHz VLA maps.  The
sum is over all sources within the luminosity range.  We then map
sources to their FIR luminosity using the results of \S~4.1.  This
procedure naturally accounts for the spectroscopic incompleteness of
our survey, assuming that our spectroscopically-identified sources are
representative of all the total population of radio-SMGs.  The {\it
spectroscopic desert} has also been taken into account, since the
volume between $z=1.5$--1.8 is not included in either our $z=0.9$ or
$z=2.5$ bins.

There are several selection effects in our samples whose influence on
the luminosity functions cannot be easily quantified. The most obvious
of these are the overlap with radio-unidentified SMGs, and differences
in radio detection rate as a function of submm flux (fainter submm
sources are likely to have fainter radio counterparts, and our
completeness in identifying radio counterparts will be reduced).  We
have discussed the probable redshift range for the $\sim$35\% of the
total SMG population without radio identifications.  These sources
probably mostly lie at higher redshifts (Fig.~\ref{nz}), but overlap
with the high-redshift tail of the radio-SMGs (\S~2.4), producing
increasing incompleteness for the faintest and coldest sources at
$z>2.5$.  We may therefore have slightly underestimated the power-law
slope of the luminosity function: and we note that incompleteness
quickly dominates the high-redshift sample at $<$10$^{13}$\,L$_\odot$.

Comparing the number counts as a function of submm flux for our sample
with those from imaging surveys (e.g., as compiled in Blain et al.\
2002; Borys et al.\ 2004) allows us to estimate our relative radio
completeness with submm flux.  SMGs with $S_{850\mu \rm m}>$5\,mJy are
detected in the radio $\sim 6\times$ more frequently than 3--5\,mJy
sources (Chapman et al.\ 2002b -- their Fig.~8).  A correction was
calculated for the effect of differential radio selection by fitting
this progressive divergence of the radio-SMG count from the total submm
count with decreasing submm flux.  This correction was applied to the
submm luminosities, before translating to FIR luminosities.  The result
is a steepening of the slopes of the FIR luminosity function.  The
correction is designed to reduce the sensitivity of our luminosity
function to the \td-dependent radio selection in our faintest bins,
while representing the pure radio-identified sample for the brighter
SMGs.

In Fig.~\ref{lbol}, we plot our luminosity functions, parametrized by
FIR luminosity (\fir).  We compare this to the equivalent distribution
for local {\it IRAS} galaxies: the local FIR luminosity function for
galaxies with S$_{60\mu m}>$1.2\,Jy from Chapman et al.\ (2003c), which
is constructed in a consistent manner to our SMG estimated function
using the Dale \& Helou (2002) SED templates.

The comparison in Fig.~\ref{lbol} reveals an evolution in number
density of three orders of magnitude for FIR-luminous galaxies locally
and at $z\sim 2.5$.  The slopes of the bright ends of our SMG
luminosity functions are very similar to the local {\it IRAS}
distribution, suggesting that the evolution from the local function
through the $z\sim1$ and $z\sim 2.5$ functions are consistent with
almost pure luminosity evolution at the bright end (although density
evolution cannot easily be distinguished from luminosity evolution for
the bright end of the LF -- e.g., Chary \& Elbaz 2001).  We note
however that density evolution is known to overproduce the submm
background (Blain et al.\ 1999a), since it would result in a very large
number of low luminosity objects.  The turnover in our lowest
luminosity bins primarily reflect incompleteness in our survey,
although we have attempted to correct for this as described above.

To better constrain the form of the faint-end of the FIR luminosity
function at $z\sim 2.5$ we need to turn to samples of galaxies selected
in other wavebands.  In particular, we can try to use limits on the FIR
emission from UV-selected galaxies to place a lower limit on the space
densities of high-redshift sources with much lower FIR luminosities
than are detectable with current submm facilities.  The submm
properties of the $z\sim3$ LBGs have been discussed by Chapman et al.\
(2000, 2001b), Peacock et al.\ (2000), and Webb et al.\ (2003b).
Chapman et al.\ (2001b) and Peacock et al.\ (2000) report $\sim3\sigma$
statistical detections of luminous LBGs at 850\,$\mu$m, while Webb et
al.\ (2003b) report a formal non-detection of the combined sample of
LBGs lying within their CFRS-03 and Westphal-14 survey fields. We place
the Chapman et al.\ (2001b) and Peacock et al.\ (2000) measurements on
Fig.~\ref{lbol}, by assuming a \td=36\,K dust template to estimate
\fir. We plot both points as lower limits owing to the unknown fraction
of the FIR-luminous population which are missed by the UV selection.
Nevertheless, it is clear that the inferred properties of the
UV-selected populations support a steep faint-end slope to the $z\sim
2.5$ FIR luminosity function which is similar in form to that seen for
{\it IRAS} galaxies in the local Universe.

We also show on Fig.~\ref{lbol} the inferred FIR luminosity function
based on FIR--UV-$\beta$ relation (Meurer et al.\ 1997) applied to a
survey of LBGs from Adelberger \& Steidel (2000), and normalized to the
effective volume containing the $z\sim3$ LBGs (Steidel et al.\ 1999).
For the $z\sim2$ {\it BX/BM} population (Steidel et al.\ 2004),
stacking analysis of their radio and X-ray emission have suggested that
the bolometric luminosity function of Adelberger \& Steidel (2000) is a
reasonable representation (Reddy \& Steidel 2004).

%
%
\begin{inlinefigure}
\vspace{6pt}
\psfig{figure=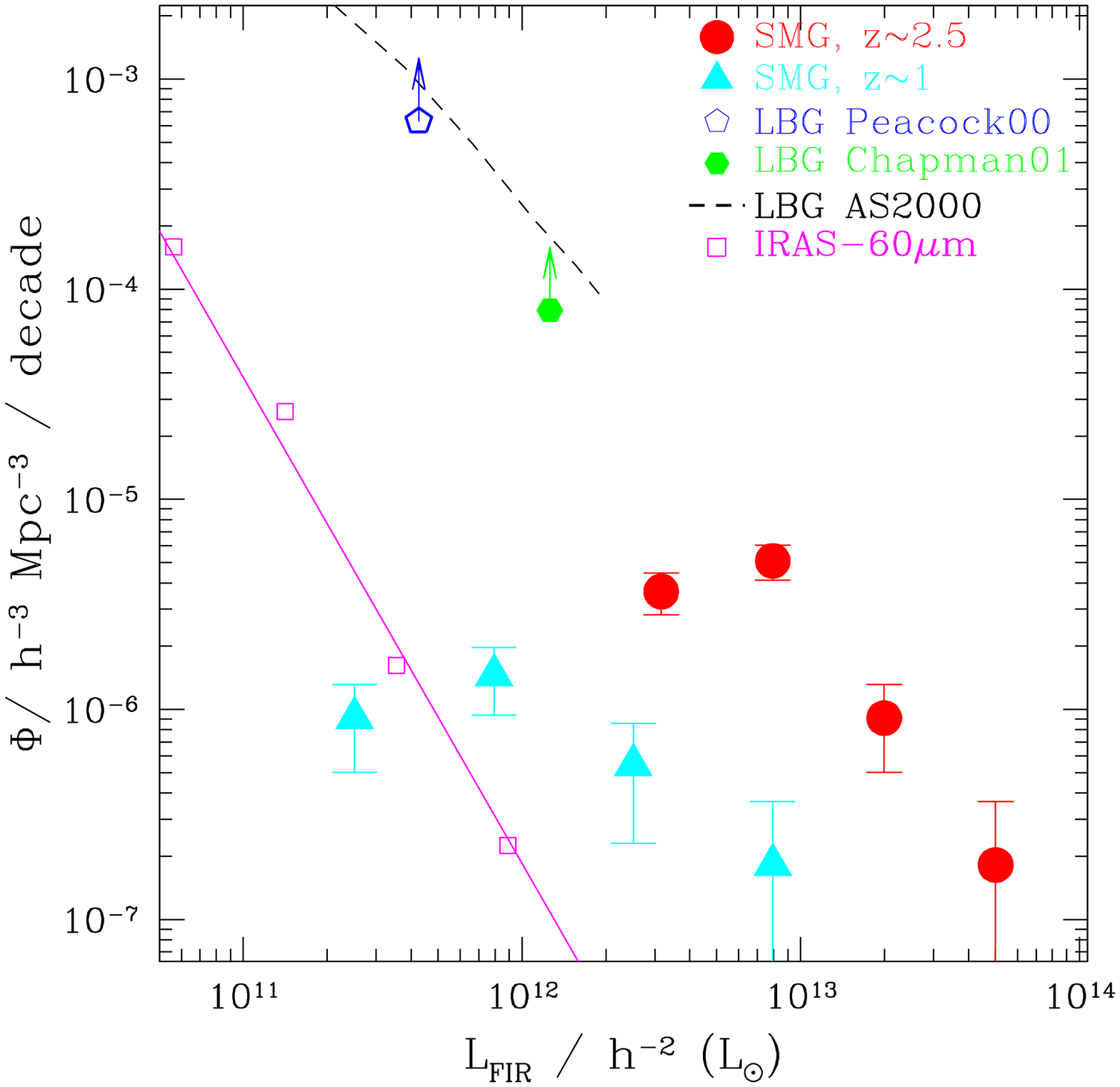,angle=0,width=3.5in}
\figurenum{10}
\vspace{6pt}
\caption{\small
The FIR luminosity functions at $z=0.9\pm0.3$ and $z=2.5\pm0.5$. \fir\
is calculated by integrating under the template SED from the Dale \&
Helou (2002) catalog which best fits the submm and radio fluxes of the
galaxy at its measured redshift.  For comparison, we show the local FIR
luminosity function from Chapman et al.\ (2003c), constructed in a
consistent manner with our $z=2.5$ function using the Dale \& Helou
(2002) templates.  Also shown are the submm detections ($>3\sigma$) of
LBGs from Peacock et al.\ (2000) for SFRs $>$1$\msun$ in the HDF, and
for those galaxies with inferred SFRs $>100\msun$ from Chapman et al.\
(2000, 2001b).  For reference we plot the predicted $\Phi$(\fir) for
LBGs from Adelberger \& Steidel (2000) which assumes the
FIR/UV--$\beta$ relation (where $\beta$ is the UV continuum slope) of
Meurer et al.\ (1997).
}
\label{lbol}
\addtolength{\baselineskip}{10pt}
\end{inlinefigure}

It is difficult to determine whether pure luminosity evolution can
explain the dramatic increase in volume density of high-redshift
galaxies with 10$^{11}$--10$^{13}\lsun$ (based on the UV- and
submm-selected samples) over similarly luminous local galaxies. The
high-redshift, FIR luminosity function remains poorly constrained in
both low- and high-luminosity regimes.  For UV-selected galaxies at
high redshift it is currently very difficult to accurately estimate
\fir\ (e.g., Adelberger \& Steidel 2000; Reddy \& Steidel 2004). In
addition, it is unclear what the completeness is in $\Phi$(\fir) for a
UV-selected sample.  The submm-estimated $\Phi$(\fir) at high
luminosities is also poorly constrained due to incompleteness effects
that are difficult to characterize (as described above).  We have
already explored the extent to which the submm-selection itself may
have significantly underestimated the total volume of luminous galaxies
at high redshifts, as galaxies with hotter characteristic dust
temperatures are missed by submm selection (Chapman et al.\ 2004b;
Blain et al.\ 2004a).  Similarly, our radio-pre-selection means we are
missing a small fraction of colder SMGs without radio counterparts
which may contribute to the number density of $z\sim2.5$ galaxies with
\fir$>10^{12}\lsun$ (Ivison et al.\ 2005).  We anticipate the deep {\it
Spitzer} observations of all these high-redshift galaxy populations
will shed additional light on this complex question.

\subsection{The restframe-UV derived \fir\ of SMGs}

By far the best-studied population of high-redshift star-forming
galaxies is that identified through their restframe UV-emission
(Steidel et al.\ 1999, 2004).  These galaxies have provided unique
insights into the evolution of the star-formation density in the
Universe and the corresponding formation of normal galaxies (Madau et
al.\ 1996).  A key question for SMGs is to understand how they fit into
the framework defined by the UV populations -- in particular, how well
does the recipe for deriving SFRs for UV-selected galaxies work on this
restframe FIR-selected population?

We can use our multi-color optical data for a subset of the SMGs to
investigate their restframe UV properties and derive SFRs in an
analogous manner to that applied to LBGs.  This analysis relies on
estimating the luminosities and spectral slopes at wavelength around
1500\AA\ in the restframe. For galaxy populations at $z\sim 2$--3, this
can be achieved using $B$- and $R$-band photometry for the HDF sources,
and using $g$- and $R$-band photometry for the SSA22 and Westphal-14
sources.

As described in \S2, we measured $BR$-band photometry for all SMGs in
our sample.  Since UV-derived luminosities and the corrections for dust
extinction, are highly susceptible to photometric errors (Adelberger \&
Steidel 2000), we need to isolate a sample of SMGs with well-measured
photometry.  Unfortunately, most SMGs at higher redshifts are faint and
as a result to obtain a reasonable sample we are required to use those
galaxies where the photometric errors are only less than 0.1 mags in
both bands.  We define the subsample which includes all 33 SMGs lying
in the HDF, Westphal-14, and SSA22 fields, with redshifts $z>1.5$ to
allow for accurate measurement of the dust-corrected UV luminosity (the
same sample used in Fig.~\ref{ugr}).  We also consider the HDF
subsample on its own (17 SMGs), since its large size and contiguous
areal coverage make it statistically representative on its own.  The
sample has a median $R_{\rm AB}$=24.9 and a 1$\sigma$ rms of 1.1, and a
median photometric error of $dR_{\rm AB}=0.04$.  The dust-corrected
luminosities are biased in a manner which is difficult to quantify.
The optically-faintest sources will have very small UV luminosities,
but may in principle have very steep continuum slopes, with large
implied dust correction factors (Adelberger \& Steidel 2000).  While we
will calculate UV luminosities for all SMGs in this subsample, we
identify those SMGs showing AGN spectra, and conservatively exclude
them from the average properties calculated below.

To estimate the SFR from the UV, we follow the prescriptions of Meurer
et al.\ (1997) and Adelberger \& Steidel (2000): the UV luminosities
are first corrected for line blanketing in the Ly$\alpha$ forest, and
then corrected for dust extinction using the UV continuum slope derived
from the $(G-R)$ color, corresponding to wavelengths between rest-frame
1000\AA\ and 2900\AA.  The transformations from the measured $(B-R)$ to
$(g-R)$ in the HDF field are described in \S~3.4.  We correct the
$(g-R)$ color for the opacity of the Ly$\alpha$ forest according to the
statistical prescription of Madau (1995).  Values of $(g-R)_{\rm corr}$
ranging from 0.0 to 1.0 correspond to a UV spectral index $\beta=-2$ to
+0.6, when the spectrum is approximated by a power law of the form
$F_\lambda \propto \lambda^\beta$.  The $(g-R)_{\rm corr}$ color is
then mapped to a color excess, E$(B-V)$, from which the dust-corrected
UV luminosities are derived.  Our median $\beta=-1.5\pm0.8$,
corresponds to E$(B-V)=0.14\pm0.15$ for a Calzetti extinction law, very
close to the distribution for LBGs presented in Adelberger \& Steidel
(2000), suggesting that the UV identifications of SMGs do not
distinguish themselves from the general LBG population with
significantly redder UV spectral slopes (c.f.\ Smail et al.\ 2004a).

The dust corrected UV luminosity translates into a SFR, following
Kennicutt (1998):
$${\rm SFR (M\odot yr^{-1})} = 1.4\times10^{-28} L_{1500} {\rm (erg
s^{-1} Hz^{-1})},$$ where the relationship applies to galaxies with
continuous star formation over time scales of 10$^8$ years or
longer. For younger stellar populations, the UV continuum luminosity is
still increasing as the number of massive stars increases, and the
above equation will underestimate the SFR.  We measure a median
SFR=13$\sfr$ from the dust corrected UV.  Finally, to compare these
estimates with those from the FIR we simply convert our UV-derived SFR
directly into a FIR luminosity (Kennicutt 1998).

From the dust-corrected UV, we predict a median FIR luminosity of
$0.055^{+0.02}_{-0.04}\times10^{12}\lsun$; this compares to the
submm/radio measurement of $L_{\rm
FIR}=5.6^{+2.1}_{-1.6}\times10^{12}\lsun$.  This corresponds to a
median radio/submm-to-UV ratio in the derived \fir\ of 100, with a
quartile range of 29--168.  We reiterate that the UV estimate has been
{\it corrected} for dust extinction in the standard manner.  The
dust-corrected UV-estimated FIR luminosities are compared directly with
the FIR luminosities measured from the radio/submm in
Fig.~\ref{FIRugr}. The relative offset between the two does not differ
significantly if we use the more statistically reliable subsample from
the HDF: $L_{\rm FIR, UV}=5.28^{+1.51}_{-3.40}\times10^{10}\lsun$
versus $L_{\rm FIR, radio}=5.79^{+1.96}_{-1.41}\times10^{12}\lsun$.  In
addition, we note that those sources in the HDF with the faintest
apparent $R$-band magnitudes do not have significantly different
UV-inferred \fir\ from those with brighter $R$ magnitudes.

As suggested above this procedure of estimating the FIR luminosity from
the UV continuum slope is meaningless for the SMGs exhibiting strong
AGN-signatures in their spectra, since the UV continuum is not
necessarily dominated by stellar emission, and so we have not included
these AGN-classified SMGs in the median calculations above.  We use
different symbols for the SMGs with AGN spectra in Fig.~\ref{FIRugr} to
identify them to the reader.

Since the FIR--radio luminosity relation is essentially linear except
at the faintest extreme (Condon 1992), the relation at high redshift
would have to vary by a similar factor to our observed discrepancy
($\sim100$) in order for the luminosity estimates to be in accord.
This is highly unlikely given the relatively well understood physics of
the relation (Condon 1992).  However, direct measurements of the
FIR-radio relation at high redshift are required to refute this
possibility. (We can begin to rule out evolution of FIR--radio at the
implied level $\sim$100$\times$, based on previous discussion in \S~4.1
-- Kovacs et al.\ in preparation.)

Clearly the {\it dust-corrected} UV luminosity using the standard
prescription rarely hints at the huge bolometric luminosities measured
in the radio/submm, underestimating the true luminosity by over two
orders of magnitude.  Three of the SMGs {\it do} have UV-inferred FIR
luminosities within a factor of three of that observed in the
radio/submm.  One of these shows hybrid SB/AGN characteristics and
complicated multi-component kinematics (SMM\,J163650.43+405734.5 --
Smail et al.\ 2003a).  The other two are apparently starbursts which
reveal the true magnitude of their bolometric luminosity in the UV.

%
%
\begin{inlinefigure}
\vspace{6pt}
\psfig{figure=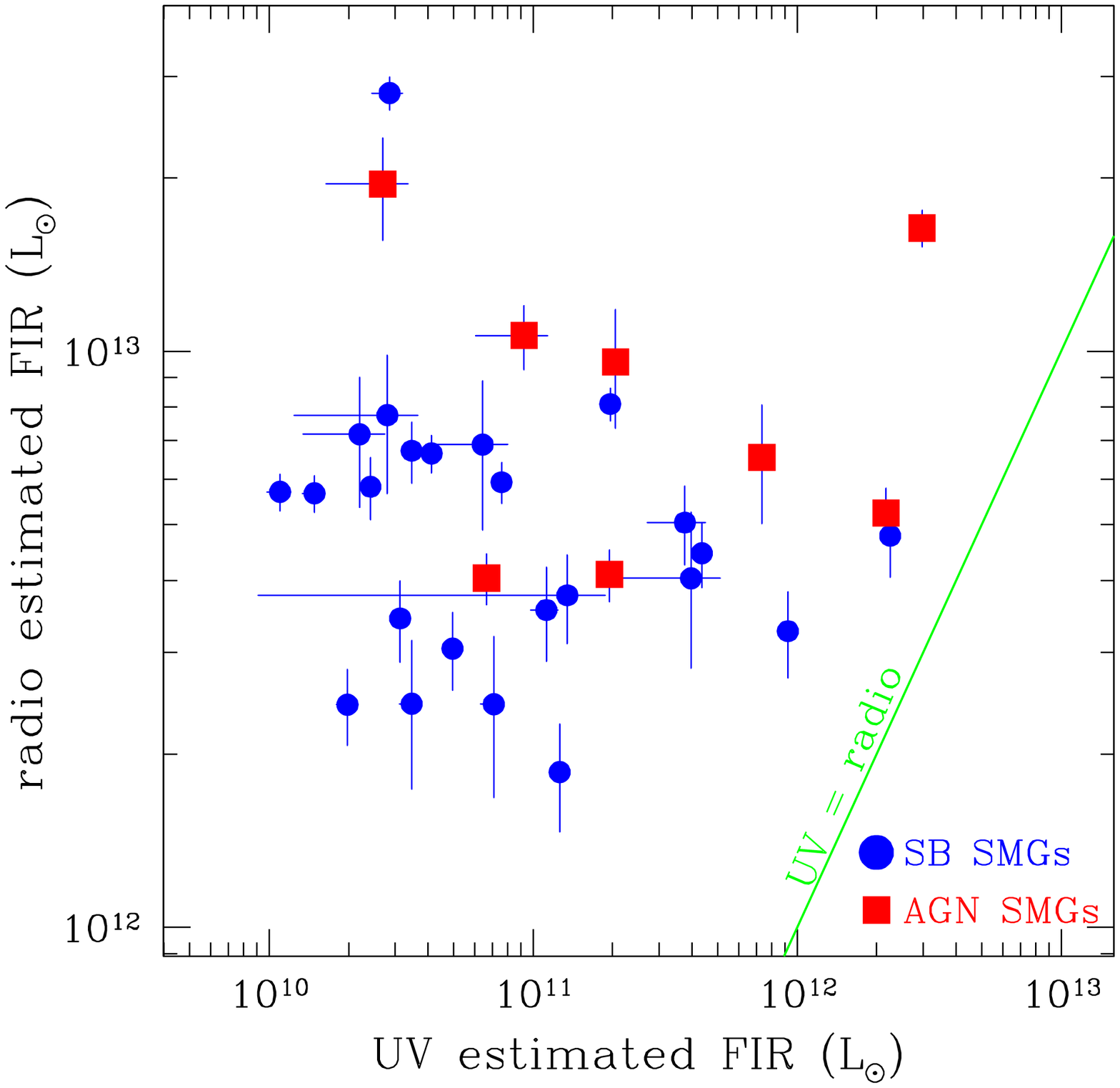,angle=0,width=3.5in}
\figurenum{11}
\vspace{6pt}
\caption{\small
The FIR luminosity of SMGs (used as a proxy for their SFRs) as measured
from the radio and submm flux density ratio, compared with the FIR
luminosity estimated from the UV luminosity and spectral slope.  The
line is a simple equality, L$_{\rm FIR}({\rm radio})$ = L$_{\rm
FIR}({\rm UV})$, the expected correlation if the dust-corrected UV
luminosity is a reliable measure of the total star formation rate in
these systems.  Note the large offsets of most SMGs from this line.
Error bars are derived from uncertainties on the radio, submm, $B$- or
$g$-, and $R$-band fluxes.  SMGs showing obvious AGN spectra are shown
by large squares.  There are no upper limits as we include only those
SMGs in the HDF, Westphal-14hr, SSA22, and ELAIS-N2 fields which are
detected in both the $B$ and $R$ bands.
}
\label{FIRugr}
\addtolength{\baselineskip}{10pt}
\end{inlinefigure}

The very large disparity we derive for the UV- and radio-estimated
luminosities of UV-detected SMGs is apparently at odds with the
conclusion of Reddy \& Steidel (2004) who find that radio (and X-ray)
estimates of the SFRs for $z\sim2$ BX galaxies on average match the
UV-derived SFRs, with an average dust correction factor of just 4.5.
How do we reconcile this with our finding that SMGs have UV-inferred
SFRs which underpredict the radio-measured SFR by a factor of 100 on
average?

One possible contributing factor is that there may be a wide range in
dust obscuration in the high-redshift galaxy population.  Chapman et
al.\ (2004b) used MERLIN radio images and {\it HST} UV images of SMGs
at $\sim$0.3\arcsec\ resolution to study the differential emission
between the two wavelengths.  They demonstrated that the radio emission
(and by implication submm emission) is always more compact than the UV
as traced by {\it HST} imagery.  In addition they found that the radio
highlights regions of low UV emission as bolometrically-luminous in
$\sim$50\% of the SMGs.  An increasing proportion of very-highly
obscured activity may therefore be present in the more active
systems. This has also been demonstrated through measurements of the
nebular H$\alpha$ emission line for our SMG sample (Swinbank et al.\
2004).  In this way the selection of flux-limited samples of galaxies
in the restframe UV and FIR would give significantly different mean
obscurations.  This is equivalent to stating that the UV luminosities
and spectral slopes are measurable for only the least-obscured regions
of the galaxies and hence are not representative of the bulk of the
emission in these galaxies.  It is not therefore surprising that they
indicate much lower bolometric emission.

We therefore conclude that although many SMGs can be detected, and
their redshifts estimated or measured in deep observations in the
restframe UV, they cannot generally be identified as bolometrically
luminous galaxies without the use of radio or submm observations.
There is a second, related important point to make before we proceed to
discuss the evolutionary history of the luminosity density contributed
by SMGs versus UV-selected galaxies over the lifetime of the
Universe. In the light of the vast mismatch in the derived bolometric
emission for this population we will assume that the contribution of
bright SMGs to the star formation density at high redshifts is not
included in current UV estimates (e.g., Madau et al.\ 1996; Steidel et
al.\ 1999) -- and so we need to derive the contribution from the highly
obscured population independently and sum this with that from the UV to
derive the total.

\subsection{Luminosity and Star Formation histories}

Using the bolometric luminosities measured above for the SMGs, we
calculate the following bolometric luminosity densities in redshift
bins from our survey: $\rho_{\rm L} (10^7\ {\rm L_\odot\ Mpc^{-3}})$ =
$2.2 \pm 2.3$, $8.0^{+4.9}_{-3.2}$ and $3.4^{+4.6}_{-2.5}$ for redshift
intervals of $z=0.5$--1.5, 1.8--2.6 and 2.6--3.5 respectively.  These
luminosity densities are calculated for the entire radio-identified SMG
population, and corrected for spectroscopic completeness in the
following way.  For the 26\% of radio-SMGs for which we did not obtain
redshifts, we corrected the clear deficit of sources in the redshift
range $z=1.5$--1.8 to result in a smooth distribution matching our
normalized model (Fig.~\ref{nz}). The remainder of the incompleteness
(17\%) was distributed uniformly over the entire $N(z)$ uncovered by
our robust redshift sample.  Our justification for this procedure is
two-fold.  First, the \ratio\ properties of sources with robust
redshifts are indistinguishable from those where we did not obtain
redshifts, suggesting a similar range in \td\ and $z$.  Secondly, the
$UBR$ colors of these two samples are similar, although with large
uncertainties as many of the spectroscopic failures are typically very
faint, and are consistent with colors of $z\sim2$ star-forming galaxies
(Steidel et al.\ 2004): see \S3.4.

In order to justify translating these luminosity densities into star
formation rate densities (SFRDs) we must first clearly identify the
signs of AGN activity in the individual galaxies, and then remove any
contribution to the luminosities of these galaxies from the AGN, either
direct contribution to the radio emission or through heating of dust by
the AGN.

We begin by assessing the possible AGN contribution in individual submm
galaxies from their UV spectral properties.  Three SMGs are identified
with QSOs in our radio-identified sample (SMM\,J123716.01+620323.3,
SMM\,J131215.27+423900.9, and SMM\,J131222.35+423814.1): these are the only
objects in which the optical luminosity is a non-negligible fraction of
the bolometric luminosity.  These sources have comparable optical and
FIR luminosities and space densities to submm-detected QSOs at $z>2$
(Omont et al.\ 2001, 2003; Carilli et al.\ 2001).  The fraction of
optically bright QSOs in the SMG population could actually be slightly
higher since $\sim$20\% of our SMGs were pre-selected with optically
faint magnitudes.  This implies that only $\sim4$\% (3/80, as a
fraction of the total radio-SMG sample observed spectroscopically,
which weren't pre-selected as optically faint) of optically-bright AGN
have \fir\ emission with comparable luminosity to that seen in the
optical waveband (since our survey would have uncovered all the QSOs
emitting strongly in the radio/submm).  This fraction of radio-SMGs are
therefore removed from consideration in determining the luminosity
density evolution.

There are several strong indications that star formation dominates the
luminosity of the majority of the 98 radio-SMGs in our sample. As
discussed in \S3.3, 30\% of our sample with the brightest UV continua
exhibit stellar and interstellar absorption lines, implying that their
continua are dominated by young, massive stars and not the non-thermal
power law spectrum of an AGN. Another 25\% of our sample have UV
continua which are too faint to detect absorption features, but remain
consistent with starbursts, while a further 25\% do not reveal enough
features to even identify a redshift.  These statistics suggest that
$\sim $\,80\% of radio-SMGs do not harbor an partially- or un-obscured,
luminous AGN.

The presence of an obscured AGN is much more difficult to determine
from the restframe UV spectra and instead we have to turn to
multiwavelength surveys of samples of SMGs.  High spatial resolution
radio observations of SMGs are one route for determining the morphology
of their radio (and by implication FIR) emission and hence search for
the presence of a strong AGN contribution to the FIR (identified as a
radio point-source).  Such studies reveal extended morphologies on
scales $>$1\arcsec\ in 65\% of cases (8/12 galaxies in Chapman et al.\
2004b).  These observations suggest that AGN do not dominate the FIR
emission of most SMGs, as star formation is the most likely route to
produce spatially-extended FIR/radio emission.  Several of the galaxies
with spatially-extended radio emission show AGN emission lines in their
rest-frame UV spectra, underlining the fact that intense starbursts are
likely to be energetically-important, even in the presence of AGN.

A second route to search for obscured AGN is to employ observations in
the X-ray waveband.  The proportion of the submm population detected in
deep {\it Chandra} and {\it XMM-Newton} X-ray surveys, which are
sensitive to even strongly dust-obscured AGN (Ivison et al.\ 2002;
Barger et al.\ 2002; Almaini et al.\ 2003; Alexander et al.\ 2003,
2005), at flux densities significantly greater than expected from star
formation alone is at most 30\%.  X-ray spectral analysis reveals that
most of the SMGs are not Compton-thick sources with QSO-like
luminosities, and instead suggest that the AGN in SMGs have modest
X-ray luminosities (Alexander et al.\ 2005).

Two more recently-pursued routes to search for the presence of an AGN
within an SMG are to use near-infrared spectroscopy and mid-infrared
imaging.  Swinbank et al.\ (2004) report on near-infrared spectroscopy
of 24 SMG's from our sample.  Of the 15 SMG's classified as
star-forming from their UV spectra, the emission line properties (line
widths and [N{\sc ii}]/H$\alpha$ flux ratios) in the restframe optical
support this classification for 60\%, a further 20\% have intermediate
classifications (H$\alpha$ line widths of 500--1000\,km\,s$^{-1}$) and
only 20\% are clear-cut AGN based on their restframe optical spectra.
This broadly supports our UV spectral classifications and implies that
only $\sim 30$\% of SMGs are likely to host luminous AGN.  A similar
rate of identification of AGN-like SEDs (20--30\%) is found in recent
{\it Spitzer} mid-infrared photometric studies of SMG's: 2/13 from the
combined sample of Ivison et al.\ (2004) and Egami et al.\ (2004) and
2/7 for the radio-identified SMGs in Frayer et al.\ (2004).

We conclude that around 20--30\% of SMGs show detectable signs of AGN
activity using a number of independent indicators.  The broad agreement
between this proportion based on such a wide-range of indicators gives
us confidence that it represents a true limit to the extent of AGN
activity in the population.  We stress that this doesn't indicate that
30\% of the FIR emission from these galaxies comes from AGN.  Even when
an obvious and strong AGN is present in an SMG, there are signs that it
does not dominate the bolometric emission (e.g., Frayer et al.\ 1998).
The AGN contribution to the luminosity of the whole SMG population may
be as low as 10\% (Alexander et al.\ 2005).  Using the results that
$\ls$\,30\% of SMGs have properties indicative of an AGN, we will
conservatively assume that 70\% of the FIR luminosity is derived from
star formation in the total SMG population.

The luminosity densities we estimated earlier can now be translated
into SFRDs by scaling them down by 30\% to account for a maximal
AGN-contribution and then using the standard calibration of $(1.9\pm
0.3) \times 10^{9}\,$L$_\odot$ (M$_\odot$ yr$^{-1})^{-1}$ (Kennicutt
1998), we plot the resulting SFRDs in Fig.~\ref{madau}.

An additional point has been plotted on Fig.~\ref{madau} to represent
the $\sim 35$\% of the total SMG population which are not detected in
our radio maps.  Their redshifts are assumed to follow the
radio-undetected fraction from our $N(z)$--normalized model
(Fig.~\ref{nz}), implying a significant overlap with the redshift range
probed by our radio-SMG sample.  The redshift range we show for the
radio-undetected sample extends to $z=5.1$, at which point less than
one SMG would lie in a total SMG sample of 151 galaxies (our targeted
98 radio-SMGs, plus an extra 35\% undetected in the radio).  This {\it
uniform population} model is the most intuitive representation of the
radio-unidentified SMGs, but we stress it is based purely on a model
and it is possible that the radio-undetected galaxies have a very
different distribution to the one we predict, extending to higher
redshifts, or even a bimodal distribution compared to our well
characterized radio-SMGs.

%
%
\begin{inlinefigure}
\vspace{6pt}
\psfig{figure=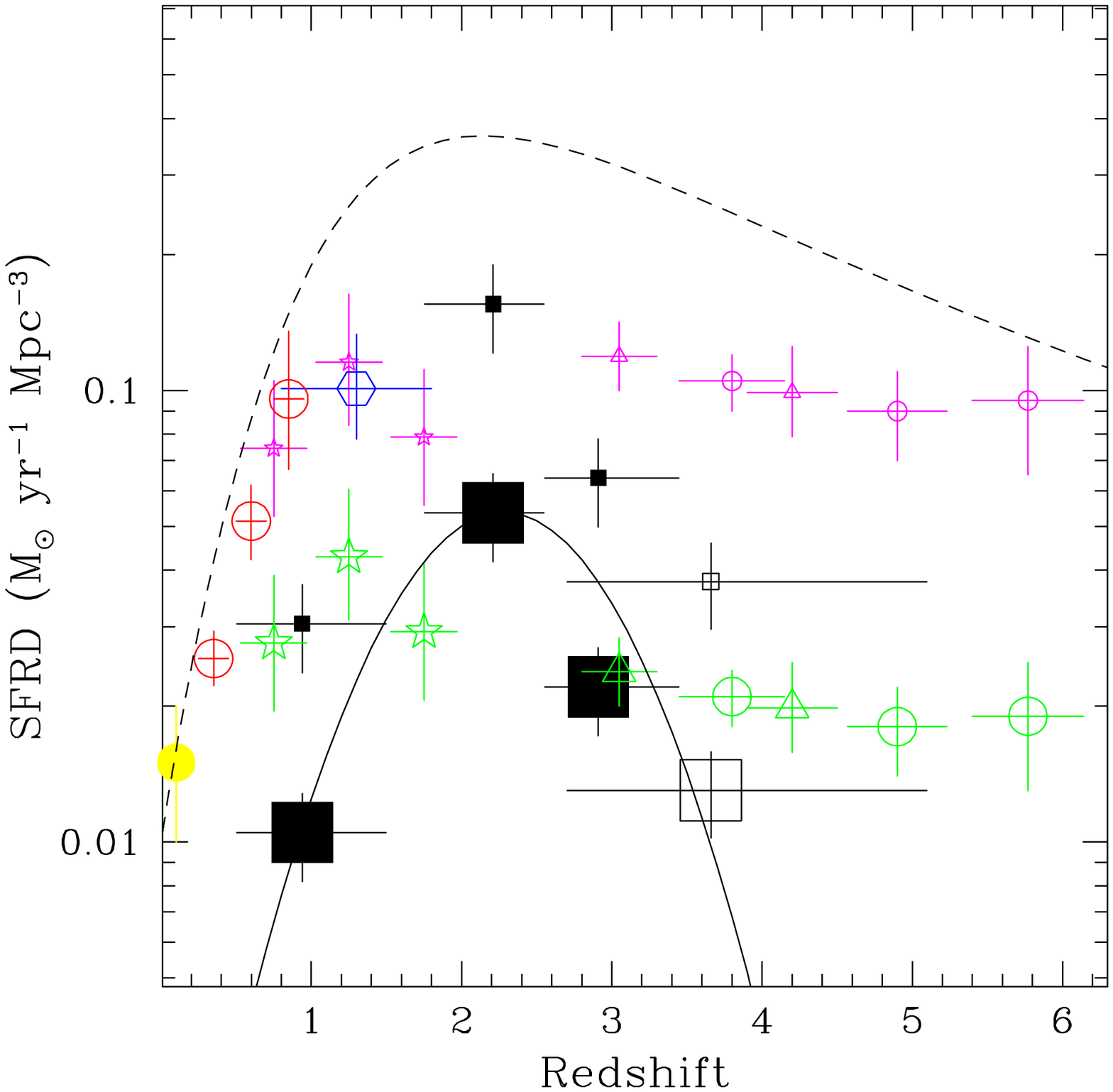,angle=0,width=3.5in}
\figurenum{12}
\vspace{6pt}
\caption{\small
The evolution of the energy density (parametrized by SFRD) in the
Universe with epoch. Our new submm measurements (large squares, shown
at the median value for each redshift bin) are compared to the
published estimates from optical/UV surveys and radio/IR tracers of the
star formation density.  The open square indicates the SMGs without
radio identification, at the median redshift derived from our modeling
of Fig.~\ref{nz}. The smaller symbols for the optical estimates
indicate dust-corrected estimates.  A Gaussian fit is shown for the
four submm galaxy points, tracing an evolution comparable to luminous
radio-selected Quasars (Shaver et al.\ 1998).  For the submm sources,
the smaller points show a simple redshift-independent correction to the
luminosity density to match the submm extragalactic background down to
1\,mJy.  The dashed line is the best fit for a simple parametric model
constrained by the counts of sources in the FIR/submm and the spectrum
of the extragalactic background (Blain et al.\ 2002).  Other UV, mid-IR
and radio derived points are from Giavalisco et al.\ (2003 --
highest-$z$ circles), Steidel et al.\ (1999 -- $z=3$--4 triangles),
Connolly et al.\ (1997 -- $z=1$--2 stars), Yan et al.\ (1999 -- $z=1.3$
hexagon), Flores et al.\ (1999 -- $z=0.3$--1 circles), Yun, Reddy \&
Condon (2001 -- low-$z$ solid circle).
}
\label{madau}
\addtolength{\baselineskip}{10pt}
\end{inlinefigure}

We can now estimate the evolution of the SFRD for all SMGs brighter
than $\sim$5\,mJy at 850\mum.  A solid curve is plotted in
Fig.~\ref{madau}, representing a Gaussian fit to the four SMG points
(after redistributing the objects in the high redshift bins into two
non-overlapping bins in redshift).  The fit is $SFRD = 1.26 \times
\exp[-(z-2.18)^2 /\sigma^2],$ with $\sigma=1.30$.  This evolutionary
behaviour is quite similar to that inferred for the luminosity density
of Quasars (e.g., Boyle et al.\ 2000; see Fig.~5).

We also plot in Fig.~\ref{madau} the SFRD estimated from a number of
UV-selected galaxy surveys at $z=1$--6 (Connolly et al.\ 1997; Steidel
et al.\ 1999; Giavalisco et al.\ 2003), and low redshift ($z<1$) radio
and mid-IR observations (Yun, Reddy \& Condon 2001; Flores et al.\
1999).  A dust-correction of a factor of $5\times$ for the $z \sim 3$
LBG population (Pettini et al.\ 2001) has been applied to the
high-redshift UV-selected populations to give their dust-corrected
estimates.  Recent analysis of the SFRD evolution via C{\sc
II}$^*\lambda$1335.7 in damped Ly$\alpha$ absorbers suggests a total
SFRD comparable to the dust-corrected UV estimates (Wolfe et al.\
2003a,b).
 
This analysis allows us to bring together the various high redshift
populations. Fig.~\ref{madau} highlights the relative importance of
different classes of high-redshift star-forming galaxy, critical for a
full understanding their relative importance galaxy evolution. We see
that although the bright, radio-detected SMGs presented here represent
only 20\% of the 850-$\mu$m background, the estimated star-formation
density at $z=2$--3 is within a factor of two of that inferred from
restframe UV observations (Steidel et al.\ 1999; Adelberger \& Steidel
2000).  Including a contribution from the more numerous, less-luminous
SMGs with 850-$\mu$m fluxes below $\sim 5$\,mJy would result in their
SFRD matching or even exceeding that seen in the UV.  Moreover, as we
argued in the previous section, the SFRD contribution of this
population is effectively missed by UV-selected surveys, and hence we
must {\it sum} the bright SMG and UV estimates to determine the {\it
total} SFRD.

We also need to account for the large fraction of the submm galaxy
population which is below our $\sim 5$\,mJy flux limit at 850\,$\mu$m.
We chose only to integrate down to 1\,mJy as this is the flux scale
where the estimated SFR of typical UV-selected galaxies become
comparable to the FIR sources.  There is a correction factor of
$\sim2.9\times$ which is required for the submm points to account for
the SMGs integrated down to 1\,mJy, comprising $\sim60$\% of the
extragalactic background in the submm waveband (Smail et al.\ 2002).
We apply this correction to Fig.~\ref{madau} assuming that these
fainter SMGs share the same redshift distribution to the brighter SMGs
whose distribution is explicitly measured in this paper.  The SFRD
measurements corrected in this manner would suggest that SMGs are the
dominant site of star formation activity in the Universe at $z\sim
2$--3. However, the discussion below suggests that our assumption of a
similarity between the redshift distributions of bright and faint SMGs
is likely to fail, even by flux densities of $S_{850}\sim1$\,mJy (see
also Lacey et al.\ 2004).

A second important point to draw from Fig.~\ref{madau} concerns the
relative evolution of the UV- and submm-selected populations.  Our
redshifts show that SMGs are coeval and energetically comparable in a
volume-average sense with the population of UV-bright, star-forming
galaxies detected at $z \sim 2$--3 (Steidel et al.\ 1999).  However,
the SMGs with $S_{\rm 850\mu m}>$5\,mJy and UV-selected galaxies (which
have a median 850-$\mu$m flux of $\ls0.5$\,mJy, Adelberger \& Steidel
2000; Reddy \& Steidel 2004) clearly don't evolve in the same manner.
SMGs appear to evolve more strongly than the UV-selected population out
to $z\sim 2$ (C03) and indeed seem to behave in a manner very similar
to luminous Quasars and X-ray selected AGN, whose luminosity density
peaks at $z\sim2.3$ (Boyle et al.\ 2000; Croom et al.\ 2004; Silvermann
et al.\ 2004).  As Fig.~\ref{madau} makes clear, this is in stark
contrast with the individually-less luminous, UV-selected galaxies
whose comoving luminosity density is approximately constant out to at
least $z\sim5$ (Lehnert et al.\ 2003; Giavalisco et al.\ 2004).  This
suggests that the properties of the bright submm population is more
closely linked with the formation and evolution of the galaxies or
galactic halos which host QSOs, than the more typical, modestly
star-forming galaxies identified from their restframe UV emission.
Nevertheless, there is likely to be an intermediate luminosity regime
where the UV- and submm-selected populations overlap significantly, and
therefore where their evolution becomes similar -- we propose that this
likely occurs at sub-mJy levels.

\subsection{Contributions to the FIR background}

Using our the long-wavelength SEDs which we fitted to the submm/radio
observations of the SMGs we can calculate their contribution to the
extragalactic background at other wavelengths in the far-infrared.  The
measured FIR background (FIRB), and the contribution per unit
wavelength from our SMGs with redshifts and well-fitted SEDs are shown
in Fig.~\ref{bkgnd}.  At 850\mum\ this is simply the sum of the flux
measurements for our SMGs normalized by the effective survey area
(assuming a radio-detected fraction of 65\%).  At all shorter
wavelengths, the curve represents the sum of the best-fit SEDs,
examples of which were shown in Fig.~\ref{seds}.  We see that the SMG
sample we are studying contributes around 20\% of the background at
$>600\mu$m and a diminishing fraction at shorter wavelengths.

To provide a simple baseline comparison we also illustrate the
contributions to the FIRB if all of the SMG's lie at $z=2.3$ (the
median redshift of our model-corrected spectroscopic sample) and have
SEDs matching that of Arp\,220, again normalized to our total 850\mum\
flux. Note that the actual peak of the background due to the SMGs is
significantly broadened by the dispersion in their redshifts and dust
temperature, compared to a single \td$\sim$45\,K-like Arp\,220 SED at
$z=2.3$.

The contribution of SMGs to the peak of the FIRB varies on a galaxy by
galaxy basis by a factor of $\sim10$. Galaxies with hotter
characteristic dust temperatures contribute more, but because they
typically lie at higher redshifts in our submm-selected sample, their
contribution is scaled down.  Splitting our sample into high and low
redshift bins, we assess the relative contribution to the emission
around the peak of the FIRB at $\sim 200\mu$m as a function of
redshift. SMGs at $z=2.8\pm0.3$ contribute three times less to the FIRB
than SMGs at $z=2.2\pm0.3$. This implies that the increase in
characteristic \td\ with redshift is not fast enough to counteract the
diminuation of restframe 200\mum\ flux from the K-correction.  If we
include plausibly-identified SMGs without radio detections (Ivison et
al.\ 2005), which are predicted to have colder dust temperatures, we
find that their contribution to the FIRB is much smaller (by a factor
$\sim$5) compared with the warmer \td\ radio-SMGs. Moreover, the
radio-unidentified SMGs are spread throughout both redshift bins, they
do not substantially affect the relative FIRB contributions as a
function of redshift.

We can also estimate the contribution of the bulk of the submm galaxy
population to the FIRB by applying a correction factor ($\times$2.9) to
our composite SMG template to include galaxies down to 1\,mJy,
comprising $\sim$60\% of the Fixsen et al.\ (1998) 850\mum\ background.
This extrapolation suggests that galaxies selected at 850-$\mu$m are
significant contributors (responsible for $\gs 30$\% of the total
emission) to the FIRB at wavelengths of $\gs 400\mu$m, assuming that
modest extrapolations up the luminosity function are composed of
galaxies similar in characteristics to those SMGs in our sample.
However, the $>$1\,mJy SMG population probably contribute no more than
6\% of the emission at the peak of the FIRB at $\sim 200\mu$m, which is
dominated by lower-redshift and/or hotter populations.  For sources
fainter than 1\,mJy, there is likely to be a wider distribution of
redshifts, and predictions based on our bright submm sample are much
more uncertain.

%
%
\begin{inlinefigure} \vspace{6pt}
\psfig{figure=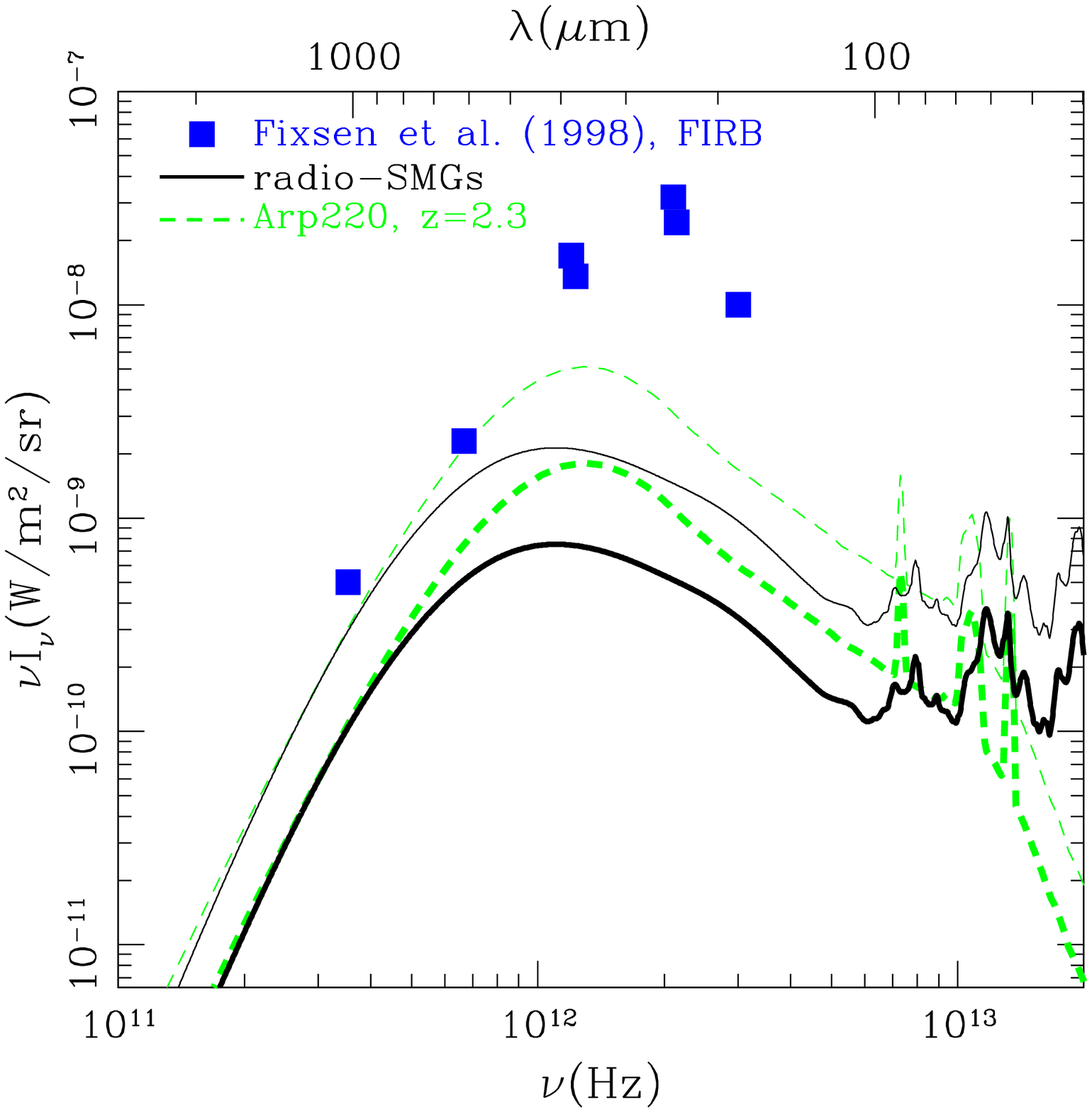,angle=0,width=3.5in}
\figurenum{13}
\vspace{6pt}
\caption{\small
Measurements of the FIR background from Fixsen et al.\ (1998) and the
contribution per unit wavelength of spectroscopically-identified SMGs
(heavy black line).  A curve representing Arp\,220 at $z=2.3$ is shown
normalized to our total 850\mum\ flux (heavy dashed line).  We also
show these two models, but corrected to account for the 850\mum\
background as resolved by submm observations down to $1$\,mJy (Blain et
al.\ 1999; Cowie et al.\ 2002) as thin curves.
}
\label{bkgnd}
\addtolength{\baselineskip}{10pt}
\end{inlinefigure}

The procedure we used in the calculation above assumes that submm
sources with flux densities fainter than our sample have a similar
redshift distribution and range of SEDs/dust temperatures.  There is
some evidence for this down to 1\,mJy from cluster lensed SMGs (e.g.,
Smail et al.\ 2002; Borys et al.\ 2004; Kneib et al.\ 2004).  We
discussed in the previous subsection why this may not be a good
approximation fainter than 1\,mJy: UV-selected galaxies appear to have
a very different evolution history from SMGs, and the UV galaxies are
likely to contribute a substantial fraction of the submm background at
$\sim$0.1\,mJy flux densities.  The corrected distribution should
therefore be assessed critically.  Specifically, if fainter sources
begin to evolve more like UV-selected galaxies, with less of a peak at
$z\sim 2.5$, then these higher redshift and less luminous sources will
have a progressively smaller contribution to the FIRB near 200\mum.

\subsection{Mid-/far-IR {\it Spitzer} fluxes and spectral diagnostics}

The {\it Spitzer} Space Telescope marks a new era in IR astronomy,
pushing significantly deeper than past space IR missions.  A pressing
question is therefore the overlap of the SMGs with the faint {\it
Spitzer} population.  By assuming the locally-calibrated SEDs of Dale
\& Helou (2002), we demonstrate that our observed factor of two range
in submm flux translates into greater than two orders of magnitude in
24\mum\ flux, with the faintest galaxies having predicted fluxes of
only 10$\mu$Jy (see Fig.~\ref{seds}).  Prior to obtaining redshifts for
SMGs, it was impossible to realistically assess the required
sensitivities for IR follow-up with {\it Spitzer}.  Several authors
have already suggested that the detection rate of SMGs by {\it Spitzer}
at 3.6\mum\ through 24\mum\ is very high (Frayer et al.\ 2004; Ivison
et al.\ 2004; Egami et al.\ 2004), with flux densities roughly in
accord with our predictions in Fig.~\ref{seds}.

Mid-IR spectral surveys with {\it ISO} (Lutz et al.\ 1999; Tran et al.\
2001) demonstrated that this wavelength regime is a powerful tool to
classify the energetics of galaxies in an independent manner from
conventional optical classification techniques (e.g., Swinbank et al.\
2004).  The main advantage of the mid-IR is that it probes the physical
conditions in the more obscured regions closer in to the
optically-thick, dust-enshrouded nuclei and molecular cloud regions of
the galaxies where the bulk of the FIR emission arises (Sturm et al.\
2000, 2002; Verma et al.\ 2003; Spoon et al.\ 2004).  Key emission
features include those from polycyclic aromatic hydrocarbon (PAH) at
6--18\,$\mu$m, high-excitation Ne emission lines and the broad silicate
absorption feature.  It has been suggested that some of these features
are diagnostic of the relative balance of dust heating by AGN and star
formation.

{\it Spitzer} spectroscopy of SMGs to determine mid-IR properties will
be important for disentangling the physical conditions in the most
active regions within these highly-obscured galaxies. The Dale \& Helou
(2002) spectral templates include PAH features, Ne emission and the
broad silicate absorption feature, all estimated as a function of
luminosity and temperature based on local correlations. Our template
fits to the SMGs with redshifts therefore allow predictions of their
mid-IR spectral properties (Figs.~\ref{seds} \& \ref{bkgnd}),
indicating that low-resolution {\it Spitzer} spectroscopy should yield
line detections for the majority of the SMGs in our sample.  Indeed,
stacking {\it Spitzer}-IRS spectra for similar classes of SMGs should
yield sufficient signal-to-noise to study the shape of the PAH feature
and probe the chemical evolution of silicates (e.g., Honda et al.\
2004).

However, it is possible that the physical properties in high-$z$ ULIRGs
(such as the gas fraction, metallicity, star-formation extent and
star-formation rate) may differ from these local systems.  For example,
a comparison of the UV and long-wavelength morphologies of SMGs
indicates that their star formation is more spatially extended than in
the local population of ULIRGs (Chapman et al.\ 2004b), likely
resulting in differences in the physical conditions experienced by dust
grains (Lu et al.\ 2003).  Mid-infrared spectroscopy of SMGs may be
able to test this claim and provide new insights into the physics of
SMGs.

\subsection{The duty-cycle of SMGs}

Having discussed the observational and inferred properties of our SMG
sample, we wish to finish by discussing their evolutionary connections
to other populations at high and low redshifts.  In order to connect
the SMG population to a likely population of descendents, we must
estimate what the duty-cycle for SMGs might be. Our greatest
uncertainty in this calculation is the duration of the submm-luminous
event. Smail et al.\ (2003a, 2004a) have estimated the past duration of
the starburst events using both fits to the UV spectra with the {\it
Starburst99} synthetic templates (Leitherer et al.\ 1999), and simple
model fits to the optical and near-IR photometry. They conclude that
the starbursts have been ongoing for $\sim 10$--100\,Myr.  Using the
gas masses from the CO observations compiled for a small sample of SMGs
by Neri et al.\ (2003) we can infer that there is enough molecular gas
present in a typical SMG to continue the starbursts at their present
rate for at the most another $\ls 100$\,Myr, and perhaps considerably
shorter. Thus the total duration of these submm-luminous events might
therefore be $\sim$\,100\,Myr.

The length of the starburst is likely to be regulated by the enormous
superwind outflows that are expected to be driven in SMGs; Greve et
al.\ (2005) and Smail et al.\ (2003, 2005) have demonstrated offsets
between Ly$\alpha$ and molecular CO or H$\alpha$ as large as
1000\,km/s, implying outflows of large velocity. The SMGs are the most
luminous galaxies in the Universe, and as such are expected to drive
some of the largest superwinds, with if the mass ejected is
proportional to the star formation rate (Heckman et al.\ 2001), would
produce typical outflows of 1000\,M$_\odot$\,yr$^{-1}$. Simple scaling
arguments suggest that the kinetic energy put into SMG winds (and hence
the local environment) would be $\sim10$ times that of LBGs, which are
already known to show evidence for strong outflows (Adelberger et al.\
2003). Given the potential power of this feedback mechanism we must
treat our estimated submm-luminous timescale as an upper limit, since
the outflows and associated turbulence could easily turn the starburst
off in a shorter time period.

The radio-identified SMG redshift distribution was well-fit by a
Gaussian with a central redshift of 2.2 and $\sigma_z\sim 1.2$
(accounting for incompleteness in the spectroscopic desert).  Modeling
of the full SMG redshift distribution suggests $\sigma_z\sim 1.3$,
although this is clearly dependent on the model adopted.  Focusing just
on the radio-identified SMGs, these populate a period of about 1\,Gyr
in length. Dividing the submm redshift distribution by the
submm-luminous timescale of 100\,Myr suggests a duty-cycle of 10. In
other words, there is an underlying population of galaxies ten times as
numerous as the SMGs that correspond to the immediate progenitors and
descendents of the SMG phenomenon. This parent population would have a
volume density at $z=1$--3 of $\sim 10^{-4}$\,Mpc$^{-3}$, and would be
expected to have the similar clustering properties to the SMG events
themselves (r$_0\sim7 h^{-1}$\,Mpc -- Blain et al.\ 2004b).  One
possible identification of the descendents of the SMG population comes
from the discovery of luminous, near-infrared galaxies at $z\sim 2$
with apparent stellar ages of 1--2.5\,Gyrs, stellar masses of
1--$5\times 10^{11}$\,M$_\odot$ and a space density a factor of several
times higher than SMGs (van Dokkum et al.\ 2003, 2004; Franx et al.\
2003; Glazebrook et al.\ 2004).

To understand the evolutionary connections of the SMGs with other
galaxy populations at similar redshifts, it is crucial to measure the
SMG clustering properties.  Surprisingly, even with our small sample,
we are already in a strong position to characterize the clustering.  We
notice in Table~3 that SMGs often come in pairs or larger associations
to about 1200\,km\,s$^{-1}$, with approximately one in five SMGs in our
sample lying in such associations.  This can be quantified into an
estimate of the clustering strength (Blain et al.\ 2004b), suggesting a
correlation amplitude comparable to low-redshift EROs (e.g., Daddi et
al.\ 2001; McCarthy et al.\ 2001).  The SMG associations seem to point
to global overdensities in the galaxy populations, typified by the
$z=3.1$ protocluster in the SSA22 field (Steidel et al.\ 1998) where 3
SMGs lie at $z=3.1$.  The five SMGs at $z=2.0$ in the HDF field suggest
that we have likely identified another overdensity of a similar scale
to the SSA22 protocluster.

The clustering scale of 2dF QSOs is similar to that of SMGs within
errors (QSOs: $r_0\sim5$\,Mpc -- Croom et al.\ 2002, SMGs:
$r_0\sim7$\,Mpc -- Blain et al.\ 2004b), and as we have noted
repeatedly in this study, the redshift distributions of both
populations are very similar.  However, the volume densities of bright
QSOs and SMGs do not match; the SMGs outnumber QSOs by a factor of
$\sim5$--10$\times$ (Chapman et al.\ 2003b).  The SMGs and QSOs can be
viewed as different phases of the same population if the timescales
compensate for this discrepancy.  Martini \& Weinberg (2001) suggest
that QSOs endure for $\sim$40\,Myr, requiring that SMGs are active for
$\sim$200--400\,Myr, not too discrepant from our adopted SMG timescale
of 100\,Myr estimated above.  Another possibility is that the QSO may
often represent a phase in the active lifetime of a subset of SMGs,
dependent on environmental conditions that facilitate the funneling of
material from the extended SMG starburst to the growing central engine.

It is worth seriously considering the possibility of an inter-relation
between the SMG and QSO phenomena.  The local {\it Magorrian} relation
(Magorrian et al.\ 1998) between black hole mass and stellar bulge mass
suggests that super-massive black holes (SMBHs) must be growing at a
rate roughly proportional to the stars.  The SMGs appear to be
prodigiously forming stars, with the modest X-ray luminosities of SMGs
suggesting either low rates of accretion onto super-massive black holes
(SMBHs) and/or moderate SMBH masses (Alexander et al.\ 2005).  Both of
these possibilities suggest that substantial SMBH growth is needed to
evolve onto a Magorrian relation.  A logical conclusion is that the
SMGs are indeed an early phase in the evolution of a massive galaxy,
forming many stars quickly over $\sim$10\,kpc spatial scales, as shown
by the extended radio emission tracing UV structures (Chapman et al.\
2004b).  As the merger proceeds, the growing SMBH will begin to accrete
material more quickly as instabilities drive material to more
concentrated configurations.  The SMBH thus grows ever more rapidly, on
a timescale which is delayed from that of the initial starburst (see
also Archibald et al.\ 2002), eventually blowing channels through the
dust and becoming visible as a QSO.  Indeed, Page et al.\ (2001) and
Stevens et al.\ (2004) have measured significant submm emission from
X-ray self-absorbed QSOs, suggesting that an intermediate stage between
SMGs and QSOs is being directly observed.  This is also consistent with
the 3/80 QSO detection rate (4\%) for our unbiased SMG sample, if it is
assumed that all submm-detected QSOs are X-ray absorbed (Stevens et
al.\ 2004).  Under this scenario, the proportion of X-ray absorbed QSOs
is 1/6 of the entire QSO population (Stevens et al.\ 2004) and hence
the ratio of QSO-to-SMG lifetimes should be 18/80, or $\sim$0.23,
consistent with the volume densities described above.

Most likely, the relationship between SMGs and other galaxy populations
spans a range of scenarios and evolutionary histories, and more
comprehensive models will be required to understand and further test
the evolution of SMGs.

\section{Conclusions}

We have assembled a detailed account of the properties of the
radio-identified SMGs.  Spectroscopic redshifts have been obtained for
73 radio-SMGs from deep optical spectroscopy of 98 targets attempted,
for a success rate of 74\%.  The redshift distribution of SMGs is now
well constrained -- showing a pronounced peak at $z\sim 2.2$ for our
radio-selected sample and a predicted median of $z\sim 2.3$ for a
purely submm-flux-limited sample, similar in form to distribution seen
for radio-, optical- and X-ray-selected QSOs. We also demonstrate that
the simple radio/submm redshift estimator, which has been used
extensively in the literature to attempt constraint the redshift
distribution of SMGs, is not reliable for individual SMG redshifts
($\Delta z \sim 1$).  For the SMGs which are detectable with good
signal-to-noise ($R<26$) in deep ground-based images, photometric
redshifts in the $UBRIK$ bands appear to provide a more accurate
constraint on the true redshifts of individual SMGs.  We find that the
radio-SMGs for which we failed to obtain robust spectroscopic redshifts
have similar $(U-g)$/$(g-R)$ colors to those with robust redshifts, and
conclude that the entire radio-SMG population is likely to have a
similar distribution to the spectroscopic redshift distribution
presented here.  We have also discuss the likely redshifts of SMGs
without radio identifications to fill in more details of the entire SMG
population brighter than 5\,mJy at 850\,$\mu$m.

We measure dust temperatures and bolometric luminosities for SMGs by
exploiting our precisely-known redshifts and assuming the local
FIR--radio correlation.  The SED variation in our SMGs likely arises
from a dispersion in dust temperature, but could also arise from a
broader scatter in the FIR--radio correlation than observed locally.
If the SED variation were caused entirely by variations in the
FIR--radio correlation then the FIR--radio scatter would have to
increase from the 0.2\,dex observed locally to $\sim0.8$\,dex.  We view
this as unlikely and so attribute most of the range in SED properties
for the SMGs as down to differences in characteristic dust temperature.

We construct a bolometric luminosity function and compare to local {\it
IRAS} galaxies as well as estimates for UV-selected galaxies at
$z\sim2$--3.  We predict the expected FIR luminosities of the SMG
population based on their UV luminosities and continuum slope, using
the standard prescription for this conversion for UV-selected
populations.  We find that the true bolometric luminosity of an SMG is
typically underestimated by $\sim100\times$ when extrapolated from
their restframe UV properties.  We therefore conclude that observations
at radio/submm wavelengths are essential to distinguish which $z\sim
2$--3 galaxies have the huge luminosities characteristic of SMGs.

We assess the SFRD evolution of SMGs, both from our observed
$\sim$5\,mJy SMG sample and from a modest extrapolation of our sample
properties down to 1\,mJy.  Our results highlight the very different
evolution of SMGs over UV-selected galaxies (and noting the close
similarity in evolution between SMGs and QSOs).  This suggests that the
properties of the bright submm population is more closely linked with
the formation and evolution of the galaxies or galactic halos which
host QSOs, than the more typical, modestly star-forming galaxies
identified from their restframe UV emission.  We emphasize that
UV-measurements of the SMGs do not typically reveal the enormous
bolometric luminosities present, and the SMGs and dust-corrected
UV-selected galaxies in the SFRD diagram should be summed together to
obtain a complete census of the SFRD evolution in the Universe.

The SMG sample we are studying contributes around 20\% of the
background at $>600\mu$m and a diminishing fraction at shorter
wavelengths.  The contribution of SMGs to the FIRB at $\sim200$\mum\
can be predicted by fitting dusty SED templates to the radio and submm
fluxes at the measured redshifts. This prediction is considerably less
than expected from assuming an Arp~220-like SED, since the SMGs have
cooler characteristic dust temperatures than Arp~220.  The {\it Spitzer
Space Telescope} is likely to provide important diagnostics of the
highly obscured central regions of the SMGs through mid-IR imaging and
spectroscopy.

We are rapidly approaching the point of being able to put all the
evidence about properties of SMGs together.  The spectroscopic
redshifts allow us to study aspects of the SMG population which cannot
be addressed using photometric redshift estimators.  Astrophysical
properties of the SMGs are diagnosed through the restframe UV and
optical spectra themselves, probing for example the excitation
conditions, metallicities, and wind outflows (Swinbank et al.\ 2004;
I.\ Smail, in preparation).  The clustering of the SMGs is clearly a
pressing question, and surprisingly, even with our small sample we are
already in a strong position to characterize the clustering through the
redshift distribution, finding a correlation length r$_0\sim7
h^{-1}$\,Mpc (Blain et al.\ 2004b).  The impact of SMGs on their
environments and the inter-galactic medium can be studied directly
using background QSOs in the same fields (S.\ Chapman, in preparation).
With precise redshifts, X-ray spectral analysis in deep {\it Chandra}
exposures is possible through the stacking of sub-groups of SMGs with
similar X-ray absorption properties, yielding significant Fe{\sc
K$\alpha$} emission line detections and characterization of the X-ray
spectral shape (D.\ Alexander, in preparation).

We conclude that with the availability of precise redshifts for large
samples of FIR luminous galaxies, these populations can now be used to
trace the evolution of the most luminous galaxies in the Universe. A
focus of our future work will be to identify the influence of this
population on their environments.

\acknowledgements
We thank the referee, Paul van der Werf, for a thoughtful reading of
the manuscript.  SCC acknowledges support from NASA through grants
GO-9174 \& GO-9856.  AWB acknowledges support from NSF grant
AST-0205937, from the Research Corporation and from the Alfred P. Sloan
Foundation.  IRS acknowledges support from a Royal Society URF.  We
thank Dr.\ Kurt Adelberger for kindly providing his suite of LRIS
reduction scripts and instruction in their use.  We would like to thank
the following people for helpful discussions: C.\ Steidel, K.\ Adelberger,
A.\ Shapley, N.\ Reddy, C.\ Borys.  The authors wish to recognize and
acknowledge the very significant cultural role and reverence that the
summit of Mauna Kea has always had within the indigenous Hawaiian
community.  We are most fortunate to have the opportunity to conduct
observations from this mountain.


\begin{deluxetable}{lccccccl}
\renewcommand\baselinestretch{1.0}
\tablewidth{0pt}
\parskip=0.2cm
\tablenum{1}
\tablecaption{Field Properties}
\small
\tablehead{
\colhead{name} & {radio center} & {radio depth$^a$} & {N$_{\rm SMG}^b$} & {N$_{\rm SMG}^b$}& {N$_{\rm spectra}^c$} & {N$_{\rm redshifts}^d$} & {optical data comments$^e$}\\
{} & {(J2000)} & {($\mu$Jy)} & {map} & {phot}& {} & {} & {}\\
}
\startdata
CFRS-03 & 03:02:40.9 00:09:08 & 10 & 9 & 0 & 9 & 6 & cfh12k, LRIS\\
Lockman & 10:52:00.1 57:18:10 & 5 & 10 & 0 & 8 & 4 & MOSAIC, cfh12k, SUPRIME\\
HDF-N & 12:37:07.2 62:14:02 & 8 & 27 & 11 & 31 & 24 & MOSAIC, SUPRIME\\
SSA-13 & 13:12:16.5 42:41:21 & 4 & 9 & 8 & 16 & 9 & MOSAIC, SUPRIME\\
CFRS-14 & 14:17:49.4 52:30:23 & 14 & 8 & 1 & 9 & 8 & cosmic, WHT, cfh12k\\
ELAIS-N2 & 16:36:50.0 40:57:35 & 15 & 8 & 3 & 10 & 10 & WHT, LFC\\
SSA-22 & 22:15:15.1 00:13:55 & 8 & 7 & 8 & 15 & 11 & cfh12k, LFC, cosmic\\
\enddata
\label{tab3}
$a$ 1$\sigma$ radio depth at phase center.\\
$b$ number of radio identified submm sources above 3$\sigma$ divided into
 SCUBA mapping and SCUBA photometry observations. Photometry observations
targeted the radio source position.\\
$c$ number of spectra attempted.\\
$d$ number of successful redshifts.\\
$e$ optical data used in the identifications:
 CFHT/CFH12K (Cuillandre et al.\ 2000),
 KPNO4m/MOSAIC (Muller et al.\ 1998),
 Subaru/SUPRIME (Komiyama et al.\ 2003),
 WHT/PFC (Tulloch\ 2000),
 Palomar200inch-LFC/cosmic (Simcoe et al.\ 2000),

\end{deluxetable}

%
%
\begin{deluxetable}{llll}
\renewcommand\baselinestretch{1.0}
\tablewidth{0pt}
\parskip=0.2cm
\tablenum{2}
\tablecaption{Observation Logs}
\small
\tablehead{
\colhead{Date} & {Field RA} & {nights clear} & {Configurations}\\
}
\startdata
ESI\\
2001 Jul 16 & ELAIS-N2/SSA22 & 1 & Echelle \\
\noalign{\smallskip}
LRIS\\
2002 Mar 18 -- Mar 21 & Lockman/HDF/SSA13/ELAIS-N2 & 2 & B400/D560/R600 \\
2002 Sep 30 -- Oct 01 & SSA22/CFRS03 & 2 & B400/Mirror \\
2002 Dec 26 -- Dec 27 & CFRS03/Lockman/HDF & 1 & B400/D560/R600 \\
2003 Mar 28 -- Mar 29 & Lockman/HDF/Westphal-14 & 0.5 & B400/D680/R400 \\
2003 May 22 -- May 24 & Lockman/HDF/Westphal-14/ELAIS-N2 & 2 & B400/D560/R600 \\
2003 Aug 22 -- Aug 23 & ELAIS-N2/SSA22 & 2 & B400/D560/R600 \\
2004 Feb 14 -- Feb 15 & Lockman/HDF/SSA13 & 2 & B400/D560/R600 \\
\enddata
\label{tab3}
\end{deluxetable}
%
%
\begin{deluxetable}{lcccccccl}
\renewcommand\baselinestretch{1.0}
\tablewidth{0pt}
\parskip=0.2cm
\tablenum{3}
\tablecaption{Properties of radio-SMGs}
\small
\tablehead{
\colhead{ID} & {S$_{\rm 1.4 GHz}$} & {$B_{\rm AB}$} & {$R_{\rm AB}$} & {S$_{\rm 850 \mu m}$} & {z} & {L$_{bol}$} & {\td} & {comment} \\
{} & {($\mu$Jy)} & {(mag)} & {(mag)} & {(mJy)} & {} & {$\times 10^{12}$ (L$_\odot$)} & {(K)} & {} \\
}
\startdata
SMM\,J030226.17+000624.5&  481.5$\pm$9& 19.1&17.8&  7.9$\pm$1.6& 0.080 &0.03 &11.4 & $M$; probable lens \\ %
SMM\,J030227.73+000653.5& 217$\pm$9&  23.8&22.9&  4.4$\pm$1.3& 1.408 & 27.8 &56.4 &$M$;  \lya, abs, H$\alpha$ (SB)\\ %
SMM\,J030231.81+001031.3&  45.1$\pm$9&  $>$26.4&$>$25.1&  5.0$\pm$1.5& 1.316 & 1.3 & 25.6 & $M$; OII (SB) \\ %
SMM\,J030236.15+000817.1&  42.1$\pm$9.1& $>$26.4 & 26.1 &  3.4$\pm$0.6& 2.435 & 6.7 & 41.3 & $M$; \hal, NII (SB)\\ %
SMM\,J030238.62+001106.3$^d$&  347.3$\pm$9& 25.6&24.3& 4.1$\pm$1.4& 0.276 & 0.21& 25.5 & $M$; OII, NeV (AGN)\\ %
SMM\,J030244.82+000632.3& 154.0$\pm$34.1&  20.5&19.1& 4.9$\pm$1.1& 0.176 & 0.12 & 13.5 & $M$; OII, \hal (SB)\\ %
{} & & & & & & & & \\
SMM\,J105151.69+572636.0&  134.4$\pm$13.0&  n/a& 25.2&  6.7$\pm$1.7&1.147 &1.6 & 25.4& $M$; OII, MgII (SB)\\ %
SMM\,J105155.47+572312.7$^d$& 46.3$\pm$10.2& 23.8&23.2& 5.7$\pm$1.4&  2.686 &11.7&41.7&$M$; \lya, IS abs, \hal\ (SB)\\ %
SMM\,J105158.02+571800.2&  98.1$\pm$11.6&  n/a& 24.1 & 7.7$\pm$1.7&  2.239 &12.3&39.2& $M$; \lya\ abs, CIV abs, \hal (SB)\\ %
SMM\,J105200.22+572420.2$^a$&  57.4$\pm$13.2& n/a& 22.1&    5.1$\pm$1.3&  0.689 &0.33 & 20.0 &$M$; Oii, FeII (SB)\\ %
SMM\,J105201.25+572445.7&  72.1$\pm$10.2& n/a&25.7& 9.9$\pm$2.2&  2.148 &8.1&33.0&$M$; \lya\ abs, \hal\ (SB)\\ %
SMM\,J105207.49+571904.0&  277.8$\pm$11.9&  n/a& 26.0&    6.2$\pm$1.6& 2.692 & 17.5& 45.4 &$M$; \lya, \hal\ (SB) \\ %
SMM\,J105225.79+571906.4$^b$& 127.4$\pm$5.1& n/a&24.7& 4.9$\pm$1.5&  2.372 & 18.7 & 49.3 &$M$; \lya, IS abs (SB)\\ %
SMM\,J105227.77+572218.2$^d$&  40.4$\pm$9.4& n/a&26.0& 7.0$\pm$2.1& 1.956 & 3.5 & 23.9 &$M$; \lya, CIV (SB)\\ %
SMM\,J105227.58+572512.4& 39.2$\pm$11.4& n/a&25.0& 4.5$\pm$1.3&  2.142 & 4.0 & 33.7 &$M$; IS abs, CIII (SB)\\ %
SMM\,J105230.73+572209.5&  86.3$\pm$15.4&  n/a& 23.3&   11$\pm$2.6& 2.611 & 14.5 & 37.1 &$M$; \lya, SiII, \hal\ (SB) \\ %
SMM\,J105238.19+571651.1$^b$& 71.1$\pm$12.6& n/a&22.7 & 5.3$\pm$1.6&  1.852 & 5.8 & 35.9 &$M$; IS abs (SB)\\ %
SMM\,J105238.30+572435.8$^{c,d}$&  61.0$\pm$22.0&  n/a& 24.6 &  10.9$\pm$2.4& 3.036 & 18.1 & 39.4 &$M$; \lya, CIV (AGN)\\ %
{} & & & & & & & & \\
SMM\,J123549.44+621536.8&  74.6$\pm$9.5& 24.2&23.7& 8.3$\pm$2.5& 2.203 & 8.9 & 35.4 & $P$; IS~abs, \hal, CO(3-2) (SB)\\  %
SMM\,J123553.26+621337.7$^d$&  58.4$\pm$9.0& 24.8&24.7& 8.8$\pm$2.1& 2.098 & 6.1 & 31.6 & $P$; IS~abs (SB)\\  %
SMM\,J123555.14+620901.7& 212.0$\pm$13.7& 24.5&24.2& 5.4$\pm$1.9& 1.875 & 16.2 & 46.8 & $P$; \lya, IS~abs (SB) \\  %
SMM\,J123600.10+620253.5& 262.0$\pm$17.1& 25.7&25.4& 6.9$\pm$2.0& 2.710 & 56.2 & 59.9 & $P$; extended \lya\ (SB) \\  %
SMM\,J123600.15+621047.2& 131.0$\pm$10.6& 25.4&25.1& 7.9$\pm$2.4& 1.994 & 12.0 & 39.0 & $P$; \lya, \hal\ (SB) \\  %
SMM\,J123606.72+621550.7&  24.0$\pm$5.9&  23.5&23.6& 4.4$\pm$1.4& 2.416 & 3.7 & 33.1 & $M$; \lya, SiIV, CIV (AGN)\\ %
SMM\,J123606.85+621021.4&  74.4$\pm$4.1& 25.6&25.2& 11.6$\pm$3.5& 2.505 & 12.8 & 35.5 & $P$; \lya, \hal\ (SB) \\  %
SMM\,J123616.15+621513.7$^c$& 53.9$\pm$8.4& 26.8&25.7& 5.8$\pm$1.1& 2.578 & 10.0 & 39.9 & $M$; \lya\ (SB)\\  %
SMM\,J123618.33+621550.5$^d$& 151.0$\pm$11.0& 26.0&25.9& 7.3$\pm$1.1& 1.865 & 11.4 & 39.4 & $M$; IS~abs (SB) \\  %
SMM\,J123621.27+621708.4$^c$& 148.0$\pm$11.0& 25.1&24.9& 7.8$\pm$1.9&  1.988 & 13.5 & 40.3 & $M$; IS~abs, \hal\ (SB) \\ %
SMM\,J123622.65+621629.7&  70.9$\pm$8.7& 25.6&25.4& 7.7$\pm$1.3& 2.466 & 11.6 & 38.6 & $M$; \lya, \hal\ (SB) \\ %
SMM\,J123629.13+621045.8&  81.4$\pm$8.7& 26.1&24.6& 5.0$\pm$1.3& 1.013 & 1.2 & 26.0 & $M$; OII, MgII (SB) \\ %
SMM\,J123632.61+620800.1&  90.6$\pm$9.3& 23.8&23.6& 5.5$\pm$1.3& 1.993 & 8.2 & 40.0 & $M$; \lya, NV, SiIV, CIV (AGN) \\  %
SMM\,J123634.51+621241.0& 230.0$\pm$13.8& 24.4&23.9& 4.3$\pm$1.4&  1.219 & 5.5 & 37.5 & $M$; OII, MgII (SB)\\ %
SMM\,J123635.59+621424.1&  87.8$\pm$8.8& 24.2&24.2&  5.5$\pm$1.4& 2.005 & 8.1 & 40.3 & $M$; \lya, CIV (AGN)\\ %
SMM\,J123636.75+621156.1&  39.0$\pm$8.0& 21.9&21.6&  7.0$\pm$2.1& 0.557 & 0.12 & 15.2 & $M$; OII, NeV (AGN)\\ %
SMM\,J123651.76+621221.3$^d$&  49.3$\pm$7.9& 21.6&21.4&  4.6$\pm$0.8& 0.298 & 0.08 & 13.3 & $M$; probable lens\\ %
SMM\,J123707.21+621408.1$^a$& 45.0$\pm$7.9& 26.9&26.0&   4.7$\pm$1.5& 2.484 & 7.5 & 39.2 & $M$; \lya, \hal\ (SB) \\ %
SMM\,J123711.98+621325.7$^a$& 53.9$\pm$8.1&   26.0&25.8&   4.2$\pm$1.4& 1.996 & 4.9 & 36.3 & $M$; \lya, \hal\ (SB) \\ %
SMM\,J123712.05+621212.3&  21.0$\pm$4.0&   27.0&25.5&  8.0$\pm$1.8& 2.914 & 5.5 & 31.3 & $M$; \lya, CIV (AGN)\\ %
SMM\,J123716.01+620323.3& 109.0$\pm$11.4&  20.3&20.2&   5.3$\pm$1.7& 2.037 & 10.5 & 41.7 & $P$; \lya, NV, SiIV, CIV (QSO)\\ %
SMM\,J123721.87+621035.3&  41.0$\pm$9.0& 24.3&23.3&  12.0$\pm$3.9& 0.979 & 0.53 & 16.9 & $M$; \hal, FeII, (AGN)\\ %
\enddata
\label{tab3}
\end{deluxetable}

%
%
\begin{deluxetable}{lcccccccl}
\renewcommand\baselinestretch{1.0}
\tablewidth{0pt}
\parskip=0.2cm
\tablenum{3}
\tablecaption{Continued,  Properties of radio-SMGs}
\small
\tablehead{
\colhead{ID} & {S$_{\rm 1.4 GHz}$} & {$B_{\rm AB}$} & {$R_{\rm AB}$} & {S$_{\rm 850 \mu m}$} & {z} & {L$_{bol}$} & {\td} & {comment} \\
\ & {($\mu$Jy)} & {(mag)} & {(mag)} & {(mJy)} & {} & {$\times 10^{12}$ (L$_\odot$)} & {(K)} & {} \\
}
\startdata
SMM\,J131201.17+424208.1&  49.1$\pm$6.0&   25.2&24.0&   6.2$\pm$1.2& 3.405 & 20.2 & 47.1 & $MP$; \lya, SiIV, CIV, OIII, CO(4-3) (AGN)\\ %
SMM\,J131208.82+424129.1&  82.4$\pm$4.8&   25.7&23.9&   4.9$\pm$1.5& 1.544 & 3.7 & 33.2 & $M$; \hal, MgII, CIV (AGN)\\ %
SMM\,J131212.69+424422.5& 102.6$\pm$7.4&   $>27$&26.7&   5.6$\pm$1.9& 2.805 & 24.3 & 50.7 & $M$; \lya, SiIV, CIV (AGN) \\ %
SMM\,J131215.27+423900.9&  69.3$\pm$4.0&   18.6&18.3&   4.4$\pm$1.0& 2.565 & 12.7 & 45.7 & $MP$; \lya, CIV, HeII (QSO) \\ %
SMM\,J131222.35+423814.1&  26.3$\pm$3.9&   21.8&20.2&   3.0$\pm$0.9& 2.565 & 4.8 & 39.2 & $MP$; \lya, CIV, HeII (AGN) \\ %
SMM\,J131225.20+424344.5&  76.4$\pm$6.8&   23.3&22.1&   2.4$\pm$0.8& 1.038 & 1.2 & 31.4 & $M$; OII, FeII (SB) \\ %
SMM\,J131225.73+423941.4 & 752.5$\pm$4.2  & 24.4 &24.2&4.1$\pm$1.3 & 1.554 & 33.8 & 62.2 & $MP$; MgII, FeII, CIV, SiII (SB) \\ %
SMM\,J131228.30+424454.8&  50.9$\pm$8.1&   25.9&24.7&   3.4$\pm$0.9& 2.931 & 13.6 & 49.7 & $M$; \lya, CIV (SB) \\ %
SMM\,J131231.07+424609.0$^d$&  39.4$\pm$8.5&   $>27$&26.9&   4.9$\pm$1.6& 2.713 & 8.5 & 39.9 & $MP$; \lya\ (SB) \\ %
SMM\,J131232.31+423949.5&  94.8$\pm$4.3&   25.7&24.6&   4.7$\pm$1.1& 2.320 & 13.1 & 45.4 & $M$; \lya, \hal\ (SB)  \\  %
SMM\,J131239.14+424155.7 &  49.8$\pm$6.6  & 26.0 & 25.7 &  7.4$\pm$1.9 & 2.242 & 6.3 & 33.2 & $M$; \lya, \hal\ (SB) \\ %
{} & & & & & & & & \\
SMM\,J141742.04+523025.7& 232$\pm$23&  19.7&19.4&  2.6$\pm$0.9&  0.661 & 1.2 & 33.7 & $M$; OII, MgII (SB) \\ %
SMM\,J141741.81+522823.0&  80$\pm$16&  23.0&21.5&  3.3$\pm$1.0&  1.150 & 1.7 & 31.1 & $M$; OII, FeII (SB) \\ %
SMM\,J141750.50+523101.0&  57$\pm$14&  25.1&25.0& 2.8$\pm$0.9&      2.128 & 5.8 & 42.1 & $M$; \lya, IS~abs (SB) \\ %
SMM\,J141800.50+512820.8& 128$\pm$19&  24.1&23.3 &  5.0$\pm$1.0&  1.913 & 11.1 & 43.2 & $M$; \lya, IS~abs (SB)\\ %
SMM\,J141802.87+523011.1& 39$\pm$14&  24.4&24.2& 3.4$\pm$0.9&      2.127 & 4.9 & 38.3 & $M$; IS~abs (SB) \\ %
SMM\,J141809.00+522803.8$^d$&  67$\pm$15&  25.8&25.7 & 4.3$\pm$1.0&  2.712 & 14.4 & 47.4 & $M$; \lya\ (SB) \\ %
SMM\,J141813.54+522923.4& 93$\pm$16 & 27.4&25.9 &3.6$\pm$1.1   & 3.484 & 39.1 & 68.5 & $MP$; \lya, CIV (AGN)\\ %
{} & & & & & & & & \\
SMM\,J163627.94+405811.2&  92$\pm$23&  25.2&24.9 &6.5$\pm$2.1&  3.180 & 31.1 & 52.1 & $M$; \lya, OI, CIV, \hal\ (AGN) \\ %
SMM\,J163631.47+405546.9$^d$&  99$\pm$23&  24.9&24.1&  6.3$\pm$1.9&  2.283 & 13.1 & 42.0 & $M$; \lya, CIV (AGN) \\ %
SMM\,J163639.01+405635.9& 159$\pm$27&  24.6&23.8&  5.1$\pm$1.4&  1.488 & 6.4 & 32.7 & $M$; CIV, HeII, CIII, FeII, \hal\ (SB) \\ %
SMM\,J163650.43+405734.5$^c$& 221$\pm$16&  23.3&22.5&  8.2$\pm$1.7&  2.384 & 32.8 & 49.9 & $M$; \lya, IS~abs, CIV, \hal, CO(3-2) (AGN/SB)\\ %
SMM\,J163658.19+410523.8&  92$\pm$16&  26.0&25.8& 10.7$\pm$2.0&  2.454 & 14.9 & 37.7 & $M$; \lya, \hal, CO(3-2) (SB)\\ %
SMM\,J163658.78+405728.1$^d$&  74$\pm$29&  23.9&22.7&  5.1$\pm$1.4&  1.190 & 1.7 & 27.5 & $M$; OII, MgII (SB)\\ %
SMM\,J163704.34+410530.3&  45$\pm$16&  24.2&23.1& 11.2$\pm$1.6&  0.840 & 0.40 & 16.5 & $M$; OII, OIII (SB)\\ %
SMM\,J163706.51+405313.8&  74$\pm$23&  n/a&24.6&   11.2$\pm$2.9&  2.375 & 10.9 & 34.3 & $P$; \lya, SiIV, CIV, \hal\ (AGN)\\ %
{} & & & & & & & & \\
SMM\,J221724.69+001242.1$^d$& 121.1$\pm$10.7&  20.4&20.0&   8.6$\pm$1.9& 0.510 & 0.25 & 15.9 & $M$; probable lens\\  %
SMM\,J221725.97+001238.9&  41.2$\pm$9.3&  $>27$&27.1& 17.4$\pm$2.9& 3.094 & 12.9 & 32.1 & \lya, IS~abs, CO(4-3) (SB) \\ %
SMM\,J221733.02+000906.0&  161.7$\pm$16.3&  24.5&23.8& 11.1$\pm$3.4& 0.926 & 1.9 & 26.4 & $MP$; FeII, H$\alpha$ (SB) \\ %
SMM\,J221733.12+001120.2&  69.2$\pm$10.3&  22.8&21.4&   6.9$\pm$2.1& 0.652 & 0.33 & 18.8 & $MP$; OII,MgII,FeII (SB)  \\ %
SMM\,J221733.91+001352.1&  44.5$\pm$13.4&  24.8&24.1& 9.1$\pm$1.1& 2.551 & 8.1 & 33.6 & $M$; IS~abs, \hal\ (SB) \\ %
SMM\,J221735.15+001537.2&  49.4$\pm$13.3&  26.1&25.7& 6.3$\pm$1.3& 3.098 & 15.3 & 43.6 & $M$; \lya, CO(3-2) (SB) \\  %
SMM\,J221735.84+001558.9&  44.3$\pm$12.8&  26.6&25.5&   4.9$\pm$1.3& 3.089 & 13.1 & 44.6 & $M$; \lya\ (SB) \\ %
SMM\,J221737.39+001025.1&  110.1$\pm$14.0&  25.1&24.6&  6.1$\pm$2.0& 2.614 & 21.4 & 44.0 & $MP$; \lya, IS~abs (SB) \\ %
SMM\,J221804.42+002154.4$^d$&  43.8$\pm$10.4&  25.2&24.7&   9.0$\pm$2.3&  2.517 & 7.6 & 33.1 & $P$; IS~abs (SB) \\ %
SMM\,J221806.77+001245.7&  241.5$\pm$11.2& 25.8&24.3&   8.4$\pm$2.3&  3.623 & 118.6 & 72.0 & $P$; \lya, OIII (SB) \\ %
\enddata
\label{tab3}

\begin{tabular}{l}
$^a$  These SMGs have double radio source identifications, both confirmed
      to lie at the same redshift.\\
$^b$  These SMGs have double radio source identifications, one lying at the
	tabulated redshift and a second lying at $z<0.5$.\\~In calculations
	and figures presented herein, we have inferred based on the
	radio luminosity that the high-redshift
	source is the\\~dominant contributor to the submm emission.\\
$^c$  These SMGs have double radio source identifications, however only one
      has a spectroscopic redshift.\\
$^d$  These SMGs have spectroscopically identified optical sources which are
offset from the radio source identifications\\~(typically  $\sim1$ arcsec),
although it is usually the case that the optical identification is 
extended and overlaps the radio source.\\~Chapman et al.\ (2004b) addresses
these issues through higher spatial resolution radio and optical imagery.\\
$^e$  The radio flux error estimates are based on the integrated radio flux, 
and do not always reflect the point source\\~detection sensitivity in the radio map.
\end{tabular}
\end{deluxetable}

\begin{deluxetable}{cc}
\renewcommand\baselinestretch{1.0}
\tablewidth{0pt}
\parskip=0.2cm
\tablenum{4}
\tablecaption{FIR luminosity function for SMGs -- $z=0.9$}
\small
\tablehead{
\colhead{$\log$(\fir)/$\lsun$} & {$\log$(N)}\\
{} & {(h$^{-3}$Mpc$^{-3}$/decade)} \\
}
\startdata
        11.4   &   $-$6.04$^{+0.16}_{-0.26}$ \\
        11.9   &   $-$5.84$^{+0.13}_{-0.19}$ \\
        12.4   &   $-$6.26$^{+0.20}_{-0.37}$ \\
        12.9   &   $-$6.74$^{+0.30}_{-1.70}$ \\
\enddata
\label{tab4}
\end{deluxetable}

\begin{deluxetable}{cc}
\renewcommand\baselinestretch{1.0}
\tablewidth{0pt}
\parskip=0.2cm
\tablenum{5}
\tablecaption{FIR luminosity function for SMGs -- $z=2.5$}
\small
\tablehead{
\colhead{$\log$(\fir/$\lsun$)} & {$\log$(N)}\\
{} & {(h$^{-3}$Mpc$^{-3}$/decade)} \\
}
\startdata
        12.5   &   $-$5.44$^{+0.09}_{-0.11}$ \\
        12.9   &   $-$5.29$^{+0.08}_{-0.09}$ \\
        13.3   &   $-$6.04$^{+0.16}_{-0.26}$ \\
        13.7   &   $-$6.74$^{+0.30}_{-1.70}$ \\
\enddata
\label{tab5}
\end{deluxetable}

\begin{deluxetable}{lcc}
\renewcommand\baselinestretch{1.0}
\tablewidth{0pt}
\parskip=0.2cm
\tablenum{6}
\tablecaption{radio-SMG Spectral Classification Statistics}
\small
\tablehead{
\colhead{class} & {number} & {percentage}\\
}
\startdata
\smallskip
{\bf successful spectroscopic IDs} & {\bf (73)} & \\
	UV-emission line & 38/56 & 68\% \\
	AGN lines (CIV etc.) & 18/73 & 25\% \\
	broad-line AGN & 3/73 & 4\% \\
\bigskip
	Identified primarily through Ly$\alpha$ & 25/73 & 34\% \\
{\bf total spectroscopic sample} & {\bf (98)} & \\
	Ly$\alpha$ emitting & 38/81 & 47\% \\
	AGN characteristics & 18/98 & 18\% \\
	Star Forming galaxies & 30/98 & 31\% \\
	Difficult to classify & 25/98 & 25\% \\
	Spectroscopically Unidentified & 25/98 & 26\% \\
\enddata
\label{tab5}
\end{deluxetable}


\begin{references}
\reference{}Adelberger, K., Steidel, C., 2000, ApJ, 544, 218

\reference{} Alexander, D., Smail, I., Bauer, F., Chapman, S.,
Blain, A., Ivison, R., 2005, Nature, submitted

\reference{alexander02}Alexander, D.,
Vignali, C., Bauer, F. E., Brandt, W. N., Hornschemeier, A. E., Garmire,
G. P., Schneider, D. P./ 2002, AJ, 123, 1149

\reference{}Almaini, O., et al., 2003, MNRAS, 338, 303

\reference{} Archibald, E., et al., 2002, MNRAS, 336, 353

\reference{}Aretxaga, I., et al., 2003, MNRAS, 342, 759
\reference{}Aretxaga, I., et al., 2004, MNRAS, in press

\reference{}Avni, Y., Bachall, J.\ N., 1980, ApJ, 235, 694

\reference{barger03}Barger, A. J., et al.\ 2003, \aj, 126, 632
\reference{barger02}Barger, A. J., et al.\ 2002, \aj, 124, 1839
\reference{barger01a}Barger, A. J., Cowie, L.  L., Mushotzky, R. F., \& Richards, E. A.\ 2001a, \aj, 121, 662

\reference{barger01b}Barger, A. J., Cowie, L. L., Steffen, A. T., Hornschemeier, A. E.,Brandt, W. N., \& Garmire, G. P.\ 2001b, \apj, 560, L23

\reference{bcr00}Barger, A. J., Cowie, L. L., \& Richards, E. A.\ 2000, \aj, 119, 2092

\reference{bcs99a}Barger, A. J., Cowie, L. L., \& Sanders, D. B.\ 1999a, \apj, 518, L5

\reference{barger99b}Barger, A. J., Cowie, L.L., Smail, I., Ivison, R. J.,Blain, A. W., \& Kneib, J.-P.\ 1999b, \aj, 117, 2656

\reference{barger98}Barger, A. J., et al.\ 1998, Nature, 394, 248

\reference{} Baugh, C. M., Lacey, C. G., Frenk, C. S., Granato, G. L.,
Silva, L., Bressan, A., Benson, A. J., Cole, S., MNRAS, submitted
(astro-ph/0406069)

\reference{}Bertoldi, F., et al., 2000, A\&A, 360, 92

\reference{blain04cluster}Blain, A., Chapman, S., Smail, I., Ivison, R.\ 2004a, ApJ, in press, astro-ph/0405035

\reference{blain04seds}Blain, A., Chapman, S., Smail, I., Ivison, R.\ 2004b, ApJ, in press, astro-ph/0404438

\reference{}Blain, A., Barnard, V., Chapman, S.\ 2003, MNRAS, 338, 733

\reference{blain02review}Blain, A., et al.\ 2002,  Phys.\ Rep., 369, 111

\reference{blain99}Blain, A.\ 1999, MNRAS, 309, 955 

\reference{blainetal99a}Blain, A. W., Smail, I., Ivison, R. J., \&
Kneib, J.-P.\ 1999a,\mnras, 302, 632

\reference{blainetal99b}Blain, A., et al., 1999b, MNRAS, 309, 715 

\reference{}Bolzonella, M., Miralles, J.-M., Pello, R., 2000, A\&A, 363, 476

\reference{}Boyle, B., et al., 2000, MNRAS, 317, 1014


\reference{borys03}Borys, C.,
Chapman, S. C., Halpern, M., \& Scott, D.\ 2003, \mnras, 344, 385


\reference{borys04}Borys, C., Chapman, S.\ C., et al.\ 2004, \mnras, in press 


\reference{}Capak P., et al., AJ, 2004, AJ, 127, 180

\reference{carilli01}Carilli, C. L., et al., 2001, AJ, 122, 1679 

\reference{cy00}Carilli, C. L. \& Yun, M.\ 2000, \apj, 539, 1024
\reference{cy99}Carilli, C. L. \& Yun, M.\ 1999, \apj, 513, 13L
(CY)

\reference{chapman00}Chapman, S.\ C., et al., 2000, MNRAS, 319, 318 

\reference{chapman01a}Chapman, S.\ C., Richards,
E.\ A., Lewis, G.\ F.,Wilson, G., \& Barger, A. J.\ 2001a, \apj, 551, L9

\reference{chapman01b}Chapman, S.\ C., et al.\ 2001b,
Deep millimeter surveys : implications for galaxy formation and evolution, Proceedings of the UMass/INAOE conference, University of Massachusetts, Amherst, MA, USA, 19-21 June 2000. Published by Singapore: World Scientific Publishing, 2001. xi, 207 p. Edited by James D. Lowenthal, and David H. Hughes.

\reference{chapman02a}Chapman, S.\ C., Scott, D., Borys, C., Fahlman, G.,
	2002a, MNRAS, 330, 92

\reference{chapman02b}Chapman, S.\ C., Lewis, G.\ F., Scott, D., Borys,
C., Richards, E.\ A.\ 2002b, \apj, 570, 557

\reference{chapman02c}Chapman S.\ C., Smail I., Ivison R., Blain A.,
	2002c, \mnras, 335, 17 

\reference{chapman02d}Chapman, S. C., Smail, I., Ivison, R.,
Helou, G., Dale, D., Lagache, G.\ 2002d, \apj, 573, 66 

\reference{}Chapman S.\ C.,  et al.\ 2003a, ApJ, 585, 57

\reference{C03}Chapman S.\ C., Blain A., Ivison R.,  Smail I.,
	2003b, Nature, 422, 695 (C03)

\reference{} Chapman, S.\ C., Helou, G., Lewis, G., Dale, D.\ 2003c, \apj, 588, 186

\reference{} Chapman, S.\ C., et al., 2004a, ApJ, in press 
\reference{} Chapman, S.\ C., Smail, I., Windhorst, R., Muxlow, T.,
	Ivison, R., 2004b, ApJ, in press

\reference{} Chary, R., Elbaz, D., 2001, ApJ, 556, 562

\reference{condon92}Condon, J. J.\ 1992, \araa, 30, 575


\reference{}Connolly, A., et al., 1997, ApJ, 486, L11

\reference{cowie04}Cowie, L. L., Barger, A. J., Fomalont, E.,
Capak, P., 2004, ApJ, 603, 69L

\reference{cowie02}Cowie, L. L., Barger, A. J., Kneib, J.-P.\ 2002, \aj, 123, 2197

\reference{} Croom, S.M., Boyle, B.J., Loaring, N., Miller, L.,
Outram, P.J., Shanks, T., Smith, R.J., 2002, MNRAS, 335, 459

\reference{} Croom, S.M., Smith, R.J., Boyle, B.J., Shanks, T., Miller, L.,
Outram, P.J., Loaring, N.S. 2004, MNRAS, 349, 1397 Silverman, J.D. et al.,
2004, ApJ, submitted

\reference{cuillandre00}Cuillandre, J.-C., Luppino, G., Starr, B.,
Isani, S.in ``Astronomical Telescopes and Instrumentation'',2000, Proc.
SPIE, 4008, 1010

\reference{dale01}Dale, D., et al.\ 2001, ApJ, 562, 142

\reference{dale02}Dale, D., Helou, G., 2002, ApJ, 576, 159

\reference{}Dannerbauer, H., Lehnert, M., Lutz, D., Tacconi, L.,
Bertoldi, F., Carilli, C., Genzel, R., Menten, K., 2002, ApJ, 573, 473

\reference{}Dannerbauer, H., et al., 2004, ApJ, 606, 664

\reference{} Dunlop, J., et al., 2004, MNRAS, 350, 769

\reference{dunne02}Dunne, L., et al.\ 2002, MNRAS, 327, 697
\reference{dunne00}Dunne, L., Clements, D., Eales S., 2000, MNRAS, 319, 813

\reference{eales03}Eales, S., Bertoldi, F., Ivison, R., Carilli, C.,
Dunne, L., Owen, F.\ 2003, MNRAS, 344, 169

\reference{eales00}Eales, S., Lilly, S., Webb, T., Dunne,
L., Gear, W.,Clements, D., \& Yun, M.\ 2000, \aj, 120, 2244

\reference{eales99}Eales, S., et al.,\ 1999, ApJ, 515, 518

\reference{}Egami, E., et al., 2004, ApJ, in press

\reference{fixsen98} Fixsen, D. J., Dwek, E., Mather, J.
C., Bennett, C. L., Shafer, R. A.\ 1998, \apj, 508, 123

\reference{flores99} Flores, H., et al., 1999, ApJ, 517, 148

\reference{} Fomalont, E., et  al., 2004, in preparation

\reference{} Franx, M., et al., 2003, ApJL, 587, 79
 
\reference{} Frayer, D., et al., 2004, ApJ, in press 
\reference{} Frayer, D., et al., 2003, AJ, 126, 73  
\reference{} Frayer, D., et al., 1999, ApJ, 514, 13L
\reference{} Frayer, D., et al., 1998, ApJ, 560, 7

\reference{} Garrett, M.,  2002, A\&A, 384, 19

\reference{} Gear, W., et al., 2000, MNRAS, 316, 51L
\reference{} Glazebrook, K., et al., 2004, Nature, 430, 181

\reference{} Honda, M., et al., 2004, ApJL, 610, 49

\reference{} Tacconi, L., et al., 2005, ApJ, submitted

\reference{}Greve, T., et al., 2004, MNRAS, in press, astro-ph/0405361

\reference{}Greve, T., et al., 2005, MNRAS, submitted

\reference{Giavalisco03} Giavalisco, M., Dickinson, M., Ferguson, H.\ C.,
Ravindranath, A., Kretchmer, C., Moustakas, L.\ A., et al., 2004, ApJL,
600, L103

\reference{haarsma01} Haarsma, D., Partridge, R. B., Richards, E. A.
\& Windhorst, R. A.\ 2000, ApJ, 544, 641

\reference{helou85}Helou, G., et al.\ 1985, ApJ 440, 35


\reference{holland99}Holland, W. S., et al.\ 1999, \mnras, 303, 659

\reference{}Hughes, D.\ H.,  et al.\ 2002, MNRAS, 335, 871

\reference{hughes98}Hughes, D.\ H., Serjeant, S., Dunlop, J.,
 Mann, R.\ G., Ivison, R., et al., 1998, Nature, 394, 241

\reference{} Hunt, M., et al., 2004, ApJ, 605, 625

\reference{ivison04}Ivison, R. J., et al., 2004, ApJ, in press
\reference{ivison02}Ivison, R. J., et al., 2002, \mnras, 337, 1
\reference{ivison00}Ivison, R. J., et al., 2001, \apj, 561, 45L
\reference{ivison00}Ivison, R. J., et al., 2000, \mnras, 315, 209 
\reference{ivison98}Ivison, R. J., et al., 1998, \mnras, 298, 583

\reference{ivison98}Ivison, R. J., Smail, I., Le Borgne, J.-F.,
 Blain, A. W., Kneib, J.-P., B{\'e}zecourt, J., Kerr, T.H.,\& Davies, J.K.\
 1998, \mnras, 298, 583

\reference{} Jenness, T., Lightfoot, J.\ F., Holland, W.\ S., 1998,
Proc. SPIE Vol. 3357, p. 548-558, Advanced Technology MMW, Radio, and Terahertz Telescopes, Thomas G. Phillips; Ed.

\reference{kennicutt} Kennicutt, R.\ C., 1998, ARA\&A, 36, 189

\reference{} Komiyama, Y., et al., 2003, SPIE, 4841, 152

\reference{} Kovacs, A., et al., 2004, ApJ, in preparation

\reference{} Kneib, J.-P., et al., 2004, MNRAS, 349, 1211

\reference{} Knudsen, K.\ K., 2004, PhD Thesis, U.\ Leiden

\reference{} Kreysa, E., Gemund, H.-P., Raccanelli, A.,
Reichertz, L. A., Siringo, G., 2002,
EXPERIMENTAL COSMOLOGY AT MILLIMETRE WAVELENGTHS: 2K1BC Workshop. Breuil-Cervinia, Valle d'Aosta, Italy, 9-13 July, 2001. Edited by Marco De Petri and Massimo Gervasi. American Institute of Physics, 2002. AIP Conference Proceedings, Volume 616, pp. 262-269

\reference{} Lacey, C., et al., 2004, in preparation

\reference{ledlow02} Ledlow, M., Smail, I., Owen, F.,
 Keel, W., Ivison, R., Morrison, G., 2002, \apj, 577, 79L

\reference{}Lehnert M., Bremer, M., 2003, ApJ, 593, 630L

\reference{}Lewis, G., Chapman, S., Helou, G., 2004, ApJ, submitted
\reference{}Lilly, S., et al., 1999, ApJ, 518, 641

\reference{}Lutz, D., et al., 2001, A\&A, 378, 70L
\reference{} Lutz, D., et al., 1999, ApSS, 266, 85

\reference{}Madau, P., 1995, ApJ, 441, 18
\reference{}Madau, P., et al.,  1996, MNRAS, 283, 1388

\reference{}
Magorrian, J.,  et al., 1998, AJ, 115, 2285
\reference{}
Martini, P., Weinberg, D., 2001, ApJ, 547, 12

\reference{} Meurer, G., et al., 1997, AJ, 114, 54
\reference{} Muller, G., Reed, R., Armandroff, T., Boroson, T., 
Jacoby, G., 1998, SPIE, 3355, 577

\reference{} Neri, R., et al., 2003, ApJ, 598, 39L

\reference{} Neufeld, D., 1991, ApJ, 370, 85L

\reference{oke95} Oke, J.B., et al.\ 1995, \pasp, 107, 375

\reference{} Omont, A., et al., 2003, A\&A, 398, 857 
\reference{} Omont, A., et al., 2001, A\&A, 374, 371 

\reference{}
Page, M., Stevens, J.A., Mittaz, J.P.D., Carrera, F.J.,
2001, Science, 294, 2516

\reference{} Peacock, J., et al., 2000, MNRAS, 318, 535

\reference{} Pettini, M., et al., 2001, ApJ, 554, 981

\reference{} Reddy, N., Steidel, C., 2004, ApJ, 603, 13L

\reference{richards00} Richards, E. A.\ 2000, \apj, 533, 611

\reference{} Efstathiou, A., Rowan-Robinson, M.,  2003, MNRAS, 343, 322


\reference{Scott02}Scott, S., et al.\ 2002, \mnras, 331, 817

\reference{}Serjeant, S., et al., 2003, MNRAS, 344, 887

\reference{}Shapley, A., et al., 2003, ApJ, 588, 65 

\reference{Shaver}Shaver, P., et al., 1998, APS Conf series 156,
Highly Redshifted Radio Lines,
ed. C.Carilli, S.Radford, K. Menten, G. Langston, (San Francisco: ASP163)

\reference{}Simcoe, R., et al., 2000, BAAS, 196, 5209
\reference{}Silverman, J.D., et al., 2004, ApJ, submitted, astro-ph/0406330

\reference{}Simpson, C., et al., 2004, MNRAS, in press 

\reference{smail04a}Smail, I., Chapman, S., Blain, A., Ivison, R.,
	2004a, ApJ, submitted 
\reference{smail04b}Smail, I., Chapman, S., Blain, A., Ivison, R.,
	2004b, ApJ, in preparation 

\reference{smail03a}Smail, I., et al, 2003a, MNRAS, 342, 1185 
\reference{smail03b}Smail, I., et al, 2003b, ApJ, 583, 551 

\reference{smail02}Smail, I., Ivison,
R. J., Blain, A. W., Kneib, J.-P., 2002, \mnras, 331, 495

\reference{smail00}Smail, I., Ivison, R. J., Owen, F. N., Blain, A. W.,
\& Kneib, J.-P.\ 2000, \apj, 528, 612

\reference{smail99}Smail, I., et al., 1999, MNRAS, 308, 1061

\reference{sib97}Smail, I., Ivison, R., Blain A., 1997, ApJ, 490, L5

\reference{} Spoon, H., et al., 2004, A\&A, 414, 873

\reference{} Steidel, C.,  et al.\ 2004, ApJ, 604, 534 
\reference{} Steidel, C.,  et al.\ 2003, ApJ, 592, 728 
\reference{} Steidel, C.,  et al.\ 2002, ApJ, 576, 653 
\reference{} Steidel, C.,  et al.\ 1999, ApJ, 519, 1 

\reference{} Stevens, J., et al., 2004, MNRAS, submitted 

\reference{}
 Sturm, et al., 2000, A\&A, 358, 481

\reference{} Sturm, et al., 2002, A\&A, 393, 821

\reference{} Swinbank, M., Smail, I., Chapman, S., Blain, A., Ivison, R.,
Keel, B., 2004, ApJ, submitted

\reference{} Tran, et al., 2001, ApJ, 552, 527

\reference{} Tulloch, S., 2000, INGN, 2, 26

\reference{} Verma, et al., 2003, A\&A, 403, 829

\reference{} Wang, W., Cowie, L., Barger, A., 2004, AJ, astro-ph/0406261

\reference{webb03a}Webb, T. M. A., Eales, S. A., Lilly, S. J., Clements,
D. L., Dunne, L.,Gear, W. K., Flores, H., \& Yun, M., 2003a, \apj, 587, 41

\reference{webb03b}Webb, T. M. A., et al., 2003b, ApJ, 582, 6 

\reference{} Wiklind, T., 2003, ApJ, 588, 736

\reference{} Wolfe, A., et al., 2003a, ApJ, 593, 235
\reference{} Wolfe, A., et al., 2003b, ApJ, 593, 215

\reference{yan99} Yan, L., et al., 1999, ApJ, 519, 47L

\reference{yun02}Yun, M., Carilli, C., 2002, ApJ, 568, 88

\reference{yun01} Yun, M., Reddy, N., Condon, J.\ 2001, ApJ, 554, 803

\end{references}
\end{document}